\documentclass[12pt]{article}

\usepackage{amsmath,amssymb,amsthm,amscd,mathrsfs}
\usepackage{latexsym,amsmath,amssymb,graphicx}
\usepackage[all]{xy}
\usepackage{float}
\usepackage{multirow}
\usepackage{color}
\usepackage{epsf}
\usepackage[vcentermath]{youngtab}
\usepackage{framed}
\xyoption{v2} \xyoption{2cell} \xyoption{dvips}
\CompileMatrices \numberwithin{equation}{section}

\numberwithin{equation}{section}
\makeatletter
\newcommand{\subalign}[1]{%
  \vcenter{%
    \Let@ \restore@math@cr \default@tag
    \baselineskip\fontdimen10 \scriptfont\tw@
    \advance\baselineskip\fontdimen12 \scriptfont\tw@
    \lineskip\thr@@\fontdimen8 \scriptfont\thr@@
    \lineskiplimit\lineskip
    \ialign{\hfil$\m@th\scriptstyle##$&$\m@th\scriptstyle{}##$\crcr
      #1\crcr
    }%
  }
}
\makeatother

\newcommand{\be}{\begin{equation}}
\newcommand{\ee}{\end{equation}}

\newcommand{\IP}{\mathbb{P}}

\newcommand\IZ{\mathbb {Z}}

\newcommand\Ione{\boldsymbol{1}}

\newcommand\IQ{\mathbb {Q}}
\newcommand{\IC}{\mathbb{C}}

\newcommand{\IR}{\mathbb{R}}
\newcommand{\CU}{{\mathcal U}}
\newcommand{\ba}{\begin{array}}
\newcommand{\ea}{\end{array}}

\newcommand{\wH}{{\widetilde H}}

\newcommand{\CS}{{\mathcal S}}

\newcommand{\CK}{{\mathcal K}}

\newcommand{\CW}{{\mathcal W}}
\newcommand{\IH}{{\mathbb H}}
\newcommand{\bal}{\begin{aligned}}
\newcommand{\eal}{\end{aligned}}

\newcommand{\ualpha}{{\underline \alpha}}

\newcommand{\longto}{\longrightarrow}

\newcommand{\ch}{{\mathrm{ch}}}

\newcommand{\CO}{{\mathcal O}}

\newcommand{\CE}{{\mathcal E}}

\newcommand{\CH}{{\mathcal H}}

\newcommand{\CM}{{\mathcal M}}

\newcommand{\CC}{{\mathcal C}}

\newcommand{\ubeta}{{{\underline \beta}}}
\newcommand{\um}{{\underline m}}

\newcommand{\CT}{{\mathcal T}}

\newcommand{\us}{{\underline s}}

\newcommand{\uQ}{{\underline Q}}

\newcommand{\sfA}{{\sf A}}
\newcommand{\sfB}{{\sf B}}
\newcommand{\sfC}{{\sf C}}

\newcommand{\sfS}{{\sf S}}
\newcommand{\sfR}{{\sf R}}
\newcommand{\sfQ}{{\sf Q}}
\newcommand{\sfT}{{\sf M}}

\newcommand{\ut}{{\underline t}}
\newcommand{\calP}{{\mathcal P}}

\newcommand{\IA}{{\mathbb A}}
\newcommand{\uxi}{{\underline \xi}}
\newcommand{\ux}{{\underline x}}

\CompileMatrices \LaTeXdiagrams \UseAllTwocells
\newdimen\tableauside\tableauside=1.0ex
\newdimen\tableaurule\tableaurule=0.4pt
\newdimen\tableaustep
\def\phantomhrule#1{\hbox{\vbox to0pt{\hrule height\tableaurule width#1\vss}}}
\def\phantomvrule#1{\vbox{\hbox to0pt{\vrule width\tableaurule height#1\hss}}}
\def\sqr{\vbox{%
  \phantomhrule\tableaustep
  \hbox{\phantomvrule\tableaustep\kern\tableaustep\phantomvrule\tableaustep}%
  \hbox{\vbox{\phantomhrule\tableauside}\kern-\tableaurule}}}
\def\squares#1{\hbox{\count0=#1\noindent\loop\sqr
  \advance\count0 by-1 \ifnum\count0>0\repeat}}
\def\tableau#1{\vcenter{\offinterlineskip
  \tableaustep=\tableauside\advance\tableaustep by-\tableaurule
  \kern\normallineskip\hbox
    {\kern\normallineskip\vbox
      {\gettableau#1 0 }%
     \kern\normallineskip\kern\tableaurule}%
  \kern\normallineskip\kern\tableaurule}}
\def\gettableau#1 {\ifnum#1=0\let\next=\null\else
  \squares{#1}\let\next=\gettableau\fi\next}
\newcommand{\Appendix}[1]{%
  \refstepcounter{section}%
  \addtocontents{toc}{\protect\setcounter{tocdepth}{1}}
  \addcontentsline{toc}{section}%
    {\bfseries\appendixname~\thesection:\ #1}%
    {\medskip\noindent \Large\bfseries\appendixname\ \thesection:\ #1}%
\sectionmark{#1}\smallskip\noindent
\renewcommand{\theequation}{{\bf 
{{\thesection}}.{\arabic{equation}}}}
}

\begin{document}

\title{BPS states, torus links and wild character varieties} 
\author{
  Duiliu-Emanuel Diaconescu \and Ron Donagi 
\and Tony Pantev}
\date{}
\maketitle

\begin{abstract} 
A string theoretic framework is constructed relating the cohomology of
wild character varieties to refined stable pair theory and torus link
invariants. Explicit conjectural formulas are derived for wild
character varieties with a unique irregular point on the projective
line. For this case the string theoretic construction leads to a conjectural colored
generalization of existing results of Hausel, Mereb and Wong as well
as Shende, Treumann and Zaslow.
\end{abstract}

\tableofcontents

\section{Introduction}
The main goal of this paper is to develop a string theoretic framework
for the cohomology of wild character varieties.  Previous such
constructions \cite{wallpairs,BPSPW,Par_ref} have been carried out for
regular and tamely ramified character varieties, leading to a physical
derivation of the main conjectures of Hausel and Rodiguez-Villegas
\cite{HRV}, respectively Hausel, Letellier and Rodiguez-Villegas
\cite{HLRV}.  Very briefly, using the $P=W$ conjecture of de Cataldo,
Hausel and Migliorini \cite{hodgechar}, the string theoretic approach
places these conjectures in the framework of motivic Donaldson-Thomas
theory developed by Kontsevich and Soibelman \cite{wallcrossing}. The
conjectural formulas of \cite{HRV,HLRV} are then identified in
\cite{BPSPW,Par_ref} with refined Gopakumar-Vafa expansions for
certain Calabi-Yau threefolds. An important part of this program,
namely the refined stable pair formula for local curves without marked
points, has been recently proven by Maulik \cite{HRV_proof}.

The main outcome of the present work is a conjectural generalization
of recent results of Hausel,
Mereb and Wong \cite{Arithmetic_wild} as well as Shende, Treumann and Zaslow \cite{Fukaya_knots} in the context of wild character varieties
with one singular point on the projective line. The string theoretic
construction employed in the process provides compelling evidence for
the wild variant of the $P=W$ conjecture of de Cataldo, Hausel, and
Migliorini \cite{hodgechar}.

For completeness, note that topological and motivic invariants of
moduli spaces of Higgs bundles and flat connections have been
intensively studied in the recent mathematical literature employing
different approaches. Arithmetic methods have been used
\cite{Ind_vb_higgs,Counting_Higgs,Higgs_ind_proj_line,Ind_obj,
  Counting_Hitchin_II,Counting_Hitchin_I},
leading to complete results for Poincar\'e polynomials of Higgs bundle
moduli spaces. Moreover, the motives of the moduli stacks of irregular
Higgs bundles, as well as irregular connections over arbitrary fields
have been recently computed in \cite{Mot_conn_higgs}.  An alternative
approach based on wallcrossing for moduli spaces of linear chains on
curves was developed in \cite{Mot_chains}, and used in
\cite{y_genus_higgs} to compute the Hirzebruch genus of moduli spaces
of $PGL(r,\IC)$ Higgs bundles. Finally, a different class of character
varieties defined using Zariski closures of conjugacy classes at the
marked points was studied in \cite{Zariski_closures}. It is not clear
at the moment if there is any conceptual relation between these
results and the physical approach developed here. This remains an
important open question for future research.

\subsection{Wild character varieties}\label{wildintro} 
In this paper wild character varieties will be moduli spaces of Stokes
data associated to singular $G$-connections on curves, where $G$ is a
complex reductive group.  Such moduli spaces were used by Witten
\cite{GL_wild} for wildly ramified geometric Langlands correspondence,
and were constructed as multiplicative symplectic quotients by Boalch
in \cite{Braiding_stokes}. To set up the stage, note that according to
\cite{Braiding_stokes}, an irregular curve consists of a smooth
projective curve $C$, a finite set of marked points on $C$, and an
irregular type assigned to each marked point. An irregular type at a
point $p\in C$ is an equivalence class of ${\bf t}$-valued meromorphic
function germs at $p$ modulo holomorphic terms, where ${\bf t}$ is a
fixed Cartan subalgebra of $G$.  Given a local coordinate $z$ on $C$
centered at $p$, an irregular type $\sfQ$ admits a representative
\[ 
Q = \sum_{k=1}^{n-1} {A_k\over z^k}, \qquad A_k \in {\bf t}
\]
for some $n\in \IZ$, $n\geq 2$.  Throughout this paper the group $G$
will be $GL(r, \IC)$ and there will be only one marked point $p\in C$,
although some of the results easily generalize to several marked
points. The common centralizer of all Cartan elements $A_k$, $1\leq k
\leq n-1$, is independent of the choice of representative and will be
denoted by $H_{\sfQ}\subset GL(r,\IC)$. Since all $A_k$ are diagonal
matrices, $H_{\sf Q}$ will be conjugation equivalent to a canonical
subgroup of the form $\prod_{i=1}^\ell GL(m_i, \IC)\subset GL(r, \IC)$
for some ordered partition $r= m_1 + \ldots+m_\ell$, $\ell\geq
1$. Without loss of generality, as explained in \cite[Remark
  10.6]{Braiding_stokes} it can be assumed that $H_\sfQ$ is a subgroup
of this form.

Given an irregular curve $(C,p,\sfQ)$, the construction of {Boalch}
\cite{Braiding_stokes} produces a smooth quasi-projective variety
parameterizing Stokes data of flat singular $GL(r,\IC)$-connections on
$C$ which are locally gauge equivalent to
\be\label{eq:irregconn}
dQ +
{\rm terms\ of\ order}\ \geq -1
\ee
at $p$. This is a holomorphic
Poisson manifold. In order to obtain a holomorphic symplectic variety,
in { Boalch's} construction one also has to fix the conjugacy class of
the formal monodromy at $p$ in the centralizer $H_{\sf Q}$, in
addition to the singular type $\sfQ$. A very detailed and explicit
discussion of formal monodromy can be found in
\cite[Sect. 2.2]{GL_wild}.  Very briefly, it may be helpful to recall
that given an irregular connection as above there is an $r$
dimensional vector space of formal solutions to the flatness equations
in the infinitesimal neighborhood of $p$, i.e. formal power series
solutions. The formal monodromy at $p$ is the monodromy transformation
acting on the space of formal flat sections. Using the same local
trivialization as in \eqref{eq:irregconn} the formal monodromy is
identified with a group element of $GL(r,\IC)$ which belongs to
centralizer $H_{\sf Q}$.  The following conditions on the data
$(\sfQ,\sfT)$ will be imposed from this point on throughout the paper.
\begin{itemize} 
\item[$(i)$] The common centralizer $H_\sfQ=\prod_{i=1}^\ell GL(m_i,
  \IC)\subset GL(r, \IC)$ of the coefficients $A_{n-1}, \ldots, A_1$
  in the Laurent expansion of $Q$ will be assumed to be the same as
  the centralizer of the leading term $A_{n-1}$.
\item[$(ii)$] The formal monodromy at $p$ will be conjugation
  equivalent in $H_{\sf Q}$ to a block diagonal matrix ${\sfT}$ of the
  form
  \be\label{eq:formalmon}
  \sfT = \left(\begin{array}{cccc} \tau_1
    {\bf 1}_{m_1} & 0 & \cdots & 0 \\ 0 & \tau_{2} {\bf 1}_{m_2} &
    \cdots & 0 \\ \vdots & \vdots & \cdots & \vdots \\ 0 & 0 & \cdots
    & \tau_\ell {\bf 1}_{m_\ell},\\
\end{array}\right)
  \ee
  where $\tau_1, \ldots, \tau_\ell$ are pairwise distinct complex
numbers and ${\bf 1}_m$ denotes the $m\times m$ identity matrix.
\end{itemize} 
 
\noindent 
As shown in Section \ref{wildhodge}, both conditions are natural
consequences of the main geometric construction used in this paper.

For future reference, given a collection of positive integers
$\um=(m_1, \ldots, m_\ell)$, let $P_+(\um)\subset GL(r,\IC)$,
$P_-(\um)\subset GL(r,\IC)$ be the subgroup of upper, respectively
lower block diagonal matrices with respect to the ordered partition
$r=m_1+\cdots + m_\ell$. Furthermore let $U_\pm (\um) \subset
P_\pm(\um)$ be the subgroups of matrices with diagonal blocks
$(\Ione_{m_1}, \ldots, \Ione_{m_\ell})$.

The moduli space of Stokes data is constructed in
\cite{Braiding_stokes} by quasi-Hamiltonian reduction. A synthetic
presentation can be found in the proof of
\cite[Thm. 8.2]{Braiding_stokes}, equation (37), as well as in the
review paper \cite{Poisson_Riemann}.  For regular centralizer $H_\sfQ$
these varieties were first constructed in \cite{Quasi_hamiltonian}.  A
very brief summary is provided below assuming conditions $(i)$, $(ii)$
above.

Choosing a base point in $C\setminus\{p\}$, each irregular connection
as above determines monodromy data $(\sfA_l,\sfB_l)\in
(GL(r,\IC)\times GL(r,\IC))^{\times g}$, a group element $\sfC\in
GL(r,\IC)$, and the Stokes matrices $\sfS_\pm^k\in U_\pm(\um)$, $1\leq
k \leq n-1$. Let $\CU_{\sfQ,\sfT}$ be the closed subvariety of all
data $(\sfA_l, \sfB_l, \sfC, \sfS_+^k, \sfS_-^k)$ satisfying the
algebraic equation \be\label{eq:stokeseq} \prod_{l=1}^g \big(\sfA_l
\sfB_l \sfA_l^{-1} \sfB_l^{-1} \big)\sfC^{-1}\sfT \sf S^{n-1}_+
\sfS^{n-1}_- \cdots \sfS^1_+\sfS^1_- \sfC = {\bf 1}_r.  \ee The
variety of Stokes data is the affine algebraic quotient
$\CS_{\sfQ,\sfT} = \CU_{\sfQ,\sfT}/GL(r,\IC) \times H_\sfQ$, where the
$GL(r,\IC)\times H_\sfQ$ action on $\CU_{\sfQ,\sfT}$ is given by
\[
(g,h) \times (\sfA_l, \sfB_l, \sfC, \sfS_+^k, \sfS_-^k)\mapsto
(g\sfA_lg^{-1}, g\sfB_lg^{-1}, h\sfC g^{-1}, h\sfS_+^kh^{-1},
h\sfS_-^kh^{-1}).
\]

Given an irregular type $\sfQ$, for fixed sufficiently generic $\sfT$
as in \eqref{eq:formalmon}, the quotient $\CS_{\sfQ,\sfT}$ is a smooth
quasiprojective variety equipped with a holomorphic symplectic
structure.
According to \cite[Remark 9.12]{Braiding_stokes}, the complex
dimension of $\CS_{\sfQ,\sfT}$ is \be\label{eq:dimformulaA} {\rm
  dim}\, \CS_{\sfQ,\sfT} = 2r^2(g-1)+n(r^2-\sum_{i=1}^\ell m_i^2) +2.
\ee For the regular case, $m_1=\cdots = m_\ell=1$, this formula was
also derived in \cite[Thm. 2.2.13]{Arithmetic_wild}.  In particular
note that the result depends only on the unordered partition $\mu$ of
$r$ determined by the multiplicities $(m_1, \ldots, m_\ell)$, the
genus $g$ of the curve and the order $n$ of the pole at $p$.  For
future reference, given any partition $\mu=(\mu_1, \ldots, \mu_\ell)$ of $r\geq 1$ and any integers $n\geq 1$, $g\geq 0$ let
\be\label{eq:dformula}
d(\mu, n,g)= 2r^2(g-1)+n(r^2-\sum_{i=1}^\ell
\mu_i^2) +2.
\ee

To conclude this brief outline, it is important to note that moduli
spaces of irregular filtered flat connections are related by
hyper-K\"ahler rotations to moduli spaces of irregular Higgs bundles
on $C$, where the Higgs field has an order $n$ pole at $p$.  This
statement, known as the wild nonabelian Hodge correspondence follows
from the results of Sabbah \cite{Harmonic_metrics} and Biquard and
Boalch \cite{Wild_curves}. The first reference establishes the
correspondence between algebraic connections and solutions to Hitchin
equations while the second proves the correspondence between solutions
to Hitchin equations and Higgs bundles and constructs hyper-K\"ahler
metrics on moduli spaces.  The first part of this correspondence  \cite{Harmonic_metrics} has been generalized to higher 
dimensional 
situations by Mochizuki \cite{Wild_harmonic}.

A very clear and explicit account of wild nonabelian Hodge
correspondence can be found in \cite{HK_wild,Wild_merom}, which will serve as our
main references for the summary in Section \ref{wildhodge}. In
particular, note that the resulting Higgs fields have fixed Laurent
tail at $p$, which is determined by the data $(\sfQ,\sfT)$ up to local
isomorphisms. Moreover, one also obtains a quasi-parabolic structure
on the reduced point $p$, which is preserved by the Higgs field, and a
set of parabolic weights determined by the formal monodromy $\sfT$.

An alternative construction for moduli spaces of irregular parabolic
Higgs bundles is presented in Section \ref{higgssection}.  Inspired by
previous work of Saito and Inaba \cite{Moduli_connections_III} and
Inaba \cite{Moduli_connections_IV}, this construction employs
meromorphic Higgs bundles on $C$ with parabolic structure of type
$\um=(m_1, \ldots, m_\ell)$ along the non-reduced divisor $D=np$.  The
Laurent part of the Higgs field is encoded in the a collection
$\uxi=(\xi_1, \ldots, \xi_\ell)$ of sections of the coefficient line
bundle $M = K_C(D)$ over $D$. In addition one has to specify parabolic
weights $\ualpha = (\alpha_1, \ldots, \alpha_\ell)$ and impose a
natural stability condition. This is explained in detail in Section
\ref{setup}. The resulting moduli stack of semistable $\uxi$-parabolic
Higgs bundles will be denoted by ${\mathfrak H}_\uxi(C,D;\ualpha, \um,
d)$, where $d\in \IZ$ is the degree of the Higgs bundles.  Although a
priori different from \cite{HK_wild}, the two constructions are in
fact equivalent for sufficiently generic local data $\uxi$, as shown
in Section \ref{wildhodge}. The construction used in the present paper
facilitates the connection to string theory and enumerative
geometry. Note that very similar Higgs bundle moduli spaces are used
by Oblomkov and Yun \cite{Geom_reps, Coh_ring} for geometric
constructions of representations of Cherednik algebras.

Finally, note that the wild nonabelian Hodge correspondence leads to
the $P=W$ conjecture of de Cataldo, Hausel and Migliorini
\cite{hodgechar}. The main claim of this conjecture is that the weight
filtration on the cohomology of the character variety is identified
with the perverse Leray filtration on the cohomology of the
corresponding Hitchin system. The latter is constructed using relative
Hodge theory \cite{Hodge_maps} for the Hitchin map. This conjecture
was proven in \cite{hodgechar} for rank two Hitchin systems on curves
without marked points. As in \cite{BPSPW,Par_ref} this identification
plays a central role in the string theoretic approach to the
cohomology of wild character varieties.

\subsection{The conjecture of Hausel, Mereb and Wong}
\label{HMWconj} 

Let $\sfQ$ be an irregular type and let ${\sfT}$ be a sufficiently
generic diagonal matrix as in equation \eqref{eq:formalmon}. The
cohomology of the smooth quasi-projective variety $\CS_{\sfQ,\sfT}$
carries a weight filtration according to \cite{TH_II,TH_III}. This
yields a weighted Poincar\'e polynomial
\be\label{eq:WPoincare}
WP(\CS_{\sfQ,\sfT}; u,v) = \sum_{i,j} {\rm dim}\, Gr^W_i
H^{j}(\CS_{\sfQ,\sfT}) u^{i/2} v^j.
\ee
The conjecture of Hausel,
Mereb and Wong \cite{Arithmetic_wild} provides explicit formulas for
all these polynomials assuming the centralizer $H_{\sfQ}$ is the
standard maximal torus i.e. $\ell=r$ and $m_1= \cdots = m_\ell=1$.  In
particular $\sfT$ is regular.  The main statement will be reviewed
below for a single marked point.

One first constructs the generating function (the {\em HMW partition function})
\[ 
Z_{HMW}(z,w) = 1+\sum_{|\lambda|>0} \Omega_{\lambda}^{g,n}(z,w)
{\wH}_\lambda({\sf x}; z^2, w^2)
\]
where:
\begin{itemize}
\item the sum in the right hand side is over all Young diagrams
  $\lambda$ with a positive number of boxes $|\lambda|>0$, and
\item 
for each such $\lambda$
\[ 
\Omega_\lambda^{g,n} = \prod_{\Box \in \lambda} {
  (-z^{2a(\Box)}w^{2l(\Box)})^{n-1} (z^{2a(\Box)+1} -
  w^{2l(\Box)+1})^{2g} \over (z^{2a(\Box)+2}
  -w^{2l(\Box)})(z^{2a(\Box)} -w^{2l(\Box)+2})},
\] 
\item while ${\wH}_\lambda({\sf x}; z^2,w^2)$ is the modified
  Macdonald polynomial in the infinite set of variables ${\sf x} =
  (x_1, x_2, \ldots)$.
\end{itemize}
Next define ${\mathbb H}_{\mu,n}(z,w)$ by
\be\label{eq:HMWa} {\rm
  ln}\, Z_{HMW}(z,w) = \sum_{k\geq 1}\sum_{\mu} {(-1)^{(n-1)|\mu|}
  w^{kd(\mu,n,g)}\, {\mathbb H}_{\mu,n}(z^k,w^k)\over
  (1-z^{2k})(w^{2k}-1) } m_\mu({\sf x}^k)
\ee
where the sum is again
over all Young diagrams, $m_{\mu}({\sf x})$ are the monomial symmetric
functions and ${\sf x}^k = (x_1^k, x_2^k, \ldots)$.  The exponent
$d(\mu,n,g)$ is defined in equation \eqref{eq:dformula}.  Then,
assuming $\ell=r$ and $m_1= \cdots = m_\ell=1$, one has the following
conjectural formula
\be\label{eq:HMWb}
\boxed{WP(\CS_{\sfQ,\sfT}; u,v) =
{\mathbb H}_{(1^r),n} (u^{1/2}, -u^{-1/2}v^{-1})}
\ee
for any $r, n
\geq 1$ and any $\sfQ$.  The $v=1$ specialization of this conjecture is
proven in \cite[Thm. 1.1]{Arithmetic_wild} using arithmetic methods.
In this specialization the weighted Poincar\'e polynomial reduces to
the $E$-polynomial.


\subsection{The formula of Shende, Treumann and Zaslow} 

A different formula for the $E$-polynomial of wild character varieties
follows from the main result of of Shende, Treumann and Zaslow
\cite{Fukaya_knots}, using subsequent work of Shende, Treumann,
Williams and Zaslow \cite{Cluster_legendrian}. Specializing
\cite[Thm. 13]{Fukaya_knots} to the present context, one obtains an
explicit formula for the $E$-polynomial of wild character varieties
$\CS_{\sfQ,\sfT}$ on the projective line under the same assumptions as
in the previous subsection. Namely $\ell=r$ and  $m_1=\cdots=m_\ell=1$ while
the formal mondromy ${\sf M}$ is assumed to be regular.  The
$E$-polynomial is then related to the leading term in the expansion of the
HOMLY polynomial of the $(r,r(n-2))$ torus link with a specific
normalization.  More precisely, let $P_{(r,r(n-2))}(a,u)$ be the
HOMFLY polynomial of this link using the normalization in which the
HOMFLY polynomial of the unknot is 1. Note that $P_{(r,r(n-2))}(a,u)$
is a Laurent polynomial in $a$ with coefficients in the field of
rational functions $\IQ(u^{1/2})$.  Let $P_{(r,r(n-2))}^{(0)}(u)$ be
the coefficient of $a^0$ in
\[
(a u^{-1/2})^{1-(r-1)((n-2)r-1)} P_{(r,r(n-2))}(a,u). 
\]
Then, according to \cite{Fukaya_knots} one has
\be\label{eq:STZa}
\boxed{WP(\CS_{\sfQ,\sfT},u,-1)=(1-u)^{-r}P_{(r,r(n-2))}^{(0)}(u).}
\ee
Note that this formula is obtained from \cite[Thm. 1.13]{Fukaya_knots}
using a particular construction \cite{Cluster_legendrian} of character
varieties as moduli spaces of lagrangian cycles in the cotangent space
$T^*\Delta$ of the disc with fixed boundary conditions. The boundary
conditions require the lagrangian cycles to end on a legendrian link
isotopic to the $(r,r(n-2))$ torus link which occurs in the above
formula. Furthermore note that the link in question differs by a full
twist from the Stokes link associated to the irregular singular point.
By definition, the latter is a link in the boundary of $T^*\Delta$
which encodes the jumping behavior of local flat sections along Stokes
lines \cite{Stokes}. For a modern treatment, the reader is referred to
\cite[Sect. 3.3]{Cluster_legendrian}. In our case the Stokes
link is isotopic to the $(r,r(n-1))$ torus link. The origin of the
full twist in this construction is explained in detail in
\cite[Prop. 6.5]{Fukaya_knots} and
\cite[Prop. 6.6]{Cluster_legendrian}.

As shown below, the present paper offers a string theoretic derivation
of formula \eqref{eq:STZa} using the nonabelian Hodge correspondence and
spectral data for irregular Higgs bundles. The string theory
perspective leads to a conjectural colored generalization formulated
in Section \ref{mainconj}, where one allows arbitrary values for the
multiplicities $m_1, \ldots, m_\ell$.

\subsection{Spectral correspondence and a Calabi-Yau threefold}
\label{spcorrespCY}

Following the strategy of \cite{wallpairs,BPSPW,Par_ref} the plan is
to construct a Calabi-Yau threefold equipped with a natural projection
map to $C$ such that the moduli space of supersymmetric D2-D0 brane
configurations on the threefold is related to an irregular parabolic
Hitchin system on $C$ via a spectral construction. Given such a
construction, the perverse Betti numbers of the Hitchin system are
identified with degeneracies of spinning BPS states in M-theory as in
\cite{GV_II}. The latter are in turn determined via the refined
Gopakumar-Vafa expansion \cite{GV_II,ref_vert} by counting D6-D2-D0
bound states on the same threefold.  Mathematically such bound states
are counted by the stable pair invariants constructed by Pandharipande
and Thomas in \cite{stabpairsI} and refined by Kontsevich and
Soibelman \cite{wallcrossing}.

The idea of the construction is based on an approach to spectral data
using holomorphic symplectic surfaces due to Kontsevich and Soibelman
\cite{structures}. Taking this construction to its logical conclusion,
the new result proven in this paper establishes an isomorphism of
moduli stacks between semis-stable torsion sheaves on a holomorphic surface
and semi-stable irregular parabolic Higgs bundles on $C$. Note that a similar
result was first proven by S. Szabo in \cite{Birational_irreg} for an
open dense subset of the moduli space of stable irregular Higgs bundles. The
present construction applies to the whole moduli space, which is
needed in order to study its global topology.

The input data for the irregular spectral construction consists of the
marked curve $(C,p)$, the order of pole $n$, and the collection of
sections $\uxi=(\xi_1, \ldots, \xi_\ell)$ of the coefficient line
bundle $M=K_C(D)$ over the nonreduced subscheme $D=np\subset C$. From
this point on it will be assumed that these sections take pairwise
distinct, nonzero values $\xi_1(p), \ldots, \xi_\ell(p)$ at the
reduced point $p$.  Such section data $\uxi$ will be called generic.
Then, as shown in Section \ref{surface}, a holomorphic symplectic
surface $S_\uxi$ is constructed by blowing up the total space of $M$
along the images of the sections $\xi_1, \ldots, \xi_\ell$ and then
removing a divisor in the anticanonical linear system.  The linear
equivalence classes of compact divisors on $S_\uxi$ are in one-to-one
correspondence with collections of positive integers $\um=(m_1,
\ldots, m_\ell)$. Any compact curve $\Sigma_\um$ belonging to such a
linear system is a finite cover of $C$ of degree $\sum_{i=1}^\ell
m_i$.  Moreover for any collection of real numbers $(\beta_1, \ldots,
\beta_\ell)$ there is a compactly supported $B$-field $\beta\in
H_c^2(S_\uxi,\IR)$ such that
\[
\beta(\Sigma_{\um}) = n\sum_{i=1}^\ell m_i \beta_i. 
\] 
for any divisor $\Sigma_\um$ in a given linear system $\um$.  As explained in detail 
in Section \ref{sheavestohiggs}, this data determines a Bridgeland
stability condition for pure dimension one sheaves on $S_\uxi$ 
with determinant $\Sigma_\um$ and Euler characteristic $c\in \IZ$. 
The moduli stack of semistable pure dimension one sheaves will be denoted by 
${\mathfrak M}_\beta^{ss}(S_\uxi; \um, c)$.

In this context, the main result of Section \ref{spectralsect} is an
identification of the moduli stack ${\mathfrak M}_\beta^{ss}(S_\uxi;
\um, c)$ of semistable pure dimension one sheaves on $S_\uxi$ and the
moduli stack of semistable irregular parabolic Higgs bundles on
$C$. More precisely, the identification is with the moduli stack
${\mathfrak H}^{ss}_{\uxi}(C,D;\ualpha, \um,d)$ of semistable
irregular $\uxi$-parabolic Higgs bundles with fixed numerical data
$(\um,d)$, with $d := c+r(g-1)$.  This stack is defined at the end of
section \ref{setup}.  The result is:

\bigskip

\noindent
{\bf Spectral Correspondence}.  {\it Let $\uxi=(\xi_1, \ldots,
  \xi_\ell)$ be an arbitrary collection of generic sections of $M_D$
  over $D$ i.e. $\xi_1(p), \ldots, \xi_\ell(p)$ are pairwise distinct
  and all different from zero.  Let $\ualpha=(\alpha_1,\ldots
  \alpha_\ell)$ be a collection of real numbers such that $1>\alpha_1
  >\cdots > \alpha_\ell>0$. Then for any $\um = (m_1, \ldots,
  m_\ell)\in (\IZ_{>0})^{\times \ell}$ and any $c\in \IZ$ there is an
  isomorphism of stacks
 \be\label{eq:spectralisom}
{\mathfrak M}^{ss}_{\beta}(S_\uxi; \um,c)\simeq 
{\mathfrak H}^{ss}_{\uxi}(C,D;\ualpha, \um, c+r(g-1)).
\ee
where $\beta\in H_c^2(S_\uxi,\IR)$ is a flat $B$-field such that 
\[
\beta(\Sigma_{\um}) = n\sum_{i=1}^\ell m_i \alpha_i,
\]
$g$ is the genus of $C$ and $r=\sum_{i=1}^\ell m_i$. }

\

\bigskip

\noindent
Since this result represents the technical backbone of the paper, a
detailed proof is given in Sections \ref{sheavestohiggs},
\ref{inverse} and \ref{isomoduli}.  It should be noted that this
identification readily generalizes to several marked points on $C$.

Finally, the Calabi-Yau threefold $Y_\uxi$ is the total space of the
canonical bundle $K_{S_\uxi}$, which is isomorphic to the product
$S_\uxi \times \IA^1$. As explained in Section \ref{threefold}, each
pair $\beta, \um$ determines a $B$-field, respectively a compact curve
class on $Y_\uxi$ in a natural way.  Then is  immediate to show that
there an isomorphism
\be\label{eq:stackisomX}
{\mathfrak M}^{s}_\beta(Y_\uxi; \um, c)\simeq 
{\mathfrak M}^{s}_\beta(S_\uxi; \um, c)\times \IA^1.  
\ee
of moduli stacks of stable pure dimension one sheaves. 
Therefore the  spectral correspondence identifies the moduli stack of 
 of stable pure dimension one sheaves on $Y_\uxi$ to the 
moduli stack of stable irregular parabolic Higgs bundles on $C$ up to 
an extra $\IA^1$ factor. 
\bigskip
\bigskip

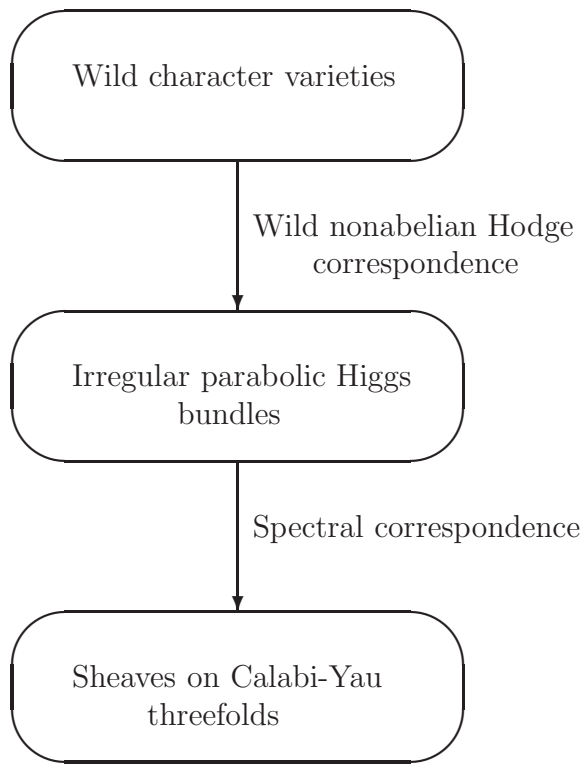
\begin{figure}[H]
\hspace{30pt}
\vspace{-10pt}
\setlength{\unitlength}{1mm}
\begin{picture}(200,100)
\thicklines
\put(70,95){\oval(60,20)}
\put(48,95){Wild character varieties}
\put(70,85){\vector(0,-1){20}}
\put(72,75){Wild nonabelian Hodge}
\put(80,70){correspondence}
\put(70,55){\oval(60,20)}
\put(48,55){Irregular parabolic Higgs}
\put(62,50){bundles}
\put(70,45){\vector(0,-1){20}}
\put(72,35){Spectral correspondence}
\put(70,15){\oval(60,20)}
\put(48,15){Sheaves on Calabi-Yau}
\put(58,10){threefolds}
\end{picture}
\caption{From wild character varieties to sheaves on threefolds.}
\label{wild_sheaves}
\end{figure}

In conclusion, the steps taken so far yield a correspondence between
wild character varieties and sheaves on Calabi-Yau threefolds
summarized in Figure \ref{wild_sheaves}. By analogy with
\cite{BPSPW,Par_ref}, this picture leads to an explicit relation
between the weighted Poincar\'e polynomials \eqref{eq:WPoincare} and
the stable pair theory of $Y_\uxi$ via Gopakumar-Vafa expansion
conjectured in \cite{GV_II,spinBH,Mtop,ref_vert}. A very detailed
physical derivation of the unrefined expansion was recently given in
\cite{GV_details}. In order to complete this program, one needs explicit formulas for the
refined stable pair invariants of the threefolds $Y_\uxi$, which is a
challenging task as discussed in the next section.

\subsection{Refined stable pair theory via torus links}\label{reflinks}
Applying the theory of Pandharipande and Thomas \cite{stabpairsI} to
the present context, a stable pair on $Y_\uxi$ consists of a compactly
supported pure dimension one sheaf $F$ on $Y_\uxi$ equipped with a
generically surjective section $s:\CO_{Y_\uxi}\to F$. These objects
form a quasiprojective moduli space equipped with a perfect
obstruction theory. Note that for a generic stable pair the support of
$F$ is a compact space curve in $Y_\uxi \simeq S_\uxi \times \IA^1$
which projects to a finite set in $\IA^1$. In contrast, the support of
a stable pure dimension one sheaf on $Y_\uxi$ projects to a single
point in $\IA^1$, as shown in equation \eqref{eq:stackisomX}. In
particular the moduli space of stable pairs on $Y_\uxi$ does not
factor as in \eqref{eq:stackisomX}.

From a physical point view  a stable pair is a supersymmetric D2-D0
configuration bound to a D6-brane.  Refined stable pair invariants
count degeneracies of such BPS states taking into account the four
dimensional spin quantum number.  Mathematically, these refined
invariants were constructed by Kontsevich and Soibelman
\cite{wallcrossing}.  The generating function for such invariants is
\be\label{eq:genfctA} Z_{Y_\uxi}(q,Q_1,\ldots, Q_\ell,y) = 1+
\sum_{\substack{c,\um\\ \um\neq(0,\ldots, 0)\\}} (-q)^c
\prod_{i=1}^\ell Q_i^{m_i}\, PT_{Y_\uxi}(m_1, \ldots, m_\ell, c; y)
\ee
where $m_1, \ldots, m_\ell$ are non-negative integers, not all
zero, encoding the curve class $\ch_1(F)$ of a D2-D0 configuration and
$c=\chi(F)$ is the D0-charge.
 
One of the main outcomes of this paper is an explicit conjectural
formula for the generating function \eqref{eq:genfctA} for genus zero
curves $C\simeq\IP^1$ with a single marked point. In
this case, the refined stable pair formula is derived in Section
\ref{stpairsect} using a compilation of mathematical conjectures and
string theoretic methods.

As explained in Sections \ref{toractsect} and \ref{stpairloc} the
first step in this derivation consists of localization with respect to
a torus action on $Y_\uxi$ preserving the holomorphic threeform.  In
more detail, there is a torus action on the total space of $M$ lifting
the natural action on $C=\IP^1$ so that the fiber $M_p$ is pointwise
fixed. Assuming the sections $\uxi=(\xi_1, \ldots, \xi_\ell)$
equivariant, this yields a torus action on the surface $S_\uxi$, which
then lifts canonically to $Y_\uxi$ imposing the condition that the
canonical class of $Y_\uxi$ be equivariantly trivial.  Using this
torus action, the stable pair theory localizes to a collection of
sections $\Sigma_1, \ldots \Sigma_\ell$ of $S_\uxi$ over $C$, all
passing through a torus fixed point ${\bf o}\in S_\uxi$.  As a divisor
on $S_\uxi$, the curve $\Sigma_1+\ldots+\Sigma_\ell$ has a singularity
of type
\be\label{eq:planesing} \prod_{i=1}^{\ell} (v-\lambda_i
w^{n-2}) =0.
\ee
where $(v,w)$ are local affine coordinates centered
at ${\bf o}$.  A theoretical framework for localization in refined
stable pair theory has been developed by Nekrasov and Okounkov
\cite{Membranes_Sheaves}, Maulik \cite{HRV_proof}, and more recently
Y. Jiang \cite{Non_arch_sheaves}. In particular, for toric threefolds,
the formalism \cite{Membranes_Sheaves} provides a mathematical theory
for the refined vertex of Iqbal, Kozcaz and Vafa \cite{ref_vert}.
This framework is reviewed and applied to the present setup in Section
\ref{stpairloc}.

General localization arguments show that the generating function 
for refined stable pair invariants admits a vertex presentation of the form
\be\label{eq:PTYa} Z_{Y_\uxi}(q,\uQ,y) = \sum_{\mu_1, \ldots,
  \mu_\ell} V^{(n)}_{\mu_1, \ldots, \mu_\ell}(q,y) \prod_{i=1}^\ell
Z_{\mu_i}(q,y) \prod_{i=1}^\ell Q_i^{|\mu_i|},
\ee
where $\uQ=(Q_1, \ldots, Q_\ell)$ are degree counting variables
associated to the curve classes $\Sigma_1, \ldots, \linebreak \Sigma_\ell$ and
$\mu_1, \ldots, \mu_\ell$ are Young diagrams. The total number of
boxes contained in such a diagram $\mu$ is denoted by $|\mu|$.  In
this formula $V^{(n)}_{\mu_1, \ldots, \mu_\ell}(q,y)$ is a
multileg  refined
vertex associated to the plane curve singularity \eqref{eq:planesing}
while $Z_{\mu_i}(q,y)$ are refined one-leg vertex factors.  The direct
localization computation of the multileg vertex $V_{\mu_1, \ldots,
  \mu_\ell}(q,y)$ turns out to be a very difficult problem.
Nevertheless, an explicit formula can be derived from the conjectures
of Oblomkov and Shende \cite{OS} and Oblomkov, Shende and Rassmusen
\cite{ORS}, which provide an enumerative geometric construction for
knot and link invariants associated to plane curve singularities.  In
string theory these conjectures have been shown to follow from large
$N$ duality for conifold transitions in \cite{DSV,DHS}. The physical
derivation leads to a colored refined generalization of these
conjectures formulated in \cite{DHS} and proven by Maulik in
\cite{Homfly_pairs} for the unrefined case.

\begin{figure}[H]
\hspace{50pt}
\vspace{0pt}
\setlength{\unitlength}{1mm}
\begin{picture}(200,180)
\thicklines
\put(70,170){\oval(60,20)}
\put(48,170){Wild character varieties}
\put(70,160){\vector(0,-1){20}}
\put(72,150){Wild nonabelian Hodge}
\put(80,145){correspondence}
\put(70,130){\oval(60,20)}
\put(48,130){Irregular parabolic Higgs}
\put(62,125){bundles}
\put(70,120){\vector(0,-1){20}}
\put(72,110){Spectral correspondence}
\put(70,90){\oval(60,20)}
\put(48,90){Sheaves on Calabi-Yau}
\put(58,85){threefolds}
\put(70,80){\vector(0,-1){20}}
\put(72,70){Gopakumar-Vafa expansion}
\put(70,50){\oval(60,20)}
\put(46,50){Stable pairs on Calabi-Yau}
\put(60,45){threefolds}
\put(70,40){\vector(0,-1){20}}
\put(72,30){Oblomkov-Shende conjecture}
\put(70,10){\oval(60,20)}
\put(58,10){Torus links}
\end{picture}
\caption{The main steps in the string theoretic approach to wild character varieties.}
\label{Chartflow}
\end{figure}
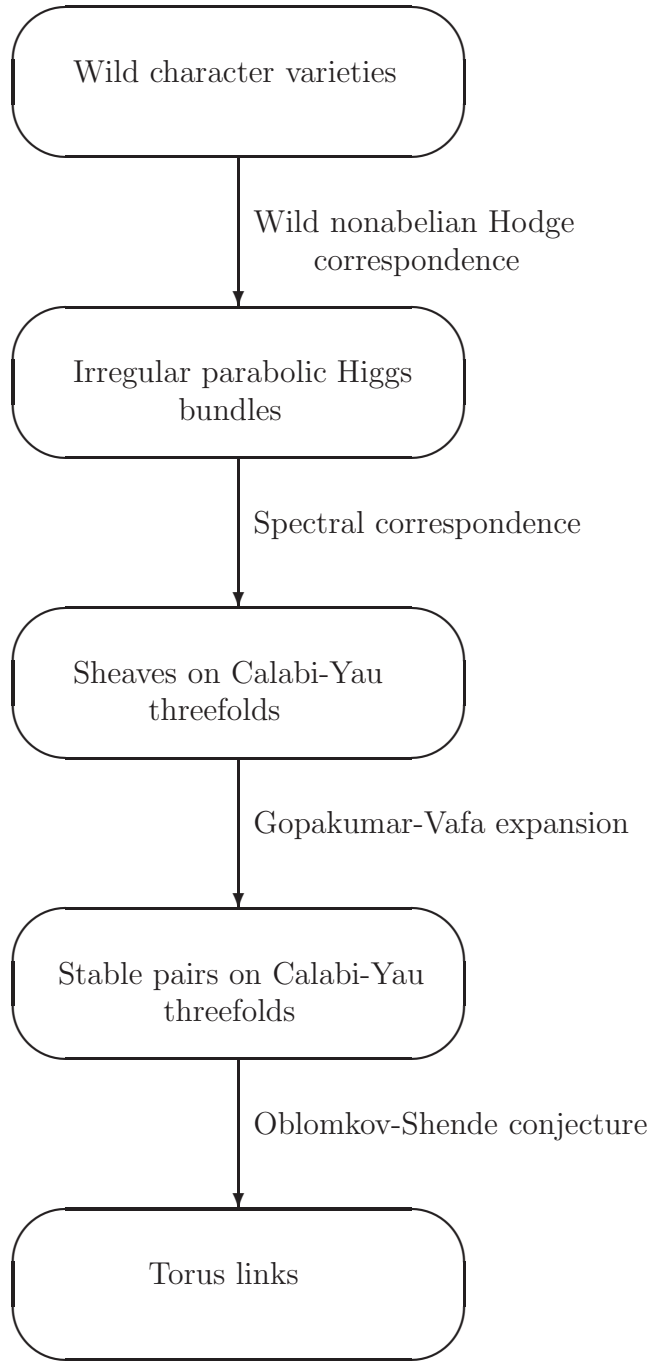

In the present context, these conjectures relate the refined stable
pair theory of $Y_\uxi$ to refined colored invariants of $(\ell,
(n-2)\ell)$-torus links. The latter can be computed in turn from the
refined Chern-Simons theory constructed by Aganagic and Shakirov
\cite{refCS} or using the constructions of Gorsky and Negut
\cite{Ref_knots_Hilb}, respectively Cherednik and Danilenko
\cite{DAHA_links}.  The approach employed in Section \ref{torlinks}
uses refined Chern-Simons theory as in \cite{Colored_HL} and some
additional large $N$ duality input. There is however a close parallel
with the formalism of \cite{Ref_knots_Hilb} as explained in Section 3
of loc. cit. Note that some explicit formulas for colored refined
invariants of certain torus links can be also found in
\cite{Link_ref_vert,Super_evol,Volume_refined,Seq_BPS}.  Moreover some
aspects of large N duality for refined torus knots and the associated
BPS states have been recently studied by Kameyama and Nawata in
\cite{Ref_N_torus}.

Finally, as explained in Section \ref{stpairloc}, coupling the refined torus
link invariants with the one leg vertex factors $Z_{\mu_i}(q,y)$ is
carried out using the refined vertex formalism constructed by Awata
and Kanno \cite{Md_vert_I,Md_vert_II}. Based on Macdonald polynomials
as opposed to Schur functions, this formalism is also manifest in the
work of Iqbal and Kozcaz \cite{Hopf_revisited}.
Collecting all the above facts, the main steps in the string theoretic
approach to the cohomology of wild character varieties are shown in
Figure \ref{Chartflow}.  
The  resulting conjectures on the cohomology
of wild character varieties are formulated below.

\subsection{The main conjectures}\label{mainconj}
 Throughout this
section it will be assumed that $C$ has genus zero and the sections
$\uxi$ are generic and equivariant under the torus action defined in
Section \ref{reflinks}.

To fix notation, for any
Young diagram $\lambda$ let  $P_{\lambda}(s,t;{\sf x})$,
with ${\sf x}=(x_1, x_2, \ldots)$, 
denote the $(s,t)$-Macdonald polynomials. 
For any triplet of Young diagrams $(\nu,\lambda, \sigma)$ let 
 $N^\sigma_{\nu, \lambda}$ denote
$(s,t)$-Littlewood-Richardson coefficients, which satisfy the 
fusion rules 
\[
P_{\nu}(s,t;{\sf x})P_{\lambda}(s,t;{\sf x})
=\sum_\sigma N^\sigma_{\nu, \lambda} P_{\sigma}(s,t;{\sf x}). 
\]
Finally, let
\[ 
f_\lambda(s,t) = \prod_{\Box\in \lambda} s^{a(\Box)} t^{-l(\Box)},
\]
be the refined framing factors defined in \cite{ref_vert} and 
set $\ut=(t^{1/2}, t^{3/2}, \ldots)$, $\us=(s^{1/2}, s^{3/2}, \ldots)$.

Then the first conjecture is:

\
 
\bigskip

\

\noindent
    {\bf Conjecture 1.}
    {\it The refined stable pairs theory of $Y_\uxi$
is given by the following formula 
\be\label{eq:PTYc}
\boxed{
{\small
\bal
& Z_{Y_\uxi}(q,\uQ,y) = 
1+ \\
& + \hspace{-2.1pc}
\sum_{\substack{\mu_1, \ldots, \mu_\ell\\ (|\mu_1|, \ldots, |\mu_\ell|)\neq (0,0,\ldots, 0)}}
\left(
{\widetilde W}^{(n-2)}_{\mu_1, \ldots, \mu_\ell}(s,t) 
\prod_{i=1}^\ell \left( Q_i^{|\mu_i|}
  f_{\mu_i}(s,t)^{n-1} P_{\mu_i^t}(t,s;\us) 
\right)\right)\bigg|_{\subalign{s&=qy\\ t&=qy^{-1}}}\\
\eal}}
\ee
where
\[
{\small
\bal
& {\widetilde W}^{(n-2)}_{\mu_1, \ldots, \mu_\ell}(s,t)
=
\sum_{\lambda_1, \ldots, \lambda_{\ell-1}} N^{\lambda_{\ell-1}}_{\mu_\ell, \lambda_{\ell-2}}
N^{\lambda_{\ell-2}}_{\mu_{\ell-1}, \lambda_{\ell-3}}\cdots N^{\lambda_2}_{\mu_3, \lambda_1} 
N^{\lambda_1}_{\mu_2, \mu_1} f_{\lambda_{\ell-1}}(s,t)^{2-n} P_{\lambda_{\ell-1}}(s,t;\ut).\\
\eal}
\]}

\

\smallskip

\noindent
The next conjecture summarizes the main points of the refined Gopakumar-Vafa 
expansion in our setting.

\

\bigskip

\

\noindent
{\bf Conjecture 2.} 

{\it 
$(i)$ 
The refined stable pair partition function in 
equation \eqref{eq:PTYc} has an expansion of the form 
\be\label{eq:refBPSa}
\boxed{
\bal
& {\rm ln}\, 
Z_{Y_\uxi}(q,\uQ,y) = \\
& -
\sum_{k\geq 1} \sum_{|\mu|\neq 0} {m_\mu(Q_1^k, \ldots , Q_\ell^k,0,\ldots)
\over k}
 {y^{-kr}
(qy^{-1})^{kd({\mu,n,0})/2} P_{\mu,n}((qy)^{-k}, -y^k)) \over 
 (1-(qy)^{-k})(1-(qy^{-1})^{k})},\\
 \eal
 }
\ee
where the sum is over all Young diagrams and $P_{\mu,n}(u,v)$ are
polynomials of $(u,v)$ with integer coefficients. Recall that
$m_\mu(x_1,x_2, \ldots)$ denotes the monomial symmetric function
function corresponding to the Young diagram $\mu$ and $d(\mu,n,g)$ is
defined in equation \eqref{eq:dformula}.

\

$(ii)$ For sufficiently generic parabolic weights $\ualpha=(\alpha_1,
\ldots, \alpha_\ell)$, and any degree $d\in \IZ$, the polynomial
$P_{\mu,n}(u,v)$ is equal to the perverse Poincar\'e polynomial
polynomials of the moduli space $\CH^s_\uxi(C,D;\ualpha, d, \um)$ of
stable irregular $\uxi$-parabolic Higgs bundles, where $\mu$ is the
unordered partition of $r$ determined by the sequence $(m_1, \ldots,
m_\ell)$.  In particular the perverse Poincar\'e polynomial of the
moduli space is independent of $(\ualpha, d)$ and of the sections
$\uxi$. }

\

\bigskip

\

Finally, using the wild non-abelian Hodge correspondence and the $P=W$
conjecture, one obtains a numerical conjecture for the weighted
Poincar\'e polynomials of $WP(\CS_{\sfQ,\sfT}; u,v)$ of wild character
varieties.  Namely note that given the generic torus invariant sections
$\uxi=(\xi_1, \ldots, \xi_\ell)$ any collection $\um=(m_1, \ldots,
m_\ell)$ of positive integers determines naturally an irregular type
$\sfQ(\uxi,\um)$ with centralizer $H_{\sfQ(\uxi,\um)} =
\prod_{i=1}^\ell GL(m_i, \IC)\subset GL(r, \IC)$. Then one has:

\

\bigskip

\

\noindent
{\bf Conjecture 3.} {\it Let $\sfT$ be a sufficiently generic diagonal
  $r\times r$ matrix of as in \eqref{eq:formalmon} and let $\mu$ be
  the partition of $r$ determined by $(m_1, \ldots, m_\ell)$. Then
\be\label{eq:Wformula}
\boxed{WP(\CS_{\sfQ(\uxi,\um),\sfT}; u,v) = P_{\mu,n}(u, v),}
\ee
where  $P_{\mu,n}(u,v)$ is determined by equation \eqref{eq:refBPSa}. }

\

\bigskip

\

Note that this conjecture applies to a larger class of wild character
varieties obtained from $\CS_{\sfQ(\uxi,\um),\sfT}$ by admissible
deformations, as defined in \cite{Braiding_stokes}.

Numerical evidence for Conjectures 2 and 3 is provided in Appendix
\ref{examples} and Section \ref{lochiggs}. Appendix \ref{examples}
summarizes some explicit predictions of formula \eqref{eq:refBPSa} for
rank two and three examples. For all examples with $\sfT$ regular, the
results are in agreement with the formula of Hausel, Mereb and Wong
\eqref{eq:HMWb}, as well as the formula of Shende, Treumann and Zaslow
\eqref{eq:STZa}.  Moreover, Section \ref{lochiggs} presents some
direct localization computations of Poincar\'e polynomials of rank
three irregular Higgs bundle moduli spaces with $\um=(2,1)$. In
particular the results obtained in Section \ref{Pexamples} for $n=5$
and $n=6$ are in agreement with the formulas in Appendix
\ref{rkthreeB}.  More precisely, the following statement holds:

\

\bigskip

\

\noindent
{\it Under the current assumptions, let $P(n,\ualpha,-1,(2,1); v)$ be
  the Poincar\'e polynomial of the moduli space
  $\CH^s_\uxi(C,np;\ualpha,-1,(2,1))$. Then the following identities
  hold for sufficiently small generic parabolic weights
  $0<\alpha_2<\alpha_1<<1$
\[
P(n,\ualpha,-1,(2,1); v) = P_{(2,1),n}(1,v), 
\]
where $n\in \{5,6\}$. }

\

\bigskip

To put Conjecture 2 in the proper perspective note that the
computations leading to this result are fairly technical, and the
complexity increases rapidly for higher rank Higgs bundles.  Moreover,
there is no known localization theorem for the perverse Poincar\'e
polynomial, which makes the direct computation of such invariants very
difficult.

\bigskip

\

\noindent
{\bf Acknowledgments.} We are very grateful to Yan Soibelman for
sharing his insights with us during the completion of this work, in
particular for pointing out the spectral construction of
\cite{structures}. We owe special thanks to Philip Boalch and Tamas
Hausel for illuminating discussions and comments on the manuscript and
Vivek Shende for very helpful explanations on the results of
\cite{Fukaya_knots, Cluster_legendrian}.  We would also like to thank
Davesh Maulik, Greg Moore, Andrei Negut, Alexei Oblomkov and Carlos
Simpson for very helpful discussions and correspondence.  The work of
Duiliu-Emanuel Diaconescu was partially supported by NSF grant
DMS-1501612.  During the preparation of this work Ron Donagi was
supported in part by NSF grant DMS 1603526 and by Simons HMS
Collaboration grant \# 390287 and Tony Pantev was supported in part by
NSF grant DMS 1601438 and by Simons HMS
Collaboration grant grant \# 347070.

\section{Irregular parabolic Higgs bundles on curves}
\label{higgssection}

This section introduces the moduli spaces of irregular parabolic Higgs
to be studied in this paper together with some basics on deformation
theory and wild non-abelian Hodge correspondence for such objects. 
Some relevant technical results are provided in Appendix 
\ref{technostuff} for completeness. 

\subsection{Setup and moduli spaces}\label{setup} 

Let $C$ be a smooth complex projective curve, $p\in C$ a fixed point
and $n\in \IZ$ a fixed positive integer such that $n \geq 3$ if $C$ is
of genus zero and $n\geq 1$ otherwise.

Let $D=np$ be the nonreduced divisor supported at $p$ with
multiplicity $n$ and $M=K_C(D)$.  For any vector bundle $F$ on $C$ let
$F_D = F\otimes_C \CO_D$ and, similarly, for any sheaf morphism
$\phi:F\to G$ between two vector bundles let $\phi_D: F_D \to G_D$
denote the restriction to $D$.  Moreover, for any two vector bundles
$F,G$ on $C$ there is a canonical isomorphism $(F\otimes_C G)_D \simeq
F_D\otimes_D G_D$ which will be implicitly used throughout this paper.

Given this setup, an irregular parabolic Higgs bundle on $C$ with a
pole of order $n$ at $p$ will consist of the following data
\begin{itemize} 
\item A pair $(E,\Phi)$  with $E$ a vector bundle on $C$ and $\Phi:
  E \to E\otimes_D K_C(D)$ a sheaf morphism.
\item A filtration 
\[ 
0=E_D^0 \subset E_D^{1} \subset \cdots E_D^{\ell-1} \subset E_D^\ell=E_D 
\]
of $E_D = E\otimes_C \CO_D$ by  locally free $\CO_D$-modules. 
(Such modules are automatically saturated, i.e. the quotient sheaves 
$E_D/E_D^i$, $0\leq i \leq \ell$, are also locally free 
$\CO_D$-modules.) 
This filtration is required to be 
preserved by the Higgs field $\Phi$, that is 
$\Phi_D(E_D^i)\subseteq E_D^i \otimes_D M_D$ for all $0\leq i \leq \ell$. 
Note that $\ell$ could be smaller than the rank $r$ of the vector bundle $E$, 
so the filtration need not necessarily be full: 
the successive quotient sheaves  $E_D^i/E_D^{i-1}$ could be vector bundles, 
rather than line bundles, over $\CO_D$.
\item A collection of parabolic weights 
$\ualpha=(\alpha_1, \ldots, \alpha_\ell)\in (0,\ 1)^\ell$
such that 
\[ 
\alpha_1>\alpha_2 > \cdots > \alpha_\ell.
\]
\end{itemize}
An irregular parabolic Higgs bundle on $C$ with a pole of order $n$ at $p$, 
i.e. a collection of data as above, 
will be denoted by $\CE=(E,\Phi, E^\bullet_D,\ualpha)$. 
For future reference note that the 
successive quotients of the above filtration fit in exact sequences of sheaves on $C$ of the form 
\be\label{eq:succquotA} 
0 \to E_D^i/E_D^{i-1} \to E_D/E_D^{i-1} \to E_D/E_D^{i} \to 0, \qquad 
1\leq i \leq \ell. 
\ee
In particular all successive quotients are locally free
$\CO_D$-modules as well. Moreover, since $\Phi$ preserves the
filtration $E_D^\bullet$, it induces morphisms of $\CO_D$-modules
\[
{\rm gr}_i{\Phi}_{D} : E_D^i/E_D^{i-1}\to E_D^i/E_D^{i-1}\otimes_D
M_D, \qquad 1\leq i \leq \ell.
\]

An extra condition will be imposed throughout this paper fixing the
polar part $\Phi_D$ of the Higgs field. Namely, let $\xi_1, \ldots,
\xi_\ell\in H^0(D, M_D)$ be arbitrary sections of the coefficient line
bundle $M= K_C(D)$ over the nonreduced divisor $D$.  Then the extra
condition requires the induced morphisms ${\Phi}_{D,i}$ to be of the
form
\be\label{eq:polarHiggsA} { \Phi}_{D,i} = \Ione_{E_D^i/E_D^{i-1}}
\otimes \xi_i, \qquad 1\leq i \leq \ell.
\ee
An irregular parabolic
Higgs bundle satisfying the above condition will be called
$\uxi$-parabolic, where $\uxi=(\xi_1, \ldots, \xi_\ell)$.

In order to construct well behaved moduli spaces, one needs a notion
of stability for irregular parabolic Higgs bundles.  There is a
natural stability condition for such objects defined in
\cite{Moduli_parabolic}, which will be employed in this paper. This stability condition has been used in a similar context in
\cite{Moduli_connections_III, Moduli_connections_IV}.

First note that the numerical invariants of an irregular parabolic
Higgs bundle $\CE$ are the flag type $\um=(m_1, \ldots, m_\ell)\in
\left(\IZ_{\geq 0}\right)^\ell$ where $m_i$ is the length of the
quotient $E_D^i/E_D^{i-1}$, $1\leq i\leq \ell$, as an $\CO_D$-module,
and the degree $d={\rm deg}(E)$. Note also that $\sum_{i=1}^\ell
m_i=r$, the rank of $E$.  Assuming $r>0$, the parabolic slope of $\CE$
is defined (in terms of the weights $\alpha_i$) by:
\[ 
\mu_{\rm par}(\CE) = {1\over r} \left(d + \sum_{i=1}^\ell 
\chi(E_D^i/E_D^{i-1}) 
\alpha_i \right)
\]
where $\chi(E_D^i/E_D^{i-1})= nm_i$ is the Euler characteristic of
$E_D^i/E_D^{i-1}$ viewed as a torsion sheaf on $C$.

The test subobjects for the stability condition will be nontrivial
saturated proper subsheaves $0\subset F \subset E$ preserved by
$\Phi$. For each such subsheaf there is an induced filtration $F_D^i
:= F_D\cap E_D^i$, $0\leq i \leq \ell$, of $F_D := F \otimes_C \CO_D$
which is preserved by $\Phi_D$.  Then (semi)-stability is defined by
\be\label{eq:stabcondA} {1 \over {\rm rk}(F)} \left({\rm deg}(F) +
\sum_{i=1}^\ell \chi(F_D^i/F_D^{i-11})\alpha_i\right) \ (\leq)
\ \mu_{\rm par}(\CE).
\ee
Again, here $\chi(F_D^i/F_D^{i-1})$ denotes
the Euler characteristic of $F_D^i/F_D^{i-1}$ as a sheaf on $C$.  Note
that this stability condition is independent on the fixed local data
$\uxi=(\xi_1, \ldots, \xi_\ell)$.

The moduli stack of semistable irregular $\uxi$-parabolic Higgs
bundles with fixed numerical data $(\um,d)$ will be denoted by
${\mathfrak H}^{ss}_{\uxi}(C,D;\ualpha, \um,d)$. This is an algebraic
stack of finite type.  The substack parameterizing stable objects is a
$\IC^\times$-gerbe over a quasi-projective coarse moduli space as
usual.

\subsection{Deformation theory}\label{deftheory}

The usual deformation theory considerations identify the complex
controlling the deformations of an $\uxi$-parabolic Higgs bundle with
its endomorphism complex shifted by one.  The local study carried out
in \cite[Sect. 3]{Dir_images} shows that in the regular case the
endomorphism complex of any parabolic (i.e.  not necessarily
$\uxi$-parabolic) Higgs bundle is a modification of the endomorphism
Dolbeault complex of the Higgs bundle. In this modification the terms
of the Dolbeault complex are modified successively by the decreasing
even steps of the monodromy weight filtration for the nilpotent part
of the residue of the Higgs field. For a $\uxi$-parabolic Higgs bundle
the associated graded with respect to the parabolic filtration of the
nilpotent part of the residue is actually zero. In this case the
$0$-th step of the weight filtration gives the parabolic endomorphisms
of the underlying parabolic bundle while the $(-2)$-nd step of the
weight filtration gives the strongly parabolic endomorphisms. This
reproduces the deformation complex of regular $\uxi$-parabolic
Higgs bundles derived in \cite{infparhiggs} and also
\cite{rankthreepar,Par_DS}. Moreover we expect that when the local
data $\uxi$ is sufficiently general so that the term in the Laurent
expansion is invertible, the analysis of \cite[Sect. 3]{Dir_images}
carries over without modification to the irregular setting. Assuming
this to be the case, the main steps in the construction will be
briefly explained below.

In the framework of Section \ref{setup}, let 
$E^\bullet=(E, E_D^\bullet,\ualpha)$, $F^\bullet=(F, F_D^\bullet,\ualpha)$ be two
bundles on $C$ equipped with parabolic structure on the 
non-reduced divisor $D=np$.  The filtrations $E_D^\bullet$, $F_D^\bullet$ are assumed of the same 
length $\ell\geq 1 $. (For simplicity we also take them to be of the same weights, though this could be generalized.)

Given an arbitrary open subset $U\subset C$ containing $p$, 
a local morphism 
$f_U: E|_U\to F|_U$ is called {\bf parabolic} if 
\[
f_U(E_D^i) \subseteq F_D^{i}, 
\]
and {\bf strongly parabolic} if 
\[
f_U(E_D^i) \subseteq F_D^{i-1},
\]
for all $1\leq i\leq \ell$.

Using these conditions, one constructs a sheaf  $PHom(E^\bullet, F^\bullet)$ of 
parabolic morphisms and a sheaf $SPHom(E^\bullet, F^\bullet)$
of 
strongly parabolic morphisms from 
$E^\bullet$ to $F^\bullet$. Note that both sheaves are locally free, and fit in exact sequences of 
$\CO_C$-modules of the form
\be\label{eq:parhomseq} 
0 \to PHom(E^\bullet, F^\bullet) \to Hom(E,F) \to 
APHom(E_D^\bullet, F_D^\bullet) \to 0
\ee
\be\label{eq:Sparhomseq} 
0 \to SPHom(E^\bullet, F^\bullet) \to Hom(E,F) \to 
ASPHom(E_D^\bullet, F_D^\bullet) \to 0
\ee
where 
\[
APHom(E_D^\bullet, F_D^\bullet) \simeq 
Hom_D(E_D^\bullet, F_D^\bullet)/ PHom_D(E_D^\bullet, F_D^\bullet)
\]
\[
ASPHom(E_D^\bullet, F_D^\bullet) \simeq 
Hom_D(E_D^\bullet, F_D^\bullet)/ SPHom_D(E_D^\bullet, F_D^\bullet).
\]

Furthermore, given a parabolic bundle $E^\bullet$ and a line bundle $L$, the tensor product $E\otimes_C L$ has a 
natural parabolic structure with filtration $E_D^i\otimes_D L_D$ and the same parabolic weights as $E^\bullet$. 
If the successive quotients 
$E_D^i/E_D^{i-1}$ are locally free on $D$ one has the following duality isomorphism \cite{infparhiggs,rankthreepar}. 
\be\label{eq:pardual} 
SPHom(E^\bullet, F^\bullet)^\vee \simeq PHom(F^\bullet, E^\bullet \otimes_C \CO_C(D)).
\ee

The infinitesimal deformation complex ${\mathcal D}(\CE)$ of an irregular $\uxi$-parabolic Higgs bundle $\CE=(E, E_D^\bullet, \Phi, \ualpha)$ is the two term complex 
\be\label{eq:defcpxA} 
\xymatrix{
PHom_C(E^\bullet, E^\bullet) 
\ar[r]^-{[\Phi,\ ]}&  SPHom_C(E^\bullet, M\otimes E^\bullet)}
\ee
of amplitude $[0,\ 1]$. As usual, it is straightforward to show that 
\[
\IH^0({\mathcal D}(\CE)) \simeq \IC. 
\]
for any stable object $\CE$. 
Moreover, since all successive quotients 
$E_D^i/E_D^{i-1}$ are locally free on $D$, using the isomorphism \eqref{eq:pardual} and Serre duality, one also obtains: 
\[
\IH^2({\mathcal D}(\CE)) \simeq \IH^0({\mathcal D}(\CE))^\vee \simeq \IC.
\]
This implies that the moduli space is smooth and its tangent 
space at the point $[\CE]$ is isomorphic to $\IH^1({\mathcal D}(\CE))$. 
Using the exact sequences \eqref{eq:parhomseq}, \eqref{eq:Sparhomseq}, 
the dimension of the moduli space is 
\be\label{eq:dimformulaB} 
2+ 2r^2(g-1) +n\big(r^2-\sum_{i=1}^\ell m_i^2\big)
\ee 
for numerical invariants $\um=(m_1, \ldots, m_\ell)$. 
Note the obvious similarity with equation \eqref{eq:dimformulaA}. In fact, this is not a coincidence, since irregular parabolic Higgs bundles are related to irregular connections by wild non-abelian Hodge correspondence, as explained next. 

\subsection{Wild non-abelian Hodge correspondence}\label{wildhodge} 

The main goal of this section is to apply the results of wild non-abelian Hodge 
theory proven by Sabbah \cite{Harmonic_metrics} and Biquard and
Boalch \cite{Wild_curves}  to the moduli spaces of irregular Higgs bundles introduced in Section \ref{setup}. 
This will yield a relation between these moduli spaces and certain moduli spaces of filtered irregular flat connections, 
which are in turn related to wild character varieties. 
The presentation will closely follow \cite{HK_wild}. 

As in Section \ref{setup} the geometric setup consists of a 
smooth projective curve $C$ with a marked point $p\in C$. 
For a fixed positive integer $n\geq 1$ let $D=np$ and 
$M = K_C(D)$. 
The numerical invariants of irregular 
parabolic Higgs bundles on $C$ are the degree $d\in \IZ$ and a collection $\um=(m_1, \ldots, m_\ell)\in \big(\IZ_{\geq 1}\big)^{\times \ell}$ encoding the flag type 
over the non-reduced divisor $D$. In the construction of the moduli space, one also chooses a collection 
$\uxi=(\xi_1, \ldots, \xi_\ell)$ of sections of $M_D$ over $D$ and imposes conditions \eqref{eq:polarHiggsA} on the 
polar part of the Higgs field. 
In this section, the sections $(\xi_1, \ldots, \xi_\ell)$ will be chosen so that 
\be\label{eq:genlocsectA}
\xi_i|_p \neq 0|_p, \quad 1\leq i \leq \ell, \quad 
{\rm and}\quad 
\xi_i|_p \neq \xi_j|_p, \quad 1\leq i,j \leq \ell, \ i\neq j.
\ee
Here $\xi|_p:p \to M_p$ is the restriction of a section 
$\xi \in H^0(D, M_D)$ to the reduced closed point $p\in D$, and $0 \in H^0(D,M)$ is the zero section. 
Such local sections will be called generic. 

As shown in Appendix \ref{loctriv}, for any irregular $\uxi$-parabolic Higgs bundle $\CE$ there exists a
trivialization $E|_D \simeq \CO_D^{\oplus r}$ such that 
\begin{itemize} 
\item[$(a)$] The  flag of $\CO_D$-modules $E_D^\bullet$ 
is identified with $V^\bullet \otimes \CO_D$, where $V^\bullet$ is the standard flag of type $(m_1, \ldots, m_\ell)$ in $\IC^r$, and 
\item[$(b)$] The restriction $\Phi_D$ of the Higgs field to $D$ is identified with the diagonal matrix  with entries 
\[
\underbrace{\xi_1, \ldots, \xi_{1}}_{m_1},  \ldots, 
\underbrace{\xi_{2}, \ldots, \xi_{2}}_{m_2},\ldots, 
\underbrace{ \xi_{\ell}, \ldots, \xi_{\ell}}_{m_\ell}.
\]
on the diagonal. 
\end{itemize}

Let ${\widehat \CO}_{C,p}$ be the ${\mathfrak m}_p$-adic completion of the local 
ring $\CO_{C,p}$ and ${\widehat \CK}_{C,p}$ denote its field 
of fractions. For any commutative ring $A$, let ${\bf t}_r(A)$ denote the ring of $r\times r$ diagonal matrices with coefficients in $A$. 
Then $\Phi_D$ determines uniquely a pair 
$({\Gamma}_\uxi, \Lambda_\uxi)$ where ${ \Gamma}_\uxi$ is an element
of the quotient 
\be\label{eq:irregtypes}
{\bf t}_r\big({\widehat \CK}_{C,p}\big)/
{\bf t}_r\big({\widehat \CO}_{C,p}\big),
\ee
and $\Lambda_\uxi\in 
{\bf t}_r$.
This can be easily seen 
by choosing a local coordinate $z$ on $C$ centered at $p$, 
in which case $\Phi_D$ has an expansion 
\[ 
\sum_{k=1}^n \Lambda_k z^{-k} dz
\]
with $\Lambda_k$, $1\leq k \leq n$ complex $r\times r$ diagonal matrices. The above local expression can be written as 
\[ 
d\Psi + \Lambda_{1} {dz \over z} 
\]
where $\Psi$ is a Laurent polynomial in $z$ with coefficients in ${\bf t}_r(\IC)$, and $\Lambda_1\in {\bf t}_r(\IC)$ is the residue of the Higgs field. Clearly both $\Psi,\Lambda_1$ are uniquely determined by the local data $\uxi$ once the local coordinate $z$ has been chosen. Moreover, the equivalence class ${\Gamma}_\uxi$ of $\Psi$ in the quotient \eqref{eq:irregtypes} and the diagonal matrix $\Lambda_\uxi=\Lambda_1$ 
are independent on the choice of local coordinate $z$. 

Next consider irregular $\uxi$-parabolic Higgs bundles with parabolic degree 
\[ 
d+ \sum_{i=1}^\ell m_i\alpha_i =0.
\]
In this case for sufficiently generic weights $\ualpha$ there are no semistable objects, and the moduli space $\CH_\uxi(C,D; \ualpha, \um)$ of stable objects is smooth and quasiprojective. 
According to the results of \cite{Wild_curves}  and  \cite{Wild_harmonic}, the moduli space of $\uxi$-parabolic Higgs bundles is related by the wild 
non-abelian Hodge correspondence to a moduli space of 
irregular filtered flat connections on $C$ constructed as 
explained below. Note that for the purposes of the present paper it suffices to specialize the construction of loc. cit. 
to connections with semisimple residue. 

Let ${\sfQ}$ be a fixed irregular type in the quotient \eqref{eq:irregtypes} and let $\sfR$ be a diagonal matrix with eigenvalues 
\[
\underbrace{\rho_1, \ldots, \rho_{1}}_{m_1},  \ldots, 
\underbrace{\rho_{2}, \ldots, \rho_{2}}_{m_2},\ldots, 
\underbrace{ \rho_{\ell}, \ldots, \rho_{\ell}}_{m_{\ell}},
\]

\noindent where $\rho_i\in \IC$, $1\leq i\leq \ell$ are 
pairwise distinct complex numbers and $m_1, \ldots, m_{\ell}\in \IZ_{\geq 1}$. As in Section \ref{wildintro}, 
it will be assumed in the following that the centralizer $H_{\sf Q}$ of $\sf Q$ in $GL(r, \IC)$ is the canonical subgroup $GL(m_1, \IC) \times \cdots \times GL(m_{\ell}, \IC)\subset GL(r,\IC)$.

Let $\ubeta=(\beta_1, \ldots, \beta_{\ell})\in \IR^{\ell}$ be a collection of real weights: 
$1 > \beta_1>\beta_2>\ldots> \beta_{\ell} \geq 0$. 
Then the moduli space parameterizes data of the form $(V, \nabla, V_p^\bullet)$ where 
\begin{itemize}
\item 
$V$ is a rank $r$ bundle on $C$ with fixed degree 
\[
{\rm deg}(V) + \sum_{i=1}^{\ell} m_i\beta_i =0,
\]
\item 
$\nabla: V \to V\otimes_C \Omega^1_C(D)$ is a meromorphic connection on $V$ with poles of order at most $n$ at $p$ and 
\item 
$V_p^\bullet$ is a flag 
\[ 
0 \subset V_1 \subset \cdots \subset V_{\ell} = V_p
\]
of type $\um$ in the fiber of $V$ at $p$. 
\end{itemize}

\noindent 
The conditions used in the construction of the moduli space 
are listed below. 
\begin{itemize} 
\item One requires the existence of a local trivialization of $V$ over an open neighborhood of $p$ in $C$ which identifies the flag $V_p^\bullet$ to the standard flag of type $(m_1, \ldots, m_{\ell})$ in $\IC^r$. Moreover, with respect to this trivialization the connection $\nabla$ is given by $d-A$ with  
\[
A = dQ + \sfR {dz\over z} + {\rm holomorphic\ terms} 
\]
where $Q\in {\bf t}_{r}(\IC[1/z])$ is a representative of $\sfQ$.

 \item  In addition, the data $(V, \nabla, V_p^\bullet)$ is subject to a parabolic stability condition which is entirely analogous to the one used in Section \ref{setup}. 
 \end{itemize}

\noindent  Note that
for sufficiently generic weights $\ubeta$ there are no strictly semistable objects, so the  resulting moduli space ${\mathcal C}_{\sfQ,\sfR}(C,D; \ubeta, \um)$ is smooth. 
(It is still non-compact 
e.g. because it maps -- in a different complex structure, as we recall below -- to the Hitchin base.)

In this framework Theorem 5 of \cite{HK_wild} states that the moduli space of irregular connections 
${\mathcal C}_{\sfQ,\sfR}(C,D; \ubeta, \um)$ is a smooth kyper-K\"ahler manifold and is naturally diffeomorphic to the moduli space ${\mathcal H}_\uxi(C,D; \ualpha, \um)$ of irregular Higgs bundles provided certain relations among the 
fixed data hold. In order to formulate a precise statement, suppose that:
\begin{itemize} 
\item[$(WH.1)$] The eigenvalues $\rho_1, \ldots, \rho_\ell$ of $\sfR$ satisfy 
\[
-1 < \rho_i \leq 0,\quad 1\leq i \leq \ell, \qquad{\rm and}\qquad  \rho_\ell > \rho_{\ell-1}> \cdots > \rho_1. 
\]
\item[$(WH.2)$] The following relations hold 
\[ 
\alpha_i = - \rho_i, \quad 1\leq i \leq \ell, \qquad 
\lambda_i = - (\rho_i + \beta_i)/2, \quad 1\leq i \leq \ell,\qquad  \sfQ= - 2\Gamma_\uxi, 
\]
where 
\[ 
\underbrace{\lambda_1, \ldots, \lambda_{1}}_{m_1},  \ldots, 
\underbrace{\lambda_{2}, \ldots, \lambda_{2}}_{m_2},\ldots, 
\underbrace{ \lambda_{\ell}, \ldots, \lambda_{\ell}}_{m_\ell}.
\]
are the eigenvalues of $\Lambda_\uxi$. 
\end{itemize} 

\noindent
If conditions $(WH.1)-(WH.2)$ above are satisfied then the moduli spaces ${\mathcal C}_{\sfQ,\sfR}(C,D; \ubeta, \um)$, ${\mathcal H}_\uxi(C,D; \ualpha, \um)$ are smooth  hyper-K\"ahler manifolds which share the  same underlying real manifold structure. Moreover the complex structures on these moduli spaces are related by a hyper-K\"ahler 
rotation.

To conclude, note that the irregular type $\sfQ$ obtained from this construction satisfies assumption $(i)$ 
above equation \eqref{eq:formalmon} in Section \ref{wildintro}. Moreover, the conjugacy class of the formal monodromy is also of the form \eqref{eq:formalmon} with $\sfT = {\rm exp}(2\pi {\sf R})$.
Therefore the associated moduli spaces of Stokes data are varieties of the form $\CS_{\sfQ,\sfT}$ as constructed below equation \eqref{eq:stokeseq}.

\section{Spectral construction}\label{spectralsect}

The main goal of this section is to provide a detailed proof 
for the spectral correspondence stated in Section \ref{spcorrespCY}. As explained there, this correspondence is based on a geometric construction carried out by Kontsevich and Soibelman in \cite[Sect 8.3]{structures}. 
As a brief overview, irregular parabolic Higgs bundles will 
be identified with pure dimension one sheaves on a complex  surface obtained by successive 
blow-ups of the total space of the coefficient line bundle $M$. Note that this spectral construction is different from 
the one employed for a similar purpose \cite{Par_ref}, as explained in more detail at the end of Section \ref{surface}.

\subsection{The holomorphic symplectic surface}\label{surface}

Recall that our group $G$ is $GL(r, \IC)$, the polar divisor $D=np$ is supported at a single point $p$, and we have fixed a partition $r=\sum_{i=1}^{\ell} m_i$.
Abusing notation, the total space of the coefficient line bundle $M= K_C(D)$ will be  
denoted also by $M$, the distinction being clear from the context. 
The natural projection to $C$ will be 
denoted by $\pi:M \to C$ and the tautological section of 
$\pi^*M$ over $M$ will be denoted by $y$. Let also $M_D$ denote $\pi^{-1}(D)$.

Now let  $\uxi=(\xi_1, \ldots, \xi_\ell)$ be a collection of 
generic sections of $M_D$. Let $\delta_i \subset M_D$ be the 
divisors defined by 
\[ 
y - \pi^*\xi_i =0, \qquad 1\leq i\leq \ell. 
\]
Then one first constructs a complex surface $T_\uxi$ 
by simultaneously blowing up the subschemes $\delta_i \subset M$, $1\leq i \leq \ell$. More concretely, this can be seen as a 
series of $\ell$ successive blow-ups of $M$ at 
 $\ell$ sequences of points determined by the local sections 
$\xi_1, \ldots, \xi_l$.  First let 
$\wp_{1,i}$, $1\leq i \leq \ell$, be the intersection points between 
$\delta_1, \ldots, \delta_\ell$ and the reduced fiber $M_p$, 
all of them transverse. Under the genericity assumption
\eqref{eq:genlocsectA}, these are pairwise distinct points not lying 
on the zero section $C_0\subset M$. Let $\Xi_{1,i}$, $1\leq i \leq \ell$ be the resulting exceptional divisors.

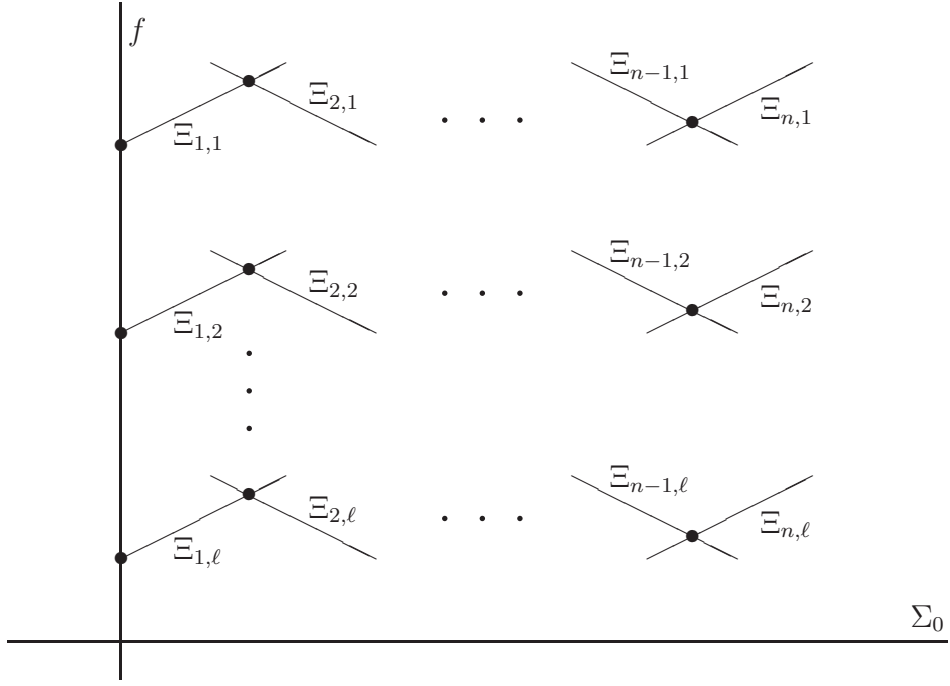
\begin{figure}
\setlength{\unitlength}{1mm}
\hspace{-20pt}
\vspace{-30pt}
\setlength{\unitlength}{1mm}
\begin{picture}(150,105)
{\thicklines\put(15,10){\line(1,0){125}}
\put(30,5){\line(0,1){90}}}
\put(135,12){$\Sigma_0$}
\put(31,90){$f$}
\put(29,20){$\bullet$}
\put(29,50){$\bullet$}
\put(29,75){$\bullet$}
\put(30,21){\line(2,1){22}}
\put(46,28.5){$\bullet$}
\put(42,32){\line(2,-1){22}}
\multiput(72,26)(5,0){3}{\huge .}
\put(90,32){\line(2,-1){22}}
\put(105,23){$\bullet$}
\put(100,21){\line(2,1){22}}
\put(30,51){\line(2,1){22}}
\put(46,58.5){$\bullet$}
\put(42,62){\line(2,-1){22}}
\multiput(72,56)(5,0){3}{\huge .}
\put(90,62){\line(2,-1){22}}
\put(105,53){$\bullet$}
\put(100,51){\line(2,1){22}}
\put(30,76){\line(2,1){22}}
\put(46,83.5){$\bullet$}
\put(42,87){\line(2,-1){22}}
\multiput(72,79)(5,0){3}{\huge .}
\put(90,87){\line(2,-1){22}}
\put(105,78){$\bullet$}
\put(100,76){\line(2,1){22}}
\multiput(46,38)(0,5){3}{\huge .}
\put(37,21){$\Xi_{1,\ell}$}
\put(37,51){$\Xi_{1,2}$}
\put(37,76){$\Xi_{1,1}$}
\put(55,26.5){$\Xi_{2,\ell}$}
\put(55,56.5){$\Xi_{2,2}$}
\put(55,81.5){$\Xi_{2,1}$}
\put(95,30.5){$\Xi_{n-1,\ell}$}
\put(95,60.5){$\Xi_{n-1,2}$}
\put(95,85.5){$\Xi_{n-1,1}$}
\put(115,24.5){$\Xi_{n,\ell}$}
\put(115,54.5){$\Xi_{n,2}$}
\put(115,79.5){$\Xi_{n,1}$}
\end{picture}
\caption{Chains of exceptional curves in the surface $T_\uxi$.}
\label{SurfaceI}
\end{figure}

\noindent
For each $1\leq i \leq \ell$, the strict transform of $\delta_i$ under the first blow-up will intersect the $i$-th exceptional divisor  transversely at 
a point $\wp_{2,i}$. All these intersection points are then blown-up again at the next step. 
Proceeding recursively, $n$ successive simultaneous 
blow-ups yield a complex 
surface $T_\uxi$ as sketched in Figure \ref{SurfaceI}. 
The exceptional locus for the blow-up map $\rho:T_\uxi \to M$ consists of $\ell$ pairwise disjoint chains of exceptional rational curves $(\Xi_{1,i}, \Xi_{2,i}, \ldots, \Xi_{n,i})$, $1\leq i \leq \ell$, on $T_\uxi$ 
with intersection matrix 
\be\label{eq:blowupintersA} 
\Xi_{a,i} \cdot \Xi_{b,j} = 
\left\{ \begin{array}{ll} -2, & {\rm for}\ i=j,\ 1\leq a=b \leq 
n-1,\\
-1, & {\rm for}\ i=j, \ a=b=n,\\
1,& {\rm for}\ i=j, \ |a-b|=1,\\
0, & {\rm otherwise}.\\
\end{array}
\right.
\ee
The intersection of any two consecutive curves $\Xi_{a,i}$, $\Xi_{a+1,i}$ in the $i$-th chain is transverse. Let
$\Sigma_0$ denote the strict transform of the zero section 
$C_0\subset M$ and let $f$ denote the strict transform of the fiber $M_p \subset M$. 
The canonical class of $T_{\uxi}$ is then:
\[ 
K_{T_\uxi} =  -nf_{} - \sum_{i=1}^\ell \sum_{a=1}^{n}(n-a)\Xi_{a,i}. 
\]
In particular, the $(-1)$ curve classes $\Xi_{n,i}$, $1\leq i\leq \ell$ occur with multiplicity 0 in the right hand side of this formula. Therefore the complement $S_\uxi \subset T_\uxi$ of 
the divisor 
\[
f+ \sum_{i=1}^\ell\sum_{a=1}^{n-1} \Xi_{a,i}
\]
in $T_{\uxi}$ 
is a holomorphic symplectic surface $S_\uxi \subset T_\uxi$. The restriction of the projection $\pi_T: T_\uxi \to C$ to $S_\uxi$ will be denoted by $\pi_S: S_\uxi  \to C$. 
As noted above, it will be assumed that the sections 
$\xi_1, \ldots, \xi_\ell$ satisfy the genericity conditions 
\eqref{eq:genlocsectA}. This implies in particular that none of the 
initial blow-up centers $\wp_{1,i}$, $1\leq i \leq \ell$ lie on the zero section $C_0\subset M$. 

For completeness we conclude this section with a brief comparison between the above construction and the one used 
for the spectral correspondence in \cite{Par_ref}.  
Given the curve $C$,the marked point $p$, and the 
line bundle $M$, the input data for the construction of 
\cite{Par_ref} consists of a collection of points 
${\underline \wp}=\wp_1, \ldots, \wp_\ell$ in the fiber of $M$ at $p$ and a collection of positive integers 
${\underline s}=(s_1, \ldots, s_\ell)$. One then obtains a 
holomorphic symplectic orbifold surface 
${\widehat S}({\underline \wp},{\underline s})$ by carrying out weighted blow-ups of the total 
space of $M$ at the points $\wp_1, \ldots, \wp_\ell$. The 
coarse moduli space of this orbifold surface admits a canonical 
crepant resolution $S({\underline \wp},{\underline s})$ also involving successive blow-ups of $M$. However, in this construction one carries out successive blow-ups  
at points on the strict transform 
of the fiber $M_p$, and the resulting surface $S({\underline \wp},{\underline s})$ is not in general isomorphic to a surface
$S_\uxi$ as above. Therefore the spectral correspondence 
proved 
in the next sections is not related to the one of \cite{Par_ref} 
by Fourier-Mukai transform. 
In fact the two 
constructions coincide only for tamely
ramified Higgs bundles with 
regular semisimple residues.

\subsection{Irregular parabolic Higgs bundles from torsion sheaves}\label{sheavestohiggs} 

As a first step of the spectral correspondence this section 
will construct stable irregular $\uxi$-parabolic Higgs bundles on $C$ from stable pure dimension one sheaves on $S_\uxi$ with compact support. 

The first task is to classify the topological invariants of such sheaves. 
Let $\Sigma_0\subset T_\uxi$ be the strict transform of the zero section $C_0\subset M$ and let $\Delta_i \in {\rm Pic}(T_\uxi)$ be 
defined by 
\[
\Delta_i=
\sum_{a=1}^n 
a \Xi_{a,i},\qquad 1\leq i \leq \ell.
\]
Then note that a compact curve class on $T_\uxi$ with support in $S_\uxi$ 
must have intersection number $0$ with $f$ and with the  $\Xi_{a,i}$ for $a<n$, i.e. it
must be of the form
\be\label{eq:strtrnsfclass}
\Sigma_{\um} = r\Sigma_0 - \sum_{i=1}^\ell m_i
\Delta_i
\ee
where $r\in \IZ$, $r\geq 1$ and $\um=(m_1, \ldots, m_\ell)$ are non-negative integers so that 
\[ 
\sum_{i=1}^\ell m_i = r. 
\]
Hence, any pure dimension one sheaf $F$ with compact support in $S_\uxi$ will have topological invariants 
\[ 
\ch_1(F) = \Sigma_{\um}, \qquad 
\chi(F) = c. 
\]
Moreover the topological support of $F$ is disjoint from the exceptional divisors $\Xi_{a,i}$, $1\leq a \leq n-1$, $1\leq i \leq \ell$ while
its intersection with each divisor $\Xi_{n,i}$, $1\leq i \leq \ell$ 
is a finite set of closed points contained in $S_\uxi$. 

Given a sheaf $F$ as above, note that $E=\pi_{S}{_*}F$ is a locally free sheaf 
on $C$. To prove this suppose $T\hookrightarrow E$ is a zero dimensional subsheaf on $C$. Then there is a nonzero morphism $\pi_S^*T \to F$ on $S_\uxi$.
If the support of  $T$ contains the point $p$, then the support of $\pi^*_ST$ contains 
the union $\cup_{i=1}^\ell \Xi_{n,i}$. 
This is a contradiction since $F$ is pure dimension one and its support does 
not contain this union of divisors. 
If the support of $T$ contains a point $q \neq p$, then
the support of $\pi^*_ST$ contains the fiber over $q$, 
which again cannot be contained in the support of $F$.
Therefore $E$ is torsion free, hence locally 
free since $C$ is a smooth curve. One can also easily show that
\[
{\rm rk}(E) =\sum_{i=1}^\ell m_i, \qquad \chi(F) = \chi(E), 
\]
and the higher direct images $R^k\pi_{S*}F$, $k\geq 1$, vanish. Furthermore, 
as shown in Appendix \ref{pwd}, the pushforward $\pi_{S*} \big(F\otimes_{S_\uxi} \CO_{k\Xi_{n,i}}\big)$ is a 
locally free sheaf on the non-reduced divisor $kp$
for any $1\leq i\leq \ell$ 
and for any $k\in \IZ$, $k\geq 1$. Therefore for 
each $1\leq i \leq \ell$ one obtains a surjective morphism  
\be\label{eq:parstrA}
E_D\twoheadrightarrow 
\pi_{S*}(F\otimes_{S_\uxi} {\CO_{n\Xi_{n,i}}})
\ee
of locally free $\CO_D$-modules, where $E=\pi_{S*}F$ is a locally free $\CO_C$-module. 
Moreover, one also obtains a Higgs field $\Phi:E\to E\otimes_C M$ by taking the direct image of the multiplication map $F\to F\otimes_{S_\uxi} \CO_{S_\uxi}(\Sigma_0)$, where $\Sigma_0\subset S_\uxi$ is the strict transform of the zero section of $M$. Note that $E\otimes_C M$ is the direct image of  $F\otimes_{S_\uxi} \CO_{S_\uxi}(\Sigma_0)$ by the projection formula, 
which can be applied to the present context since $F$ has compact support in $S_\uxi$.
Since no blowups occur on the zero section, 
this is the same as the total transform, $\Sigma_0=\rho^*C_0$, 
where $\rho:T\to M$ is the blow-up map. 

So far this construction does not yet define a parabolic Higgs bundle 
on $C$ since one needs to pick an ordering of the resulting quotients. While there is no natural ordering determined 
by the geometry of the surface $S_\uxi$, there will be one 
once one picks up a Bridgeland stability condition for 
pure dimension one sheaves on $S_\uxi$. 
More precisely, it suffices to consider a subspace of the 
moduli space of stability conditions
parameterized 
by a compactly supported $B$-field $\beta\in H^2_c(S_\uxi, \IR)$ such that 
the only non-zero periods of $\beta$ are 
\[ 
\beta(\Xi_{n,i}) = -\beta_i, \qquad 1\leq i \leq \ell. 
\]
Such a $B$-field $\beta$ defines a slope function $\mu_\beta$ 
for compactly supported pure dimension one sheaves on $S_\uxi$. Note that the scheme theoretic support of any such sheaf $F$  must be an $r_F:1$ cover of $C$ for some $r_F\geq 1$. Then one sets 
\[
\mu_\beta(F) = {\chi(F) + \beta(\ch_1(F)) \over r_F}. 
\]
and defines $\beta$-stability in the usual way. 

Now suppose the stability parameters $\beta_i$, $1\leq i\leq \ell$ are chosen so that 
\[ 
n > \beta_1 > \ldots > \beta_\ell  >0
\]
Then one can construct a filtration of $E_D = (\pi_{S*}F)_D$ 
as follows. 
Let $Q_{\ell-1} = \pi_{S*}(F\otimes_{S_\uxi} {\CO_{n\Xi_{n,\ell}}})$. As shown above equation \eqref{eq:parstrA}, $Q_{\ell-1}$ is a locally free $\CO_D$-module and
there is a surjective morphism $q_{\ell-1}:E_D \twoheadrightarrow Q_{\ell-1}$. The kernel $E_D^{\ell-1} = {\rm Ker}(q_{\ell-1})$
is also a locally free $\CO_D$-module, hence this yields a one-step filtration $E_D^{\ell-1}\subset E_D$. 

In order to construct the next step, let $Q_{\ell-2} = \pi_{S*}(F\otimes_{S_\uxi} {\CO_{n\Xi_{n,{\ell-1}}}})$
and note that there is a surjective morphism $q_{\ell-2}:E_D^{\ell-1}\twoheadrightarrow Q_{\ell-2}$. 
To prove this,
let $F_{\ell-1}\subset F$ be the kernel of the surjective morphism 
$F\twoheadrightarrow F\otimes_{S_\uxi} \CO_{n\Xi_{n,\ell}}$. 
Since $\Xi_{n,\ell}, \Xi_{n,\ell-1}$ are disjoint and not contained in  the support of $F$, there is an epimorphism 
$F_{\ell-1} \twoheadrightarrow F\otimes_{S_\uxi} \CO_{n\Xi_{n,\ell-1}}$. Let $E_{\ell-1} = \pi_{S*}F_{\ell-1}$, 
and note that $E_{\ell-1}$ is a subsheaf of $E$ such that $E_D^{\ell-1}$ is the image of the induced morphism 
$\big(E_{\ell-1}\big)_D \to E_D$. Moreover, the surjective morphism $F_{\ell-1} \twoheadrightarrow F\otimes_{S_\uxi} \CO_{n\Xi_{n,\ell-1}}$
yields a surjective morphism $(E_{\ell-1})_D \twoheadrightarrow Q_{\ell-2}$. 
At the same time by construction there is a commutative diagram of $\CO_D$-modules 
\[ 
\xymatrix{ 
(E_{\ell-1})_D \ar[rr] \ar[dr] & & E_D\ar[dl] \\
& Q_{\ell-1} &  \\}
\]
where the top horizontal arrow is the natural inclusion. This implies that the epimorphism $ \big(E_{\ell-1}\big)_D \twoheadrightarrow Q_{\ell-1}$ 
factors through $E_D^{\ell-1}$. The kernel $E_D^{\ell-2} = {\rm Ker}\left(E_{D}^{\ell-1} \twoheadrightarrow Q_{\ell-1}\right)\subset E_D^{\ell-1}$ is again locally free, providing the second step of the filtration. 

Iterating this construction, one obtains a filtration 
\[
0\subset E_D^1\subset \cdots \subset E_D^{\ell-1} \subset E_D
\]
by locally free $\CO_D$-submodules. (Again, all quotients 
$E_D/E_D^i$ are also locally free.)
This filtration is naturally preserved by the Higgs field $\Phi$, which also satisfies the 
conditions \eqref{eq:polarHiggsA} by construction. 
Finally,  $$\alpha_i= \beta_i/n$$ 
are a set of parabolic weights for the flag $E_D^\bullet$, hence 
the data $(E, E_D^\bullet, \Phi)$ is an irregular $\uxi$-parabolic 
Higgs bundle.

To summarize, the above construction assigns an irregular 
$\uxi$-parabolic Higgs bundle $(E, E_D^\bullet, \Phi)$ with weights 
$\alpha_i = \beta_i /n$, $1\leq i \leq \ell$, 
to any pair $(F,\beta)$. 
In order to check stability, one has to further study the direct image of nonzero subsheaves $F'\subset F$ such that $F/F'$ is a pure dimension one sheaf on $S_\uxi$. The main observation is that for any such sheaf one has 
\[ 
{\mathcal Tor}_1^{S_\uxi}(F/F', \CO_{n\Xi_{n,i}})=0 
\]
for any $1\leq i \leq \ell$. This follows from the fact 
that $F/F'$ is pure dimension one and its scheme theoretic 
support has no components along $\Xi_{n,i}$, $1\leq i \leq n$. 
Therefore the injection $F'\hookrightarrow F$ yields by restriction an injection 
\be\label{eq:subquotA}
F'\otimes_{S_\uxi} \CO_{n\Xi_{n,i}}\hookrightarrow F\otimes_{S_\uxi} \CO_{n\Xi_{n,i}}. 
\ee

Moreover, the direct image $E'=\pi_{S*}F'$ is a saturated subsheaf of $E$ of positive rank, hence there is an induced filtration $(E')_D^{i}= E'_D\cap E_D^i$, $1\leq i \leq \ell$.
 Using the above construction for the filtration $E_D^\bullet$ and the injectivity of the morphisms \eqref{eq:subquotA}, it is immediate that 
\[
(E')_D^i/ (E')_D^{i-1} \simeq  \pi_{S*} F'\otimes_{S_\uxi} \CO_{n\Xi_{n,i}}
\]
for all $1\leq i \leq \ell$. 
This implies that 
\[ 
\mu_\beta(F') = { {\rm deg}(E') + \sum_{i=1}^\ell \alpha_i 
\chi((E')_D^i/(E')_D^{i-1}) \over {\rm rk}(E')}. 
\]

\subsection{The inverse construction}\label{inverse}

In order to reverse the above construction, suppose $(E, E_D^\bullet, \Phi)$ is an irregular $\uxi$-parabolic 
Higgs bundle on $C$. Recall that $\pi: M \to C$ is the projection 
to $C$ and  $y \in H^0(M, \pi^*M)$ is the tautological 
section. Then the inverse construction proceeds as follows. 

{\bf Step 1.} One first constructs a pure dimension one sheaf on $M$ using a two-term monad complex. Namely, note that 
 the morphism of $\CO_M$-modules 
\[
y{\Ione}_{\pi^*E} - \pi^*\Phi : \pi^*E\otimes_M \pi^*M^{-1} \to \pi^*E
\]
is injective and its cokernel $G = {\rm Coker}\left(y{\Ione}_{\pi^*E} - \pi^*\Phi\right)$ is a pure dimension one sheaf on $M$ such that 
$\pi_*G \simeq E$.  

{\bf Step 2.} Next, using the parabolic structure of $E$ one constructs a filtration of the $\CO_{T_\uxi}$-module $\rho^*G$, 
where $\rho: T_\uxi \to M$ is the blow-up map. This construction 
is recursive. 
The first subobject is obtained as follows. 

{\bf Step 2.a.} Let $Q_{\ell-1}= E_D/E^{\ell-1}_D$, which is a locally free $\CO_D$-module by assumption. Then, given conditions 
\eqref{eq:polarHiggsA},
 one has a commutative 
diagram 
\[
\xymatrix{ 
\pi^*E \otimes_M \pi^*M^{-1}\ar[rrr]^-{y\Ione_{\pi^*E} - \pi^*\Phi} \ar[d] & & & \pi^*E  \ar[d] \\
\pi^*Q_{\ell-1} \otimes_M \pi^*M^{-1}\ar[rrr]^-{(y-\pi^*\xi_\ell) \Ione_{\pi^*{Q_{\ell-1}}}} & & &
\pi^*Q_{\ell-1} \\
}
\]
 where the vertical maps are surjective. This yields an epimorphism $G\twoheadrightarrow R_{\ell-1}$, where $R_{\ell-1}={\rm Coker}((y-\xi_\ell) \Ione_{\pi^*{Q_{\ell-1}}})$. 

 Note that $\pi^*Q_{\ell-1}$ is a locally free sheaf on the non-reduced subscheme $M_D=\pi^{-1}(D) = n M_p$ in $M$, and the cokernel $R_{\ell-1}$ 
is scheme theoretically supported on the divisor $\delta_\ell\subset M_D$ determined by $y-\pi^*\xi_\ell=0$. 
Moreover since $\xi_\ell:D \to M_D$ is a section, $R_{\ell-1}\simeq \xi_{\ell*}Q_{\ell-1}$ is locally free of rank $m_\ell$ on its 
scheme theoretic support. 

Next note that the 
pull-back $\rho^*G$ is a pure dimension one sheaf on $T$.
To prove this note that for any  nontrivial zero dimensional subsheaf $\CT\subset \rho^*G$, the direct image $\rho_*\CT \subset \rho_*\rho^*G$ is also a nontrivial zero dimensional subsheaf. Using the projection 
formula, $\rho_*\rho^*G \simeq G$, hence one obtains a contradiction since $G$ is pure of dimension one. 
Moreover, there is an epimorphism $\rho^*G \twoheadrightarrow \rho^*R_{\ell-1}$. Let $F_{\ell-1}$ be its kernel, which is obviously of pure dimension one. 

{\bf Step 2.b.} 
The next claim is that the set theoretic support of $F_{\ell-1}$ must be disjoint from $\Xi_{1,\ell}, \ldots, \Xi_{n-1,\ell}$ while its intersection with  $\Xi_{n,\ell}$ is a finite set of closed points. 
By construction, the determinant of $G$ is the spectral curve 
$\Sigma_G$ given by 
\[
{\rm det}(y \Ione_{\pi^*E}-\Phi)=0 
\]
in $M$, which belongs to the linear system $|rC_0|$. 
Given the local form of $\Phi$, the scheme theoretic intersection 
of $\Sigma_G$ with $M_D = \pi^{-1}(D)$ is the divisor $m_1\delta_1 + \cdots + m_\ell \delta_\ell$ where $\delta_i \subset M_D$ is determined by $y -\pi^*\xi_i=0$.

Moreover, a local computation shows that 
\be\label{eq:pullbackA}
\rho^{-1}(\delta_i) = {\Delta_i}
\ee
where $\Delta_i \subset T$ is the divisor given by 
\[
\Delta_i = \sum_{a=1}^n a \Xi_{a,i}.
\]
Therefore $\rho^*G$ has determinant 
\[
\rho^*\Sigma_G = \sum_{i=1}^\ell m_i \Delta_i + \Sigma'_G
\]
where $\Sigma'_G$ is the strict transform of $\Sigma_G$. 
In particular, the support of $\Sigma_G'$ satisfies the intersection conditions formulated above. 
Equation \eqref{eq:pullbackA} also implies that 
$\rho^*R_{\ell-1}$ is isomorphic to a locally free sheaf of rank $m_\ell$ on $\Delta_\ell$. Therefore its determinant is 
$m_\ell\Delta_\ell$. To finish the proof of the above claim, note that by construction 
the determinant of $F_{\ell-1}$ is $\Sigma_G'$.

{\bf Step 3.} This is the second iteration of the above construction.  Note that the exact sequence 
\be\label{eq:kernelone}
0\to F_{\ell-1} \to \rho^*G \to \rho^*R_{\ell-1}\to 0 
\ee
yields by push-forward the exact sequence of $\CO_M$-modules 
\[
0\to \rho_*F_{\ell-1} \to G \to R_{\ell-1} \to 0. 
\]
Let $G_{\ell-1} = \rho_*F_{\ell-1}$ and note that $G_{\ell-1}$ is isomorphic to $G$ on the complement of $\delta_\ell$ in $M$. Moreover, there is an exact sequence of sheaves on $C$ 
\[
0\to \pi_*G_{\ell-1} \to E \to Q_{\ell-1} \to 0
\]
which implies that $\pi_*G_{\ell-1} \simeq E_{\ell-1}\subset E$, where $E_{\ell-1} = {\rm Ker}(E\twoheadrightarrow E_D/E_{D}^{\ell-1})$. This further 
implies that $G_{\ell-1}$ is isomorphic to the cokernel of the following monad complex on $M$
\[
y{\Ione}_{\pi^*E_{\ell-1}} - \pi^*\Phi_1: \pi^*E_{\ell-1}\otimes_M \pi^*M^{-1} \to \pi^*E_{\ell-1}
\]
where $\Phi_1=\Phi|_{E_1}$. Moreover by construction there is a surjective 
morphism $E_{\ell-1} \twoheadrightarrow E_D^{\ell-1}$ on $C$, which yields a second surjective morphism 
 $E_{\ell-1} \to Q_{\ell-2}$ with $Q_{\ell-2} = E_D^{\ell-1}/E_D^{\ell-2}$. 

In complete analogy with { Step 2.a}, it then follows 
that there is an epimorphism $G_{\ell-1} \twoheadrightarrow R_{\ell-2}$ 
where $R_{\ell-2}$ is the cokernel of the morphism 
\[
(y-\pi^*\xi_{\ell-1}){\Ione}_{\pi^*Q_{\ell-2}}: \pi^*Q_{\ell-2}\otimes_M \pi^*M^{-1} \to 
\pi^*Q_{\ell-2}.
\]
Again, $R_{\ell-2}$ is scheme theoretically supported on the 
divisor $\delta_{\ell-1} \subset M_D$ given by $y-\pi^*\xi_{\ell-1}=0$. Moreover, there is an isomorphism $R_{\ell-2} \simeq \xi_{{\ell-1}*}Q_{\ell-2}$, 
hence $R_{\ell-2}$ is locally free of rank $m_{\ell-1}$ on $\delta_{\ell-1}$. 

Next let $U_{\ell-1}\subset M$ be the complement of $\delta_\ell$. 
Then $\rho^{-1}(U_{\ell-1})\subset T$ is the complement of $\Delta_\ell$ in $T$, which contains all other 
exceptional divisors of $\rho$. 
Using the sequence \eqref{eq:kernelone}, it follows that 
$F_{\ell-1}|_{\rho^{-1}(U_{\ell-1})} \simeq \rho^*G|_{\rho^{-1}(U_{\ell-1})}$. Hence also $G_{\ell-1}|_{U_{\ell-1}} \simeq G|_{U_{\ell-1}}$. This implies that 
there is a surjective morphism 
$F_{\ell-1}\twoheadrightarrow \rho^*R_{\ell-2}$. As in Step 2.b,  one 
argues again that the kernel 
$F_{\ell-2} = {\rm Ker}(F_{\ell-1}\twoheadrightarrow \rho^*R_{\ell-2})$ is a pure dimension one sheaf on $T$ with determinant $\rho^*\Sigma_G -m_\ell\Delta_\ell-m_{\ell-1}\Delta_{\ell-1}$. In particular, the support of $F_{\ell-2}$ 
is disjoint from $\Xi_{1,i}, \ldots, \Xi_{n-1,i}$, with $\ell-1\leq i \leq \ell$, while its intersection with each of $\Xi_{n,\ell-1}, \Xi_{n,\ell-2}$ is a finite set of closed points.

{\bf Step 4.} 
Proceeding recursively, one then constructs a sequence $F_{\ell-1}, F_{\ell-2}, \ldots, F_0$ of pure dimension one sheaves on $T$ which fit in exact sequences 
\[
0\to F_{i-1} \to F_i \to \rho^*R_{i-1} \to 0,\qquad 1\leq i \leq \ell
\]
where $F_\ell = \rho^*G$ and $R_{i-1} \simeq \xi_{i*}(E_D^{i}/E_D^{i-1})$ is a locally free 
$\CO_{\delta_{i}}$-module of rank $m_{i}$. The sheaf $F_0$ obtained at the last step has determinant $\rho^*\Sigma_G - 
\sum_{i=1}^\ell m_i \Delta_i$. Its support is disjoint from $\Xi_{1,i}, \ldots, \Xi_{n-1,i}$, with $1\leq i \leq \ell$, while its intersection with each of $\Xi_{n,i}$, $1\leq i \leq \ell$ is a finite set of closed points. In particular $F_0$ is a compactly supported $\CO_{S_\uxi}$-module. 
Moreover, it is straightforward to check that the irregular parabolic Higgs bundle determined by $F_0$ as in Section \ref{sheavestohiggs} is $\CE(-D) = \big(E(-D), E_D^\bullet(-D), \Phi\otimes \Ione_{\CO_C(-D)}\big)$. 

{\bf Step 5.}
Finally, in order to check stability, let $E'\subset E$ be a $\Phi$-invariant, saturated, non-trivial subsheaf of $E$. Recall that the induced filtration of $E'_D$ is defined by $(E')_D^i = E'_D\cap E_D^i$, $1\leq i \leq \ell$. According to the second claim proven in Section \ref{technostuff}, the successive quotients $(E')_D^i/(E')_D^{i-1}$ are locally free $\CO_D$-modules. 
Moreover, for each $1\leq i \leq \ell$ there is an injective morphism  $(E')_D^i/(E')_D^{i-1}\hookrightarrow E_D^i/E_D^{i-1}$. 

Then, applying the above recursive construction, one obtains a 
subsheaf $F'_0 \subset F_0$ with determinant 
\[
{\rm rk}(E') \Sigma_0 - \sum_{i=1}^\ell m_i' \Delta_i 
\]
where $m_i' ={\rm length}_{\CO_D}(E')_D^i/(E')_D^{i-1}$. 
This implies that 
\[
\mu_\beta(F'_\ell) = { {\rm deg}(E') + \sum_{i=1}^\ell \alpha_i 
\chi((E')_D^i/(E')_D^{i-1}) \over {\rm rk}(E')},
\]
and hence Bridgeland stability for the sheaf $F_0$ on $S_\uxi$ is equivalent to 
parabolic Higgs bundle stability for the data $(E,E_D^\bullet,\Phi)$ on $C$.

\subsection{Isomorphism of moduli stacks}\label{isomoduli}

This section concludes the proof of the spectral correspondence statement in Section \ref{spcorrespCY}. To summarize, recall that one has 
to choose a flat $B$-field background $\beta$ on $S_\uxi$ such that the only non-zero periods are
\[ 
\beta(\Xi_{n,i})= -\beta_i, \qquad 1\leq i \leq \ell
\]
with $\beta_i\in \IR$, $0< \beta_\ell < \cdots < \beta_1 < 1$. 
Then on one side of the correspondence one has $\beta$-semistable compactly supported pure dimension one sheaves $F$
 on $S_\uxi$ with topological invariants 
\[
\ch_1(F) = \Sigma_{\um}, \qquad \chi(F) =c,
\]
where $\um=(m_1, \ldots, m_\ell)\in \big( \IZ_{\geq 1}\big)^{\times \ell}$ and $c\in \IZ$ and 
\[
\Sigma_{\um}=r \Sigma_0 - \sum_{i=1}^\ell m_i \Delta_i
\]
is a curve class as in \eqref{eq:strtrnsfclass}. 
Let ${\mathfrak M}_\beta^{ss}(S_\uxi; \um,c)$ denote the moduli stack of such objects.  

On the other side one has semistable irregular $\uxi$-parabolic 
Higgs bundles on $C$ with numerical invariants $\um$, $d= c + 
r(g-1)$, where $r =\sum_{i=1}^\ell m_i$. The parabolic weights are given by $\alpha_i = \beta_i$, $1\leq i \leq \ell$. As in Section \ref{setup}, let 
${\mathfrak H}_{\uxi}(C,D;\ualpha, \um, d)$ denote the moduli stack of such semistable objects. 

The spectral correspondence states that there is an isomorphism 
\be
{\mathfrak M}^{ss}_{\beta}(S_\uxi; \um,c)\simeq 
{\mathfrak H}_{\uxi}(C,D;\ualpha, \um, d).
\ee
The results proven in the previous sections yield an isomorphism 
between the sets of closed points. In order to conclude the proof, one has to show that this correspondence holds for flat families. This is a fairly straightforward, although tedious, exercise which proceeds in close analogy to \cite[Section 7]{modADHM}. 
The details will be omitted.

\subsection{The Calabi-Yau threefold}\label{threefold} 

In order to make the connection with Donaldson-Thomas theory, let $Y_\uxi$ be the total space of the canonical 
bundle of $S_\uxi$. Hence $Y_\uxi$ is a smooth quasi-projective Calabi-Yau threefold. The flat $B$-field background $\beta\in H^2_c(S_\uxi, \IR)$ lifts to a $B$-field 
on $Y_\uxi$ by pull-back, and any compactly supported curve class on $S_\uxi$ yields a curve class on $Y_\uxi$ by 
pushforward via the zero section. In particular one 
can define a natural notion of $\beta$-stability for pure dimension one sheaves $F$ on $Y$ with topological 
invariants 
\[
\ch_2(F) = \Sigma_m, \qquad \chi(F) =c . 
\]
Let ${\mathfrak M}^{s}_\beta(Y_\uxi; \um, c)$ be the moduli stack of such $\beta$-stable objects. Then it is straightforward to show that there is an isomorphism 
\[
{\mathfrak M}^{s}_\beta(Y_\uxi; \um, c)\simeq 
{\mathfrak M}^{s}_\beta(S_\uxi; \um, c)\times \IA^1.  
\]
In close analogy with \cite{BPSPW, Par_ref}, the refined Gopakumar-Vafa formula will yield an explicit relation  between the cohomology of the moduli spaces of irregular $\uxi$-parabolic Higgs bundles and the refined stable pair theory 
of $Y_\uxi$. This will be spelled out in detail in the next sections.

\section{Refined stable pairs and torus links}\label{stpairsect}

The next goal is to derive an explicit conjectural formula  for 
refined stable pair invariants on the threefolds $Y_\uxi$ 
constructed in the previous section. This section 
provides such a formula for genus zero curves $C$, in which case the threefold $Y_\uxi$ admits a torus action. 
This action localizes the stable pair theory to planar configurations of curves on $S_\uxi$ which are related in the framework of Oblomkov-Shende  \cite{OS,ORS} to 
$(\ell, (n-2)\ell)$-torus links. Explicit conjectural results for the 
refined stable pair theory can then be derived  from refined 
Chern-Simons theory \cite{refCS}, as shown in detail below.

\subsection{Torus action and invariant curves}\label{toractsect}

In this section $C\simeq \IP^1$, and $M=K_C(np)$, 
$n\geq 3$,  
for a fixed point $p\in C$. Let $(U,z)$, $(V,w)$ be standard affine coordinates on $\IP^1$ with transition function 
$w=1/z$. Suppose $p\in U$ is given by $z=0$. Let 
${\bf T}\times C \to C$ be the $\IC^\times$ action on $\IP^1$
given locally by $(t,z)\mapsto tz$. 
Note that there is a unique lift of the $\IC^\times$ action 
to the total space of $M$ which leaves the fiber $M_p$ pointwise fixed. Suppose the sections $\uxi=(\xi_1, \ldots, \xi_\ell)$ 
are equivariant for this torus action, and otherwise generic
as in equation \eqref{eq:genlocsectA}. 
Then the action on $M$ lifts further to a torus action 
${\bf T}\times S_\uxi \to S_\uxi$. Furthermore there is also a unique lift ${\bf T}\times Y_\uxi \to Y_\uxi$ such that the canonical 
class of $Y_\uxi$ is equivariantly trivial. 
This torus action will be used to localize the compactly 
supported stable pair theory of $Y_\uxi$ to a planar configuration of rational curves contained in $S_\uxi$. 

In order to determine the configuration of 
${\bf T}-$invariant rational curves in $S_\uxi$ it will be 
helpful to first do so on the total space of $M$. One has 
two affine open charts $U$ with coordinates $(z,u)$ and 
$V$ with coordinates $(w,v)$ and transition functions 
\[ 
w=1/z, \qquad v = z^{2-n}u. 
\]
The torus action is locally given by 
\[
(t, (z,u)) \mapsto (tz, u) \quad {\rm and}\quad 
(t, (w,v)) \mapsto (t^{-1}w, t^{2-n}v). 
\]
Therefore there is a one parameter family of ${\bf T}-$invariant sections of $M$ over $C$ given by the local 
equations 
\[ 
u=\lambda, \qquad v=\lambda w^{n-2}
\]
where $\lambda$ is a complex parameter. 

Next recall that $S_\uxi$ is obtained by blowing up the 
divisors $\delta_i \subset M_D \subset M$ given by 
\[ 
y - \pi^*\xi_i=0, \qquad 1\leq i\leq \ell,
\]
where $y\in H^0(M, \pi^*M)$ is the tautological section. 
Assuming the sections $\xi_1, \ldots, \xi_\ell$ generic, 
the divisors $\delta_i$ will intersect the reduced fiber $M_p$ 
at $\ell$ distinct points $\wp_{1,i}$, $1\leq i \leq \ell$. 
In local coordinates, these points will be given by 
\[ 
z=0, \quad u=\lambda_i, \qquad 1\leq i \leq \ell
\]
for some nonzero, pairwise distinct, complex numbers $\lambda_1, \ldots, \lambda_\ell$. Since $S_\uxi\subset T_\uxi$ is the complement of the total transform of $M_p$, the only ${\bf T}$-invariant compact 
rational curves 
on ${S}_\uxi$ will be the strict transforms ${\Sigma}_i$ of 
the sections  $C_i\subset M$ given locally by 
\[ 
u=\lambda_i, \qquad 1\leq i \leq \ell. 
\]
It is straightforward to check that
\[
\rho^*C_i = \Sigma_i + \sum_{a=1}^n a\Xi_{a,i}, \qquad 
1\leq i \leq \ell, 
\]
where $\Xi_{a,i}$ are the exceptional divisors of the blow-up 
map $\rho: T_\uxi \to M$. Each strict transform $\Sigma_i$ 
intersects the last exceptional divisor $\Xi_{n,i}$ in the $i$-th 
chain transversely at an isolated ${\bf T}$-fixed point $\wp_i$, 
and does not meet any other exceptional divisors. 
By construction, the affine open subset $V_\uxi= \rho^{-1}(V)\subset S_\uxi$ is isomorphic to $V\subset M$. 
In this chart, the curves $\Sigma_j\subset S_\delta$ will be given by 
\[ 
v =\lambda_i w^{n-2}. 
\]

\begin{figure}
\setlength{\unitlength}{1mm}
\hspace{-70pt}
\vspace{-30pt}
\begin{picture}(250,100)
{\thicklines\put(15,10){\line(1,0){180}}
\put(30,5){\line(0,1){90}}}
\put(185,6){$\Sigma_0$}
\put(151,7){${\bf o}$}
\put(27,90){$f$}
\put(29,20){$\bullet$}
\put(29,50){$\bullet$}
\put(29,75){$\bullet$}
\put(30,21){\line(2,1){22}}
\put(46,28.5){$\bullet$}
\put(42,32){\line(2,-1){22}}
\multiput(72,26)(5,0){3}{\huge .}
\put(90,32){\line(2,-1){22}}
\put(105,23){$\bullet$}
\put(100,21){\line(2,1){22}}\put(122,31.5){$\bullet$}
{\color{blue}\qbezier(123,32)(152,-12)(183,32)}
\put(184,32){$\Sigma_\ell$}
\put(30,51){\line(2,1){22}}
\put(46,58.5){$\bullet$}
\put(42,62){\line(2,-1){22}}
\multiput(72,56)(5,0){3}{\huge .}
\put(90,62){\line(2,-1){22}}
\put(105,53){$\bullet$}
\put(100,51){\line(2,1){22}}\put(122,61.5){$\bullet$}
{\color{blue}\qbezier(123,62)(152,-42)(183,62)}
\put(184,62){$\Sigma_2$}
\put(30,76){\line(2,1){22}}
\put(46,83.5){$\bullet$}
\put(42,87){\line(2,-1){22}}
\multiput(72,79)(5,0){3}{\huge .}
\put(90,87){\line(2,-1){22}}
\put(105,78){$\bullet$}
\put(100,76){\line(2,1){22}}\put(122,86.5){$\bullet$}
{\color{blue}\qbezier(123,88)(152,-68)(183,88)}
\put(184,88){$\Sigma_1$}
\multiput(46,38)(0,5){3}{\huge .}
\put(37,21){$\Xi_{1,\ell}$}
\put(37,51){$\Xi_{1,2}$}
\put(37,76){$\Xi_{1,1}$}
\put(55,26.5){$\Xi_{2,\ell}$}
\put(55,56.5){$\Xi_{2,2}$}
\put(55,81.5){$\Xi_{2,1}$}
\put(95,30.5){$\Xi_{n-1,\ell}$}
\put(95,60.5){$\Xi_{n-1,2}$}
\put(95,85.5){$\Xi_{n-1,1}$}
\put(115,24.5){$\Xi_{n,\ell}$}\put(120,36){$\wp_{n,\ell}$}
\put(115,54.5){$\Xi_{n,2}$}\put(120,66){$\wp_{n,2}$}
\put(115,79.5){$\Xi_{n,1}$}\put(120,91){$\wp_{n,1}$}
\multiput(184.5,42)(0,5){3}{\huge .}
\end{picture}
\caption{Torus invariant curves on $S_\uxi$.}
\label{SurfaceII}
\end{figure}

\noindent
The resulting curve configuration is schematically represented in Figure \ref{SurfaceII}. 
Note that all intersection points between exceptional divisors, 
as well as the intersection points between exceptional divisors 
and the invariant sections $\Sigma_1, \ldots, \Sigma_\ell$ are 
isolated fixed points of the torus action. Moreover, each section 
$\Sigma_i$ has self-intersection 
\[
\Sigma_i^2=n-2
\]
on $S_\uxi$ and intersects the strict transform $\Sigma_0$ 
of the zero section at the point ${\bf o}$ given by $v=0, w=0$. 
The intersection multiplicity at ${\bf o}$ is 
\[ 
\Sigma_i \cdot \Sigma_0 = n-2 
\]
for all $1\leq i \leq \ell$. Finally note that the torus fixed 
locus on the surface $S_\uxi$ is the finite set $\{ \wp_{n,1}, \ldots, \wp_{n,\ell}, {\bf o}\}$ where 
$\wp_{n,i}$ is the intersection point between $\Sigma_i$ and 
$\Xi_{n,i}$.

\subsection{Stable pair theory and localization}\label{stpairloc}
Now recall that the stable pair theory \cite{stabpairsI}
counts 
complexes of the form  $\CO_{Y_\uxi} {\buildrel s\over \longto}F$, where $F$ is a pure dimension one sheaf on $Y_\uxi$
and $s$ a 
generically surjective section. Since $Y_\uxi$ is noncompact, $F$ will be required to have compact support. This may still result in a noncompact moduli space, but, as shown below the fixed loci under the above torus action are compact. 
Therefore one can define stable pair invariants by residual localization. There are two ways one could approach this in the refined theory, namely using the motivic construction due to Kontsevich and Soibelman \cite{wallcrossing} or the via the 
$K$-theoretic index as in the work of Nekrasov and Okounkov \cite{Membranes_Sheaves}. These two constructions in fact lead to identical refined invariants according to \cite{HRV_proof}. 
The 
approach employed in this paper will be that of  \cite{Membranes_Sheaves}, which has been shown in loc. cit. to
be the mathematical theory of 
the refined vertex of \cite{ref_vert} in a toric framework. 
Specializing to the present context, the main points of this construction will be summarized below.

First, in addition to the torus action ${\bf T}\times Y_\uxi\to Y_\uxi$ constructed in Section \ref{toractsect}, there is a second torus action $\IC^\times \times Y_\uxi \to Y_\uxi$ which scales the fibers of $Y_\uxi\to S_\uxi$ leaving the zero section pointwise fixed. These two torus actions commute, hence one 
has a two dimensional torus action ${\bf G}\times Y_\uxi \to Y_\uxi$ with ${\bf G} = {\bf T}\times \IC^\times$. Moreover, 
the canonical subtorus ${\bf T}\subset {\bf G}$, $t\mapsto (t,1)$ is the subgroup of ${\bf G}$ which acts trivially on the holomorphic three-form on $Y_\uxi$. 

Now let 
${\mathcal P}(Y_\uxi,\um,c)$ be the moduli space of stable  
pairs on $Y_\uxi$ with topological invariants 
\[ 
\ch_2(F) = \Sigma_\um, \qquad \chi(F)=c.
\]
as in Section \ref{isomoduli}. 
This moduli space has a perfect obstruction theory 
$\CE^\bullet = \big(\CE^1\to \CE^2\big)$ of amplitude $[0,\ 1]$, which 
yields a virtual structure sheaf ${ \CO}^{\sf vir}$ and a virtual canonical bundle ${\CK}$. 
All these constructions are naturally ${\bf G}$-equivariant, 
but it is 
essential to note that the 
prefect obstruction theory is not {\bf G}-equivariantly perfect. 
This means that one has an isomorphism 
\[ 
\big(\CE^\bullet\big)^\vee \simeq \CE^\bullet[-1]
\]
in the derived category of the moduli space, but this isomorphism is not ${\bf G}$-equivariant. 
In fact, as shown in \cite[Section 7.1.3]{Membranes_Sheaves}, 
one has an isomorphism 
\[
\big(\CE^\bullet\big)^\vee \simeq R\otimes\CE^\bullet[-1]
\]
in the ${\bf G}$-equivariant derived category, where 
$R$ is the one dimensional representation of ${\bf G}$ with character $(t,\zeta) \mapsto \zeta$. For future reference, this character of $R$ will be denoted by $\kappa$.

In the construction of refined invariants of \cite{Membranes_Sheaves} one has to choose a 
square root ${\CK}^{1/2}$ which is moreover
equivariant under  the double cover 
${\widetilde{\bf G}}=\IC^\times \times \IC^\times 
\mapsto {\bf G}$ given by $(t, {\tilde \zeta}) \to (t, {\tilde \zeta}^2)$. The representation of ${\widetilde{\bf G}}$ with character $(t,{\tilde \zeta}) \mapsto {\tilde \zeta}$ will be denoted by $R^{1/2}$, while its character will be denoted by 
$\kappa^{1/2}$. 

In situations where the moduli space of stable pairs is compact, the refined 
invariants are defined as the equivariant Euler characteristic 
$\chi_{{\widetilde{\bf G}}}({\widetilde \CO})$,
where ${\widetilde \CO} = \CO^{\sf vir} \otimes \CK^{1/2}$. 
According to \cite[Section 7.1.3, Thm 1]{Membranes_Sheaves}, this is an 
element of $\IZ[\kappa^{1/2}]$ if the compactness assumption is satisfied. Here $\kappa^{1/2}=y$ serves as the refined variable. 

For non-compact moduli spaces, refined invariants are defined by adding contributions of the {\bf T}-fixed loci, 
provided the latter are compact. In the present case, this will be shown to be the case below. Then the local contribution of a 
connected component of the fixed locus is given in 
 \cite[Section 7.2.4, Prop. 7.3]{Membranes_Sheaves}. 
The details will be omitted since the formula proved in 
loc. cit. will not be used for explicit computations in this paper. 
The strategy employed in the following will be to derive 
an explicit conjectural formula based on correspondence with 
link invariants and refined Chern-Simons theory. 

The first step in this direction is to understand the combinatorial 
classification of the fixed loci.  
All ${\bf T}$-invariant pure dimension one sheaves will be 
set theoretically supported on the ${\bf T}$-invariant curve 
$\sum_{i=1}^\ell \Sigma_i$ in $S_\uxi$. 
Since the section 
$s:\CO_Y \to F$ is generically surjective, there is an exact sequence 
\be\label{eq:pairseq}
0\longto \CO_Z \longto F \longto Q\longto 0 
\ee
where $Z$ is the scheme theoretic support of $F$ and $Q$ is the cokernel of $s$. As shown in \cite{stabpairsI} $Z$ has to be a Cauhen-Macaulay scheme of pure dimension one while the support of 
$Q$ is zero dimensional. 
For torus invariant configurations, $Z$ will have $\ell$ irreducible components, 
$Z_1, \ldots, Z_\ell$ set theoretically supported on $C_1, \ldots, C_j$ respectively with generic multiplicities $m_1, \ldots, m_\ell$.  Moreover, there is a direct sum decomposition $Q= \oplus_{i=1}^\ell Q_i \oplus Q_o$ 
where $Q_j$ are zero dimensional sheaves supported at the torus fixed points $\wp_i$ while $Q_{\bf o}$ is a zero dimensional sheaf supported at ${\bf o}$. 

Now the main observation is that the restriction of $Z$ to the open subset $\rho^{-1}(V)\subset S_\uxi$ has a very simple form. 
Namely, if $\eta$ is the natural normal linear coordinate to $S_\uxi$ in $Y_\uxi$ defined over $V$, the defining ideal of $Z\cap V$ is generated by polynomials of the form 
\[
\eta^{\mu_{i,a}}(v-\lambda_i z^{n})^a
\]
where $a\in \IZ$, $1\leq a \leq b_i$ for some fixed positive integer $b_i$ and 
$\mu_{i,a}\in \IZ$ satisfy 
\[ 
\mu_{i,1} \geq \mu_{i,2} \geq \cdots \geq \mu_{i,b_j} \geq 1.
\]
In particular the scheme structure of $Z\cap V$
is completely determined by the Young diagram 
$\mu_1 = (\mu_{i,1}, \ldots, \mu_{i,b_j})$ where 
$\mu_{i,a}$ denotes the number of boxes in the $a$-th row. 
Moreover, given a collection of partitions $(\mu_1, \ldots, \mu_\ell)$ there is a unique minimal subscheme $Z$ which agrees with which agrees with the 
above presentation over $V$.

This implies that the torus fixed locus in the moduli space
of stable pairs will be a disjoint union of subloci 
$\calP_{\mu_1,\ldots, \mu_\ell}$ labelled by ordered collections 
of $\ell$ diagrams. For a fixed collection $(\mu_1, \ldots, \mu_\ell)$ the structure sheaf $\CO_Z$ in \eqref{eq:pairseq} is fixed while the cokernel $Q$ must be set theoretically supported on the finite set $S_\uxi^{\bf T}$.  
This implies in particular that the fixed loci in the moduli space of pairs are compact, hence one could in principle apply the formalism of \cite{Membranes_Sheaves}
to compute the local contributions. 
Given the combinatorial structure of the fixed loci, the refined stable pair partition function 
is expected to have a vertex expression of the form 
\be\label{eq:PTYa}
Z_{Y_\uxi}(q,y,\ux_1, \ldots, \ux_\ell) = 
\sum_{\mu_1, \ldots, \mu_\ell} V^{(n)}_{\mu_1, \ldots, \mu_\ell}(q,y) \prod_{i=1}^\ell Z_{\mu_i}(q,y)
\prod_{i=1}^\ell Q_i^{|\mu_i|},
\ee
where $Q_i^{|\mu_i|}$ are degree counting variables associated to the curves $\Sigma_i$, $1\leq i \leq \ell$. By convention, throughout this section the 
leading term in all vertex expansions corresponding to empty partitions will always be $1$. 

As shown in \cite[Section 8]{Membranes_Sheaves}, each factor 
$Z_{\mu_i}(q,y)$ is equal to a one leg refined vertex where the 
partition $\mu_i$ labels an unpreferred leg.
Therefore one is left with the multileg vertex  $V_{\mu_1, \ldots, \mu_\ell}(q,y)$. Since the direct localization computation is very difficult, a more effective strategy
is to derive a conjectural formula for  $V_{\mu_1, \ldots, \mu_\ell}(q,y)$ using the framework of \cite{OS,ORS,DHS} 
which related stable pair invariants to link invariants. 
 In particular a close relation is expected between the vertex 
$V_{\mu_1, \ldots, \mu_\ell}(q,y)$ and the refined $(\mu_1, \ldots, \mu_\ell)$-colored invariant of 
the $(\ell, (n-2)\ell)$ torus 
link. Although no rigorous computations of such invariants are available, a conjectural expression can be obtained 
from refined Chern-Simons theory \cite{refCS}. 

Before explaining the details, one should note that this approach involves an important subtlety in coupling the 
one leg vertices $Z_{\mu_i}(q,y)$ to the refined Chern-Simons expression for $V_{\mu_1, \ldots, \mu_\ell}(q,y)$. Namely, as shown 
in Section \ref{torlinks} below, the refined link invariantsin Chern-Simons theory are written in the Macdonald basis, while the  geometric approach explained above uses the Schur basis. Therefore coupling this quantities requires 
a reformulation of the toric refined vertex formalism  in terms of Macdonald polynomials. This is already manifest 
in the large $N$ duality treatment of the refined Hopf link in \cite{Hopf_revisited}.  Fortunately, such a Macdonald formalism has been already developed in  \cite{Md_vert_I,Md_vert_II}, some details being explained 
below for a particular example. 

Consider a toric Calabi-Yau threefold containing a planar configuration of two 
$(-1,-1)$ curves. The Delzant polytope of such a toric variety is 
shown in Figure \ref{Toric}.
\begin{figure}
\setlength{\unitlength}{1mm}
\hspace{80pt}
\setlength{\unitlength}{1mm}
\begin{picture}(120,80)
\put(10,10){\line(1,1){20}}
\put(30,30){\line(1,-1){15}}
\put(36,25){$Q_1$}
\put(45,15){\line(1,1){15}}
\put(60,30){\line(1,-1){20}}
\put(60,30){\line(0,1){20}}
\put(30,30){\line(0,1){20}}
\put(49,25){$Q_2$}
\end{picture}
\caption{}
\label{Toric}
\end{figure}
As shown for example in \cite{Iqbal:2008ra}, the refined vertex partition 
function for such a configuration is given by 
\be\label{eq:refvertA}
\sum_{\mu_1, \mu_2} (-Q_1)^{|\mu_1|} (-Q_2)^{|\mu_2|} 
C_{\varnothing, \mu_1^t, \varnothing}(t,s) 
C_{\mu_2^t,\mu_1,\varnothing}(s,t) C_{\mu_2,\varnothing,\varnothing}(t,s) 
\ee
where $Q_1,Q_2$ are formal counting variable associated to the two $(-1,-1)$ curves and the sum is over all  pairs of Young diagrams $(\mu_1, \mu_2)$. 
The refined vertex expressions in the above formula are 
\[ 
C_{\varnothing, \mu_1^t, \varnothing}(t,s) = t^{-n(\mu_1^t)} s^{n(\mu_1)} S_{\mu_1^t}(\us), \qquad 
\us = (s^{1/2}, s^{3/2}, \ldots) 
\]
\[
C_{\mu_2, \varnothing, \varnothing}(t,s) = (s/t)^{|\mu_2|/2} S_{\mu_2^t}(\ut), \qquad 
\ut=(t^{1/2}, t^{3/2}, \ldots) 
\]
\[
C_{\mu_2^t,\mu_1,\varnothing}(s,t) = t^{n(\mu_1^t)} s^{-n(\mu_1)} 
\sum_{\eta} (t/s)^{(|\mu_2|+|\eta|)/2} S_{\mu_2/\eta}(\us) S_{\mu_1/\eta}(\ut)
\]
Employing standard notation, $S_\nu(x_1, x_2, \ldots)$ denotes the Schur function associated to the partition $\nu$ and $S_{\nu/\rho}(x_1,x_2, \ldots)$ denotes the skew Schur function associated to a pair of partitions $(\nu,\rho)$. Moreover, 
\[ 
n(\nu) = \sum_{\Box\in \nu} l(\Box), 
\]
where $l(\Box)$ is the leg length of a box in $\nu$. 

By convention, the Young diagram corresponding to a partition 
$\nu=(\nu_1\geq \nu_2 \geq \cdots \geq \nu_{l(\nu)})$ consists 
of $l(\nu)$ left-aligned horizontal rows such that the $i$-th row contains 
$\nu_i$ boxes. For example the partition $(7,4,3,1)$ is represented as follows $$\yng(7,4,3,1)$$ The leg length of a box $\box\in \nu$ is the number of boxes in the same vertical column and laying strictly below the given box in the Young diagram. The arm length of a box $\box\in \nu$ is the number of boxes on the same horizontal row and laying strictly to the right of the given box in the diagram.

In  the above partition function both compact curves 
correspond to ordinary, unpreferred legs of the refined vertex. 
The sums over partitions in 
\eqref{eq:refvertA} can be easily computed using Schur function identities, obtaining 
\be\label{eq:refvertB}
\prod_{i,j=1}^\infty (1-Q_1 s^{i-1/2} t^{j-1/2}) 
(1-Q_2 s^{i-1/2} t^{j-1/2}) 
(1- Q_1 Q_2 s^{i-1} t^{j})^{-1}.
\ee
Using symmetric function identities, this formula can be alternatively written in terms of Macdonald polynomials as 
\be\label{eq:MDvertA}
\sum_{\mu_1,\mu_2} (-Q_1)^{|\mu_1|}(-Q_2)^{|\mu|}
P_{\mu_1^t}(t,s,\us) P_{\mu_1}(s,t;\ut) P_{\mu_2}(s,t; s^{-\mu_1}\ut) P_{\mu_2^t}(t,s;\us),
\ee
where for any Young diagram $\nu$ with row lengths 
$\nu_1\geq \nu_2 \geq \cdots$,  
\[ 
s^{\nu} \ut = (s^{-\nu_1} t^{1/2}, s^{-\lambda_2} t^{3/2},\ldots). 
\]
Note that it may seem natural to 
think of the product 
$P_{\mu_1}(s,t;\ut)P_{\mu_2}(s,t; s^{-\mu_1}\ut)$ in the right hand side of \eqref{eq:MDvertA} 
as a refined vertex with two preferred legs. However one should keep in mind that no such direct enumerative interpretation is possible in the 
formalism of \cite{Membranes_Sheaves}. 
Finally, in order to make a concrete connection with refined stable pair theory, note that the refined vertex variables $(s,t)$  
are related to $(q,y)$ by $s = qy$, $t=qy^{-1}$. 
This was observed for example in \cite{BPSPW, CKK,  Par_ref}. 

Collecting all the  facts, one is then led to conjecture that the partition function \eqref{eq:PTYa} admits an alternative expansion of the form 
\be\label{eq:PTYb} 
Z_{Y_\uxi}(q,y,\ux_1, \ldots, \ux_\ell) = 
\sum_{\mu_1, \ldots, \mu_\ell} {\widetilde V}^{(n)}_{\mu_1, \ldots, \mu_\ell}(q,y) \prod_{i=1}^\ell 
{\widetilde Z}_{\mu_i}(q,y) \prod_{i=1}^\ell Q_i^{|\mu_i|} 
\ee
where 
\[
{\widetilde Z}_{\mu_i}(q,y) = P_{\mu_i^t}(t,s;\us)
\big|_{s=qy,\ t=qy^{-1}}
\]
and 
${\widetilde V}^{(n)}_{\mu_1, \ldots, \mu_\ell}(q,y)$ is 
directly related to colored refined link in invariants in refined 
Chern-Simons theory. An explicit conjectural formula for these invariants is derived in the next subsection, also using some 
large $N$ duality input.

\subsection{Torus links in refined Chern-Simons theory}\label{torlinks}

Colored torus knot invariants in refined Chern-Simons theory 
have been studied in detail in \cite{refCS,Colored_HL}. Following 
Section 2 of loc. cit., the main elements in their construction are the following. 

\begin{itemize}
\item 
The Hilbert space $\CH_{N,k}$ of rank $N$ level $k$ refined Chern-Simons theory, which is a subspace of the algebra $\Lambda_N$ 
of class functions on the $SU(N)$ group manifold. This space 
has a basis $\{|P_\lambda\rangle\}$ consisting of Macdonald 
polynomials labelled by partitions $\lambda=\big(\lambda_1 \geq \lambda_2\geq \cdots \lambda_{l(\lambda)}\big)$ of length 
$l(\lambda) \leq N-1$ with $\lambda_1 \leq k$. 
For 
\be\label{eq:stformula}
s = e^{{2\pi\sqrt{-1}\over k + \beta N}}, \qquad 
t = e^{{2\pi\sqrt{-1}\beta \over k + \beta N}}, \qquad 
\beta \in \IC^\times, 
\ee
this space
carries a linear representation $\rho:SL(2, \IZ)\to GL(\CH_{N,k})$ which will not be written in detail here. 

\item For each partition $\mu$ as above, a linear operator 
$\CO_\mu: \CH_{N,k}\to \CH_{N,k}$ 
defined by 
\[ 
\CO_\mu(|P_\lambda\rangle ) = \sum_{\sigma} N^\sigma_{\mu, \lambda}(s,t) |P_\sigma\rangle 
\]
where $N^\sigma_{\mu, \lambda}$ are the 
$(s,t)$-Littlewood-Richardson coefficients. 
\end{itemize} 
For the purpose of large $N$ duality it suffices to take the stable limit of the theory, as in Section 3 of \cite{Colored_HL}. 
This amounts to 
sending $N,k\to \infty$ in equation
\eqref{eq:stformula} while keeping $s,t$  fixed. This limit effectively removes all 
constraints on partitions, hence one has to work with all 
Young diagrams in the following. Accordingly, the large $N$ Hilbert 
space will be denoted by $\CH_\infty$. 

The large $N$ refined invariant of the $(\ell, p\ell)$ torus 
link is then  given by 
\[ 
W^{(p)}_{\mu_1, \ldots, \mu_\ell}(s,t) = 
\langle \varnothing | S \CW_{\mu_1, \ldots, \mu_\ell} |\varnothing \rangle 
\]
where $\CW_{\mu_1, \ldots, \mu_\ell}: \CH_\infty \to \CH_\infty$ is the linear operator 
\[ 
\CW_{\mu_1, \ldots, \mu_\ell} = \rho(U_{n,1}) \CO_{\mu_\ell} 
\cdots \CO_{\mu_1} \rho(U_{1,n}^{-1}), \qquad 
U_{p,1} = \left( \begin{array}{cc} 1 & 1 \\ p & p+1 \end{array}\right).
\] 
Using the definition of $\CO_\mu$, this can be written as 
\[ 
W^{(p)}_{\mu_1, \ldots, \mu_\ell}(s,t) = \sum_{\lambda_1, \ldots, \lambda_{\ell-1}} N^{\lambda_{\ell-1}}_{\mu_\ell, \lambda_{\ell-2}} N^{\lambda_{\ell-2}}_{\mu_{\ell-1}, \lambda_{\ell-3}}\cdots N^{\lambda_2}_{\mu_3, \lambda_1} 
N^{\lambda_1}_{\mu_2, \mu_1} W_{\lambda_{\ell -1}}(s,t)
\] 
where $W_{\lambda_{\ell-1}}(s,t)$ is the refined invariant 
of the $(1, p)$-knot colored by $\lambda_{\ell-1}$. 

Computing this quantity directly in refined Chern-Simons theory leads to rather complicated symmetric function 
identities, as shown in Section 4 of \cite{Colored_HL}. 
However, one can infer the final answer from large 
$N$ duality for conifold transitions as in \cite{DSV,DHS}. 
At large $N$, the colored refined invariant $W_{\lambda_{\ell-1}}(s,t)$ is identified up to normalization 
factors to the framed refined stable pair theory of a plane 
curve of the form $y = x^p$ embedded in a fiber of 
the resolved conifold over $\IP^1$. This curve is preserved by 
a torus action, which reads locally 
\[ 
t\times (x,y,z) \mapsto (tx,t^py, t^{-p-1}z)
\]
where $z$ is a normal coordinate to the chosen fiber. Hence the refined stable pair theory can be computed by localization. However, since the curve $y = x^p$ is smooth and isomorphic to the complex line, the result is the same as 
the framed refined stable pair theory of the curve $y=0$. 
The latter is the leading term in the $a$-expansion of the colored invariant of the $(1,0)$ torus knot, commonly referred to as the unknot. There is one slight subtlety in this argument, 
namely the above choice of torus action translates into $p$ units of framing for the refined unknot \cite{Open_GW}. Therefore large $N$ duality leads to the conjectural formula 
\[ 
W_\lambda(s,t) = {\langle \varnothing | T^p S|\varnothing \rangle 
\over \langle \varnothing | S|\varnothing \rangle } 
\]
Using the explicit expressions for $S,T$ in the stable limit, 
this yields 
\be\label{eq:ref_unknot} 
W_\lambda(s,t) = \left({t/s}\right)^{p|\lambda|/2}f_\lambda(s,t)^{-p} P_\lambda(s,t; \ut) 
\ee
where $\ut=(t^{1/2}, t^{3/2}, \ldots)$ and 
\[ 
f_\lambda(s,t) = \prod_{\Box\in \lambda} s^{a(\Box)} t^{-l(\Box)}.
\]
Therefore
\[
W^{(p)}_{\mu_1, \ldots, \mu_\ell}(s,t) = \sum_{\lambda_1, \ldots, \lambda_{\ell-1}} N^{\lambda_{\ell-1}}_{\mu_\ell, \lambda_{\ell-2}} N^{\lambda_{\ell-2}}_{\mu_{\ell-1}, \lambda_{\ell-3}}\cdots N^{\lambda_2}_{\mu_3, \lambda_1} 
N^{\lambda_1}_{\mu_2, \mu_1} \left({t/s}\right)^{p|\lambda_{\ell-1}|/2}f_{\lambda_{\ell-1}}(s,t)^{-p} P_{\lambda_{\ell-1}}(s,t; \ut)
\]
Note that $|\lambda_{\ell-1}| = |\mu_1|+\cdots + |\mu_\ell|$ 
for all nonzero term in the right hand side of the above equation. Hence one further obtains 
\be\label{eq:refCSvert}
\bal
& W^{(p)}_{\mu_1, \ldots, \mu_\ell}(s,t) =\\
& \left({t/s}\right)^{p(|\mu_1|+\cdots +|\mu_\ell|)/2}\sum_{\lambda_1, \ldots, \lambda_{\ell-1}} N^{\lambda_{\ell-1}}_{\mu_\ell, \lambda_{\ell-2}} N^{\lambda_{\ell-2}}_{\mu_{\ell-1}, \lambda_{\ell-3}}\cdots N^{\lambda_2}_{\mu_3, \lambda_1} 
N^{\lambda_1}_{\mu_2, \mu_1} f_{\lambda_{\ell-1}}(s,t)^{-p} P_{\lambda_{\ell-1}}(s,t;\ut).\\
\eal
\ee
For consistency, note that the unrefined specialization, $s=t$, of this formula is in agreement with the formulas obtained in \cite{Hecke_colored} for colored Homfly polynomials of torus links. 

\subsection{The final formula}\label{finalform}

Based on the colored refined generalization of the 
conjecture of \cite{OS}, the  vertex ${\widetilde V}_{\mu_1, \ldots, \mu_\ell}(s,t)$ 
is expected to be related to $W^{(n-2)}_{\mu_1, \ldots, \mu_\ell}(s,t)$ by a change of variables, up to a normalization factor. 
As shown in \cite{Hopf_revisited}, the variables $(s,t)$ used in refined 
Chern-Simons theory are the same as those used in the refined 
vertex formalism \cite{ref_vert}, hence they are related to $(q,y)$ by 
$s=qy$ and $t=qy^{-1}$. Therefore, as conjectured in \cite{DHS},  one expects a relation of the form 
\be\label{eq:largeNrelA}
{\widetilde V}^{(n)}_{\mu_1, \ldots, \mu_\ell}(q,y)= w^{(n)}_{\mu_1, \ldots, \mu_\ell}(s,t)
W^{(n-2)}_{\mu_1, \ldots, \mu_\ell}(s,t)\big|_{s=qy,\ t=qy^{-1}} 
\ee
where $w^{(n)}_{\mu_1, \ldots, \mu_\ell}(s,t)$ is a monomial in $(q,y)$. 
Moreover, by analogy with previous known large $N$ duality results, 
$w_{\mu_1, \ldots, \mu_\ell}(s,t)$ is expected to be a product  
$\prod_{j=1}^\ell w^{(n)}_{\mu_j}(s,t)$ of individual factors associated 
to the $\ell$ legs of the vertex. 

In order to determine these individual factors, note that for $\ell=1$ and any value of $n\geq 3$, the partition function 
$Z_{Y_\uxi}(q,y)$ reduces to the refined partition function of a 
local $(0,-2)$ curve, which can be computed using the refined vertex. 
Specializing equations \eqref{eq:PTYb},
 \eqref{eq:largeNrelA} to $\ell=1$,
one obtains 
\[ 
\bal
Z_{Y_\uxi}(q,y) & = \sum_{\mu}  Q_1^{|\mu|}w^{(n)}_{\mu}(s,t)W^{(n-2)}_{\mu}(s,t) P_{\mu^t}(t,s;\us)\big|_{s=qy,\ t=qy^{-1}}. \\
& = \sum_{\mu}  Q_1^{|\mu|} w^{(n)}_{\mu}(s,t) 
(ts^{-1})^{(n-2)|\mu|/2}
f_\mu(s,t)^{2-n}P_{\mu}(s,t; \ut) 
P_{\mu^t}(t,s;\us)\big|_{s=qy,\ t=qy^{-1}}.
\eal 
\]
The above expression reduces precisely to the refined vertex 
partition function of a $(0,-2)$ curve provided that 
\[
w^{(n)}_{\mu}(s,t) = (ts^{-1})^{(2-n)|\mu|/2} f_\mu(s,t)^{n-1}. 
\]
In this case one recovers the expression obtained in 
\cite[Section 5.3]{ref_vert} choosing the preferred direction of the refined vertex along the $(0,-2)$ curve.

In conclusion we are naturally led to the following conjectural expression for the refined stable pair partition function of 
$Y_\uxi$, 
\be\label{eq:PTYc}
\bal
& Z_{Y_\uxi}(q,y) = 
\sum_{\mu_1, \ldots, \mu_\ell} 
{\widetilde W}^{(n-2)}_{\mu_1, \ldots, \mu_\ell}(s,t) 
\prod_{i=1}^\ell \left( Q_i^{|\mu_i|} f_{\mu_i}(s,t)^{n-1} P_{\mu_i^t}(t,s;\us) \right)\bigg|_{s=qy,\ t=qy^{-1}}\\
\eal
\ee
where
\[
\bal
& {\widetilde W}^{(n-2)}_{\mu_1, \ldots, \mu_\ell}(s,t) =
 \sum_{\lambda_1, \ldots, \lambda_{\ell-1}} N^{\lambda_{\ell-1}}_{\mu_\ell, \lambda_{\ell-2}} N^{\lambda_{\ell-2}}_{\mu_{\ell-1}, \lambda_{\ell-3}}\cdots N^{\lambda_2}_{\mu_3, \lambda_1} 
N^{\lambda_1}_{\mu_2, \mu_1} f_{\lambda_{\ell-1}}(s,t)^{2-n} P_{\lambda_{\ell-1}}(s,t;\ut).\\
\eal
\]
Recall that the leading term in this formula corresponding to empty partitions is $1$ by convention. Therefore this completes the physical derivation of 
Conjecture 1.

\subsection{Refined Gopakumar-Vafa expansion}\label{refGV}
 
According to \cite{GV_II, KKV, Mtop, ref_vert}, string/M-theory arguments imply that the refined stable pair theory of $Y_\uxi$ has a conjectural refined Gopakumar-Vafa expansion. The BPS 
numbers present in this expansion are degeneracies of supersymmetric membrane bound states wrapping compact holomorphic curves in $Y_\uxi$ in $M$-theory. A mathematical theory of these invariants has been recently developed in 
\cite{GV_vanishing} generalizing the previous construction of 
\cite{HST}. For local models related to Hitchin systems as in the current paper, this construction identifies BPS degeneracies with 
perverse Betti numbers of moduli spaces of Higgs bundles, as 
explained in detail in \cite{BPSPW}. 

In string theoretic terms, the spectral correspondence proven in Section \ref{spectralsect} identifies moduli spaces supersymmetric 
D2-D0 configurations on $Y_\uxi$ with moduli spaces $\CH_\uxi^{s}(C,D; \ualpha, \um, d)$ of stable 
irregular $\uxi$-parabolic Higgs bundles. Throughout this section it will be assumed that the parabolic weights are fixed to some generic values such that all semistable objects are stable. 
Since the data $(C,D)$, $(\uxi, \ualpha)$ is fixed, in order to simplify the notation, 
the moduli spaces will be denoted by $\CH^s(\um, d)$, keeping track only of numerical invariants. These consist of 
a collection $\um = (m_1, \ldots, m_\ell)$ of positive integers 
encoding the D2-brane charge and the degree $d\in \IZ$ encoding the D0-brane charge. Proceeding by analogy with 
\cite{hodgechar}, one constructs a perverse Leray filtration 
on the cohomology of the moduli space $\CH^s(\um, d)$.
The BPS degeneracies in M-theory are the dimensions 
of the successive quotients $Gr^P_jH^k(\CH^s(\um, d))$ where 
$j\geq 1$ is the perverse degree and $k\geq 0$ is the cohomological degree. On physics grounds these numbers are 
expected to be independent of the degree $d$ for fixed 
$\um$, and also invariant under permutations of $(m_1, \ldots, m_\ell)$. 
Granting this fact, the perverse Poincar\'e polynomial 
of the moduli space $\CH^s(\um, d)$ will be denoted by
$P_{\mu,n}(u,v)$ where $\mu$ is the partition of $r$ determined 
by $(m_1, \ldots, m_\ell)$. 
 Hence for any $d$
\[ 
P_{{\mu},n}(u,v) = \sum_{j,k} {\rm dim}\, Gr^P_jH^k(\CH^s(\um, d))u^jv^k. 
\]

To conclude, by analogy with \cite{BPSPW}, one is then led to conjecture 
the following local BPS expansion of the refined stable pair theory of $Y_\uxi$:
\be\label{eq:refBPSb}
Z_{Y_\uxi}(q,\uQ,y) = {\rm exp}\,\left(- 
\sum_{k\geq 1} \sum_{|\mu|\neq 0} 
{m_\mu(Q_1^k, \ldots, Q_\ell^k,0,\ldots)\over k}
 {y^{-kr}
(qy^{-1})^{kd_{\mu,n}/2} P_{\mu,n}((qy)^{-k}, -y^k) \over 
 (1-(qy)^{-k})(1-(qy^{-1})^{k}} \right)
\ee
where $m_\mu({\sf x})$ are the monomial symmetric functions and 
$d_{\mu,n}$ 
is the dimension of the moduli space given in equation \eqref{eq:dimformulaB}.

\section{Localization of irregular Higgs bundles}\label{lochiggs}

In order to provide numerical evidence for formula \eqref{eq:refBPSa}, this section presents some explicit computations of Poincar\'e polynomials of moduli 
space of irregular Higgs bundles by localization. The computations will rely in part on the spectral correspondence 
stated in Section \ref{isomoduli}. The class of examples 
considered in this section will have $\ell =2$ and numerical invariants $\um = (2, 1)$, $c=1$. Therefore there are only two torus invariants connected rational curves, $\Sigma_1, \Sigma_2$ on $S_\uxi$ as shown in Figure \ref{SurfaceIII}. The first task is to 
find some explicit  necessary and sufficient stability conditions for pure dimension one sheaves on $S_{\uxi}$ supported on a divisor of 
the form $m_1\Sigma_1+\Sigma_2$.

\begin{figure}
\setlength{\unitlength}{1mm}
\hspace{-70pt}
\begin{picture}(250,80)
{\thicklines\put(15,10){\line(1,0){180}}
\put(30,5){\line(0,1){70}}}
\put(185,6){$\Sigma_0$}
\put(151,7){${\bf o}$}
\put(27,70){$f$}
\put(29,20){$\bullet$}
\put(29,50){$\bullet$}
\put(30,21){\line(2,1){22}}
\put(46,28.5){$\bullet$}
\put(42,32){\line(2,-1){22}}
\multiput(72,26)(5,0){3}{\huge .}
\put(90,32){\line(2,-1){22}}
\put(105,23){$\bullet$}
\put(100,21){\line(2,1){22}}\put(122,31.5){$\bullet$}
{\color{blue}\qbezier(123,32)(152,-12)(183,32)}
\put(184,32){$\Sigma_1$}
\put(30,51){\line(2,1){22}}
\put(46,58.5){$\bullet$}
\put(42,62){\line(2,-1){22}}
\multiput(72,56)(5,0){3}{\huge .}
\put(90,62){\line(2,-1){22}}
\put(105,53){$\bullet$}
\put(100,51){\line(2,1){22}}\put(122,61.5){$\bullet$}
{\color{blue}\qbezier(123,62)(152,-42)(183,62)}
\put(184,62){$\Sigma_2$}
\put(37,21){$\Xi_{1,1}$}
\put(37,51){$\Xi_{1,2}$}
\put(55,26.5){$\Xi_{2,1}$}
\put(55,56.5){$\Xi_{2,2}$}
\put(95,30.5){$\Xi_{n-1,1}$}
\put(95,60.5){$\Xi_{n-1,2}$}
\put(115,24.5){$\Xi_{n,1}$}
\put(115,54.5){$\Xi_{n,2}$}
\end{picture}
\caption{}
\label{SurfaceIII}
\end{figure}

\subsection{A stability criterion} 
For simplicity let $S,Y$ denote $S_\uxi, Y_\uxi$ in this section, keeping $\uxi$ fixed. 
Recall that $Y$ is the total space of the canonical bundle $K_S$, 
which is isomorphic to $S \times \IA^1$. 
Any torus invariant pure dimension one sheaf $F$ on $Y$ with 
\[
\ch_2(F) = m_1 \Sigma_1 + \Sigma_2 
\]
is set theoretically supported on the union $\Sigma_1\cup \Sigma_2$. Therefore it will fit in an exact sequence of $\CO_{Y}$-modules 
\be\label{eq:shextA}
0\to F_1 \to F \to F_2 \to 0
\ee
where $F_1, F_2$ are pure dimension sheaves with set theoretic 
support on $\Sigma_1, \Sigma_2$ respectively. Moreover 
\[
\ch_2(F_1)=m_1 \Sigma_1, \qquad \ch_2(F_2) = \Sigma_2. 
\]
This implies that $F_2$ is in fact scheme theoretically supported on $\Sigma_2$. Since $F$ is assumed to be stable one can easily  prove that $F$, hence also $F_1$, has to be scheme theoretically supported on $S$. The main task is then to derive an efficient stability criterion for such sheaf extensions on $S$. 
The main technical result needed in this analysis is the following. 
\bigskip

\noindent
{\bf Extension Lemma}. {\it Suppose $F_1, F_2$ are two pure dimension one sheaves on $S$ 
such that $F_1$ is set theoretically supported on $\Sigma_1$ and $F_2$ is scheme theoretically supported on  $\Sigma_2$.
Then there is an isomorphism 
\[ 
\tau: {\rm Ext}^1_S(F_2, F_1) \simeq 
{\rm Hom}_{\Sigma_2}(F_2, F_1 \otimes_S 
\CO_{\Sigma_2}(\Sigma_2)). 
\]
Moreover suppose there is a commutative diagram 
\be\label{eq:shcommdiagA}
\xymatrix{
0\ar[r] & F_1' \ar[r] \ar[d]^-{f_1} & F' \ar[r] \ar[d]^-{f}& 
F_2' \ar[r] \ar[d]^-{f_2} & 0 \\
0\ar[r] & F_1 \ar[r] & F \ar[r] & F_2 \ar[r] & 0}
\ee
with exact rows such that 
$F_1',F_2'$ are set theoretically supported on $\Sigma_1, \Sigma_2$. Let $\epsilon\in {\rm Ext}^1_S(F_2,F_1)$, 
$\epsilon'\in {\rm Ext}^1_S(F'_2,F'_1)$ be the extension 
classes determined by the rows, and let 
$\phi= \tau(\epsilon)\in {\rm Hom}_{\Sigma_2}(F_2, F_1 \otimes_S \CO_{\Sigma_2}(\Sigma_2))$, 
$\phi'= \tau'(\epsilon')\in {\rm Hom}_{\Sigma_2}(F'_2, F'_1 \otimes_S \CO_{\Sigma_2}(\Sigma_2))$. 
Then there is a commutative diagram 
\be\label{eq:shcommdiagC} 
\xymatrix{ 
F_2' \ar[r]^-{\varphi'} \ar[d]^-{f_2}& F_1' \otimes_S \CO_{\Sigma_2}(\Sigma_2)\ar[d]^-{f_1\otimes 1} \\ 
F_2\ar[r]^-{\varphi} & F_1 \otimes_S \CO_{\Sigma_2}(\Sigma_2).\\ }
\ee}
\bigskip 

This is proven by the same reasoning as in \cite[Lemma 2.6]{DHS}, hence the details will be omitted. 
Below are some useful consequences of this result. 

{\it 
${\bf E.1}.$ Let $\epsilon\in {\rm Ext}^1_S(F_2, F_1)$ be an extension class, and 
$\varphi=\tau(\epsilon)\in {\rm Hom}_{\Sigma_2}(F_2, F_1 \otimes_S \CO_{\Sigma_2}(\Sigma_2))$. Let $f_2:G_2\hookrightarrow F_2$ be a saturated subsheaf and let 
\[
f_2^*:{\rm Ext}^1_S(F_2,F_1) \to {\rm Ext}^1_S(G_2,F_1)
\]
be the naturally induced map. Then $f_2^*(\epsilon)=0$ 
if and only if $G_2\subset {\rm Ker}(\varphi)$. 

${\bf E.2}.$ Again, let $\epsilon\in {\rm Ext}^1_S(F_2, F_1)$ be an extension class, and 
$\varphi=\tau(\epsilon)\in {\rm Hom}_{\Sigma_2}(F_2, F_1 \otimes_S \CO_{\Sigma_2}(\Sigma_2))$. Let $q_1: F_1\twoheadrightarrow G_1$ be a surjective morphism, with $G_1$ a sheaf of pure dimension one, and let 
\[
q_{1*} : {\rm Ext}^1_S(F_2,F_1) \to {\rm Ext}^1_S(F_2, G_1)
\]
be the naturally induced map. Then $q_{1*}(\epsilon)=0$ 
if and only if 
\[
\big(q_1\otimes 1_{\CO_{\Sigma_2}(\Sigma_2)}\big) \circ \varphi =0.
\]
}

{\bf E.3}.
{\it  Under the same conditions as in {\bf E.2} 
let
 \[
F_2'= {\rm Ker}\left(\left(q_1\otimes 1_{\CO_{\Sigma_2}(\Sigma_2)}\right)\circ \varphi\right), \qquad 
F_1'={\rm Ker}(q_1)
\]
and let $f_i: F_i'\hookrightarrow F_i$, $1\leq i \leq 2$ be the natural inclusions.
Then there is a
sheaf $F'$ on $S$ which fits in a commutative diagram 
\be\label{eq:shcommdiagD}
\xymatrix{
0\ar[r] & F_1' \ar[r] \ar[d]^-{f_1} & F' \ar[r] \ar[d]^-{f}& 
F_2' \ar[r] \ar[d]^-{f_2} & 0 \\
0\ar[r] & F_1 \ar[r] & F \ar[r] & F_2 \ar[r] & 0}
\ee
with 
exact rows and injective columns. 

{\bf E.4}. Conversely, given a commutative diagram \eqref{eq:shcommdiagD} with exact rows and injective columns, 
let $G_1 = F_1/F_1'$ and let $q_1: F_1\twoheadrightarrow G_1$ 
be the natural projection. Then 
\[
F_2' \subseteq {\rm Ker}\left(\left(q_1\otimes 1_{\CO_{\Sigma_2}(\Sigma_2)}\right)\circ \varphi\right).
\]}

Statements {\bf E.1}, {\bf E.2} and {\bf E.4} are easy corollaries of the 
extension lemma. 
To prove {\bf E.3}, note that $F_1'$ is a subsheaf of $F$ and let $F''=F/F_1'$. Then there is a commutative diagram 
with exact rows and surjective columns
\be\label{eq:shcommdiagB}
\xymatrix{
0\ar[r] & F_1 \ar[r] \ar[d]^{q_1} & F \ar[r] \ar[d]^-{g}& 
F_2 \ar[r] \ar[d]^-{1_{F_2}} & 0 \\
0\ar[r] & G_1 \ar[r] & F''\ar[r] & F_2 \ar[r]^{q_2} & 0}
\ee
Moreover statement {\bf E.2} implies that
the
injective morphism $f_2:F_2'\hookrightarrow F_2$ factors through the projection  $q_2:F''\twoheadrightarrow F_2$ 
i.e. there is an injective morphism $h: F_2'\to F''$ such that 
$f_2 = q_2 \circ h$. Let $F'=g^{-1}({\rm Im}(h)) \subset F$.
Then it is straightforward to check that $F'$ fits
into a commutative diagram of the form \eqref{eq:shcommdiagD}.

As explained below, statements {\bf E.3} and {\bf E.4} yield a handy stability criterion for extensions 
\[
0 \to F_1 \to F \to F_2 \to 0.
\]
with $F_1,F_2$ set theoretically supported on $\Sigma_1, \Sigma_2$
and $\ch_2(F_2)=\Sigma_2$. 
 
Suppose 
\[
\ch_2(F_1) = m_1 \Sigma_1, \qquad \chi(F_2)=\Sigma_2, 
\qquad \chi(F) =1 
\]
with $m_1\geq 1$. Note that any nonzero saturated 
subsheaf $F'\subset F$ 
fits into an exact diagram of the form \eqref{eq:shcommdiagD} 
with $F_1', F_2'$ pure of dimension one. Moreover, 
\[ 
\ch_2(F_1') = m_1'\Sigma_1, \qquad \ch_2(F_2') = m_2'\Sigma_2
\]
with $0 \leq m_1'\leq m_1$, $0\leq m_2' \leq 1$ and 
$(m_1', m_2') \neq (0,0)$. 
Then $F$ is $\beta$-stable if and only 
\be\label{eq:stabextA}
{\chi(F_1') + \chi(F_2') +m_1'\beta_1+ m_2'\beta_2 \over 
m_1'+m_2'} < { 1 + m_1\beta_1 + \beta_2\over m_1+1}. 
\ee

Next note the existence of a special stability chamber in the space of parameters $(\beta_1, \beta_2)$. This is a standard result and the proof will be omitted. 
\bigskip

\noindent
{\bf Stability chamber}.
{\it For fixed $m_1\geq 1$, there exist $\beta_{1,0}, \beta_{2,0}>0$ such that for any \linebreak $0 < \beta_1 < \beta_{1,0}$,
$0<\beta_2 < \beta_{2,0}$ a sheaf $F$ as above is $\beta$-stable if and only if it is stable for $\beta=(0,0)$.} 
\bigskip

Then one can easily derive a stability criterion in this chamber 
using statements {\bf E.3} and {\bf E.4}  above. Let $\varphi: F_2 \to F_1 \otimes_S \CO_{\Sigma_2}(\Sigma_2)$ 
be the morphism corresponding to the given extension class. 
Let $K={\rm Ker}(\varphi)$. Given a nonzero saturated subsheaf $f_1 : F_1'\hookrightarrow F_1$, possibly identical with $F_1$, let 
$G_1 = F_1/F_1'$ and let $q_1: F_1\twoheadrightarrow G_1$ 
be the natural projection. Let also\[
 F_2' = {\rm Ker}\left(\left(q_1\otimes 1_{\CO_{\Sigma_2}(\Sigma_2)}\right)\circ \varphi\right).\]

\noindent
{\bf Stability Lemma}. 
{\it The extension $F$, with $\chi(F)=1$, is stable  if and only if 
\begin{itemize} 
\item[{\bf S.1.}] For any nonzero saturated subsheaf $f_1 : F_1'\hookrightarrow F_1$, possibly identical with $F_1$, 
\[
\chi(F_1') \leq 0,\qquad {\rm and}\qquad \chi(F_1') + \chi(F_2') \leq 0,
\]
and 
\item[{\bf S.2.}] $\chi(K) \leq 0$. 
\end{itemize}
}

\subsection{Torus fixed points with $(m_1,m_2)=(2,1)$} 

The moduli space of stable sheaves $F$ with numerical invariants as above is smooth. Moreover it has a torus action induced by the torus action on $S$. Using the notation in 
Section \ref{toractsect}, recall that the curves 
$\Sigma_1, \Sigma_2$ are given by the local equations 
\[ 
v= \lambda_1 w^{n-2}, \qquad v = \lambda_2 w^{n-2}
\]
in the affine chart $V\subset S$, where 
$\lambda_1,\lambda_2$ are distinct nonzero complex parameters.
In particular both are rational curves and admit the local  parameterizations  
\be\label{eq:locpar}
w=x_i, \qquad v = \lambda_i x_i^{n-2}, \qquad 1\leq i \leq 2. 
\ee
In the same chart the torus 
action reads 
\[ 
(t,(w,v)) \mapsto (t^{-1}w, t^{2-n}v)
\] 

For the remaining part of this section, set $m_1=2$. Then $F_1$ 
itself is an extension 
\be\label{eq:F1extA}
0 \to L_1 \to F_1 \to  L_2 \to 0
\ee
where $L_1, L_2$ are line bundles on $\Sigma_1$ extended  by zero to $S$. 
Let ${\bf o}$ be the intersection point $v=w=0$. To simplify the formulas let also $k=n-2$. Then note that there is an isomorphism of $\CO_{S}$-modules 
\[
\CO_{\Sigma_1}\otimes_S \CO_{\Sigma_2}
\simeq \CO_{k{\bf o}},
\]
where $k{\bf o}\subset S$ denotes the scheme theoretical intersection of  $\Sigma_1, \Sigma_2$,  given by $v=0,\ w^k=0$. 
Then the exact sequence \eqref{eq:F1extA} yields an exact sequence of $\CO_{\Sigma_2}$-modules
\be\label{eq:F1extB} 
0 \to L_1(k{\bf o})\otimes_S \CO_{\Sigma_2} \to F_1 \otimes_S \CO_{\Sigma_2}(\Sigma_2) \to  L_2(k{\bf o})\otimes_S \CO_{\Sigma_2} \to 0.
\ee
If $F$ is preserved by the torus action up to isomorphism, 
the sheaves $L_1,L_2$, $F_2$ will carry ${\bf T}$-equivariant structures. 
Hence, as $\Sigma_1,\Sigma_2$ are rational, there are isomorphisms of the form  
\be\label{eq:equivisomB}
L_i\simeq T^{a_i}\CO_{\Sigma_i}(d_i {\bf o}), \quad 1\leq i \leq 2,\qquad F_2 \simeq T^{b} \CO_{\Sigma_2}(d_3{\bf o}) 
\ee
for some weights $a_1,a_2,b \in \IZ$ and degrees 
$d_1, d_2,d_3 \in \IZ$. Moreover, the weights $(a_1,a_2,b)$ are 
uniquely defined up to a common shift, hence one can fix $b=0$ without loss of generality. 
The above relations further yield isomorphisms
\be\label{eq:equivisomC}
L_i(k{\bf o})\otimes_S \CO_{\Sigma_2} \simeq 
T^{a_i}\CO_{k{\bf o}}((d_i+k){\bf o}),  \qquad 1\leq i \leq 2
\ee
of equivariant $\CO_{\Sigma_2}$-modules.
Therefore there is a third exact sequence 
\be\label{eq:homseq}
\bal
0 \to T^{a_1} Hom_{\Sigma_2}(\CO_{k{\bf o}}(d_3{\bf o}),\, & \CO_{k{\bf o}} ((d_1+k){\bf o})) 
 \to {\rm Hom}_{\Sigma_2}(F_2, F_1 \otimes_S \CO_{\Sigma_2}(\Sigma_2)) \\
& \to 
 T^{a_2}Hom_{\Sigma_2}(\CO_{k{\bf o}}(d_3{\bf o}), \CO_{k{\bf o}} ((d_2+k){\bf o})) \to 0.\\
\eal
\ee
of ${\bf T}$-modules. 
In order to classify the torus fixed points in the moduli space of stable sheaves, one has to consider several cases. 

\subsubsection{Trivial extension on $\Sigma_1$}
First suppose the extension \eqref{eq:F1extA} is trivial i.e. 
$F_1 \simeq \CO_{\Sigma_1}(d_1{\bf o}) \oplus 
\CO_{\Sigma_1}(d_2{\bf o})$. Such sheaves will be referred to as 
Type $(I)$ sheaves in the following. 
In this case the exact sequence \eqref{eq:homseq} has a natural splitting. 
Using the local parameterization \eqref{eq:locpar}, 
one further obtains an isomorphism 
\[
{\rm Hom}_{\Sigma_2}(F_2, F_1 \otimes_S \CO_{\Sigma_2}(\Sigma_2)) \simeq 
T^{a_1}{\rm Hom}_R(x_2^{-d_3}R, x_2^{-d_1-k}R) \oplus 
T^{a_2}{\rm Hom}_R(x_2^{-d_3}R, x_2^{-d_2-k}R),
\]
where $R = \IC[x_2]/(x_2^k)$. 
For sheaves preserved by the torus action, the morphism 
$\varphi\in {\rm Hom}_{\Sigma_2}(F_2, F_1 \otimes_S \CO_{\Sigma_2}(\Sigma_2))$ encoding the extension class will be determined by a morphism of $R$-modules ${\overline \varphi}: 
x_2^{-d_3}R \to x_2^{-d_1-k}R\oplus x_2^{-d_2-k}R$ of the form 
\be\label{eq:extclassA}
{\overline \varphi}(x_2^{-d_3}) = (x_2^{l_1-d_1-k}, 
x_2^{l_2-d_2-k}) 
\ee
for some $0\leq l_1, \l_2 \leq k-1$. Then 
\[ 
{\rm Ker}({\overline \varphi}) \simeq x_2^{-d_3}(x_2^{k-l_1}) \cap x_2^{-d_3}(x_2^{k-l_2}) \simeq 
x_2^{-d_3}(x_2^{k-l})
\]
where $l = {\rm min}\{l_1,l_2\}$. 
Hence $\chi({\rm Ker}({\overline \varphi})) = l$, which implies 
\[
\chi({\rm Im}({\overline \varphi})) = \chi({\rm Im}(\varphi)) = 
k-l, \qquad \chi({\rm Ker}(\varphi)) = \chi(F_2) -k+l. 
\]
Then, using the stability lemma proven in the previous subsection, one finds the following necessary conditions for stability
\be\label{eq:stabcondA} 
1-k + l_i \leq \chi(L_i) \leq 0, \qquad 1\leq i \leq 2,\qquad 
1-k+l\leq \chi(L_1) + \chi(L_2).
\ee
In addition, in the classification of fixed points one can assume 
without loss of generality that $\chi(L_1) \leq\chi(L_2)$. 
Using conditions \eqref{eq:stabcondA}, this yields 
\be\label{eq:stabcondB}
{1-k+l\over 2} \leq \chi(L_2).
\ee
Moreover torus invariance yields the relations 
\be\label{eq:torusweightsA} 
a_i = l_i -k -\chi(L_i) + \chi(L_3), \qquad 1\leq i \leq 2.
\ee

In addition, criterion $(vii)$ yields the following necessary 
condition for stability. 
Let $e\in \IZ$, $e\leq {\rm min}\{d_1,d_2\}$. For any pair of sections $\zeta_i\in H^0(\CO_{\Sigma_i}((d_i-e){\bf o})$, $1\leq i \leq 2$, with no common zeroes on $\Sigma_2$ there is an exact sequence 
\[
\xymatrix{
0\ar[r] & \CO_{\Sigma_2}(e{\bf o}) \ar[r]^-{{\zeta_1}\choose{\zeta_2}} 
&  \CO_{\Sigma_2}(d_1{\bf o}) \oplus \CO_{\Sigma_2}(d_2{\bf o})
\ar[r]^-{q_1}  & \CO_{\Sigma_2}((d_1+d_2-e){\bf o})
\ar[r] & 0.
\\}
\]
where $q_1=(\zeta_2, -\zeta_1)$. Let $I_2$ be the image of the 
morphism
\[
\CO_{k{\bf o}}(d_3{\bf o})  \to \CO_{k{\bf o}}((d_1+d_2-e+k){\bf o})
\]
given by 
\[ 
z^{-d_3} \mapsto \zeta_2 z^{-d_1-k+l_1} - \zeta_1 z^{-d_2-k+l_2}.
\]
Then one requires 
\be\label{eq:stabcondBB}
e+1 + d_3+1 -\chi(I_2) \leq 0. 
\ee

In conclusion, conditions \eqref{eq:stabcondA}, \eqref{eq:stabcondB} and 
\eqref{eq:stabcondBB} are necessary and sufficient for stability. 

\subsubsection{Nontrivial extension on $\Sigma_1$} 
Next suppose the extension \eqref{eq:F1extA} is nontrivial. 
Such sheaves will be called type $(II)$ in the following. 

In this case the first task is to compute the extension group ${\rm Ext}^1_S(L_2,L_1)$. Recall that the restriction $\pi|_{\Sigma_i}: \Sigma_i \to C$ is an isomorphism mapping ${\bf o}\in \Sigma_i$ to $\infty \in C$. Therefore there are isomorphisms of {\bf T}-equivariant sheaves on $S$
\[
L_i \simeq T^{a_i}\pi^*\CO_{C}(d_i \infty)\big|_{\Sigma_i}, \qquad 1\leq i \leq 2. 
\]
This implies that there is also an exact sequence of {\bf T}-equivariant sheaves on $S$
\be\label{eq:L2resA}
\xymatrix{ 
0 \ar[r] & T^{a_2}\pi^*\CO_{C}(d_2 \infty)(-\Sigma_1) \ar[r]
& T^{a_2}\pi^*\CO_{C}(d_2 \infty) \ar[r] & L_2 \ar[r] & 0.
}
\ee 
This yields the isomorphisms
\be\label{eq:locextisomA}
\bal
{\mathcal Hom}_{S}(L_2, L_1) & \simeq T^{-a_2}
{\mathcal Hom}_S(\pi^*\CO_{C}(d_2 \infty),L_1)\\
 {\mathcal Ext}^1_S(L_2, L_1) & \simeq 
T^{-a_2} {\mathcal Hom}_S(\pi^*\CO_C(d_2\infty), L_1(\Sigma_1)).\\
\eal
\ee
 Using the local to global spectral sequence, one further obtains an isomorphism 
\be\label{eq:extisomA}
{\rm Ext}_S^1(L_2,L_1) \simeq 
{\rm Hom}_{\Sigma_1} (L_2,
L_1\otimes_{\Sigma_1}N_{\Sigma_1/S})\oplus 
{\rm Ext}^1_{\Sigma_1}(L_2,L_1),
\ee
where $N_{\Sigma_1/S}$ is the normal bundle of $\Sigma_1$ in $S$, which is a line bundle of degree $(-2)$. 
For any extension class in the second summand, one has 
$F_1 \simeq L_1'\oplus L_2'$ for some 
line bundles $L_1',L_2'$ on $\Sigma$. This leads back to case $(I)$. Therefore it suffices to consider extension classes with 
nonzero projection onto the first direct summand. This requires 
\be\label{eq:stabcondC}
\chi(L_1) -\chi(L_2) -2 \geq 0,
\ee
which implies that ${\rm Ext}^1_{\Sigma_1}(L_2,L_1)=0$.
Therefore in this case the extensions 
\be\label{eq:F1extAA}
0\to L_1 \to F_1 \to L_2 \to 0
\ee
will be parameterized by elements of 
${\rm Hom}_{\Sigma_1} (L_2,L_1\otimes_{\Sigma_1}N_{\Sigma_1/S} )$. Furthermore note that the normal bundle 
$N_{\Sigma_1/S}$ has the natural local frame $\partial / \partial v$ over the open subset $\Sigma_1\cap V$. 
Assuming that condition \eqref{eq:stabcondC} is satisfied, this yields an  isomorphism of ${\bf T}$-modules 
\be\label{eq:extisomC}
{\rm Hom}_{\Sigma_1} (L_2,L_1\otimes_{S} \CO_{\Sigma_1}(\Sigma_1) )\simeq T^{a_1-a_2+k} \IC\langle x_1^{d_2-d_1} , \ldots, 
x^{-2}\rangle \simeq T^{a_1-a_2+k} \left(\oplus_{j=2}^{d_1-d_2}T^j\right). 
\ee
For sheaves $F_1$ preserved by the torus action the extension  \eqref{eq:F1extAA} will be parameterized by a monomial 
 $x_1^{e}$, with 
$d_2-d_1 \leq e \leq -2$ such that
\[ 
a_2 = a_1 +k-e. 
\]

On the other hand, by restriction, there is an exact sequence 
of $\CO_{\Sigma_2}$-modules
\be\label{eq:F1extBB}
0 \to L_1\otimes_S \CO_{\Sigma_2} \to F_1 \otimes_S \CO_{\Sigma_2} \to  L_2\otimes_S \CO_{\Sigma_2} \to 0.
\ee
The exact sequence \eqref{eq:L2resA} restricts to an 
exact sequence of sheaves on $\Sigma_2$
\be\label{eq:L2resB}
\xymatrix{ 
0\ar[r] & T^{a_2}\CO_{\Sigma_2}((d_2-k){\bf o}) 
\ar[r] & 
T^{a_2}\CO_{\Sigma_2}(d_2{\bf o}) 
\ar[r] & L_2\otimes_S \CO_{\Sigma_2} \ar[r] & 0.\\}
\ee
This yields an isomorphism 
\be\label{eq:extisomB}
{\mathcal Ext}^1_{\Sigma_2}\left(
L_2\otimes_S \CO_{\Sigma_2} , 
L_1\otimes_S\CO_{\Sigma_2} \right)
\simeq T^{-a_2} 
{\mathcal Hom}_{\Sigma_2}\left(\CO_{k{\bf o}}(d_2{\bf o}), 
\CO_{k{\bf o}}((d_1+k){\bf o})\right). 
\ee
Moreover there is a commutative diagram 
\be\label{eq:extdiagA} 
\xymatrix{ 
{\mathcal Ext}^1_S(L_2, L_1) \ar[r]^-{\sim} \ar[d] & 
T^{-a_2} {\mathcal Hom}_S(\pi^*\CO_C(d_2\infty), L_1\otimes_{\Sigma_1}N_{\Sigma_1/S})
\ar[d] \\
{\mathcal Ext}^1_{\Sigma_2}\left(
L_2\otimes_S \CO_{\Sigma_2} , 
L_1\otimes_S\CO_{\Sigma_2} \right)
\ar[r]^-{\sim}  & 
 T^{-a_2} 
{\mathcal Hom}_{\Sigma_2}\left(\CO_{k{\bf o}}(d_2{\bf o}), 
\CO_{k{\bf o}}((d_1+k){\bf o})\right) \\}
\ee
where the vertical arrows are naturally induced by 
restriction to $\Sigma_2$. 
Finally, using the local to global spectral sequence there is also an 
isomorphism 
\be\label{eq:extisomB}
{\rm Ext}^1_{\Sigma_2} \left(
L_2\otimes_S \CO_{\Sigma_2} , 
L_1\otimes_S\CO_{\Sigma_2} \right) \simeq T^{a_1-a_2}
{\rm Hom}_{\Sigma_2}\left(\CO_{k{\bf o}}(d_2{\bf o}), 
\CO_{k{\bf o}}((d_1+k){\bf o})\right),
\ee
while the local parameterization \eqref{eq:locpar} yields 
a second isomorphism 
\be\label{eq:extisomC}
{\rm Hom}_{\Sigma_2}\left(\CO_{k{\bf o}}(d_2{\bf o}), 
\CO_{k{\bf o}}((d_1+k){\bf o})\right) \simeq 
{\rm Hom}_R (x_2^{-d_2}R,  x_2^{-d_1-k}R),
\ee
where $R=\IC[x_2]/(x_2^k)$.

Now suppose $\xi: L_2\to
L_1\otimes_{\Sigma_1} N_{\Sigma_1/S}$ is a morphism corresponding to the extension 
\eqref{eq:F1extAA} under isomorphism \eqref{eq:extisomA}. 
Then, using the commutative diagram \eqref{eq:extdiagA}, 
the restriction 
\[
\xi|_{\Sigma_2} : T^{a_2} \CO_{k{\bf o}}(d_2{\bf o}) \to 
T^{a_1} \CO_{k{\bf o}}((d_1+k){\bf o})
\]
is identified with the extension class corresponding to the exact sequence 
\eqref{eq:F1extBB}. 
In particular, if $\xi$ corresponds to a  monomial 
$x^e$ under the isomorphism \eqref{eq:extisomC}, then 
the restriction $\xi|_{\Sigma_2}$ is given by 
\[
\xi|_{\Sigma_2} (x^{-d_2}) = x^{-k-d_2+e} = x^{-(d_1+k)+j},
\]
where 
\[
j=d_1-d_2+e \geq 0. 
\]
Hence, using isomorphisms \eqref{eq:extisomB}, \eqref{eq:extisomC}, the extension class of  
 \eqref{eq:F1extBB} is determined by the integer $j \geq 0$.

Next consider an extension 
\[ 
0 \to F_1 \to F \to F_2 \to 0
\]
parameterized by a morphism $\varphi: F_2 \to F_1 \otimes_S \CO_{\Sigma_2}(\Sigma_2)$. Let $\varphi_2: F_2 \to L_2 \otimes_S 
\CO_{\Sigma_2}(\Sigma_2)$ be the induced morphism via the projection $F_1\twoheadrightarrow L_2$. 
Let $I_2= {\rm Im}(\varphi_2)$ and note that criterion $(vii)$ yields the following necessary conditions for stability  \be\label{eq:stabcondE}
\chi(L_1)\leq 0, \qquad 1-\chi(L_1)-\chi(L_2) -\chi(I)
 \leq 0,\qquad
1-\chi(L_2) - \chi(I_2) \leq 0. 
\ee

Now the main observation is that the morphism $\varphi$ is uniquely determined by ${\varphi}_2$ and the morphism 
$\xi:L_2 \to L_1\otimes_{\Sigma_1} N_{\Sigma_1/S}$
using stability and ${\bf T}$-invariance. 
In order to prove this statement first recall that  one has isomorphims of $\IC[x_2]$-modules 
\[ 
\Gamma(\CO_{k{\bf o}}(d_i{\bf o})) \simeq x_2^{-d_i} R, \qquad 
\] 
for $1\leq i \leq 2$, where $R = \IC[x_2]/(x_2^k)$. 
Therefore there is an isomorphism 
\[
{\rm Hom}_{\Sigma_2}(F_2, L_2 \otimes_S 
\CO_{\Sigma_2}(\Sigma_2)) \simeq
 T^{a_2}{\rm Hom}_R(x_2^{-d_3}R, x_2^{-d_2-k}R).
\]
For torus invariant extensions, the morphism ${\varphi}_2$ will be 
determined by 
\[
{\overline \varphi}_2 \in {\rm Hom}_R(x_2^{-d_3}R, x_2^{-d_2-k}R), 
\qquad {\overline \varphi}_2(x_2^{-d_3})= x_2^{l-d_2-k},
\]
for some $0\leq l \leq k-1$ such that 
\[ 
a_2 + d_2 - d_3 + k-l =0.
\]
Moreover, the stability conditions \eqref{eq:stabcondE} 
imply
\be\label{eq:stabcondF}
k-l+d_2\geq 0.
\ee

For the next step one has to consider two cases.
If $j \geq k$ the extension \eqref{eq:F1extBB} is trivial, hence  there is an isomorphism 
\[ 
F_1 \otimes_S \CO_{\Sigma_2}(\Sigma_2) \simeq T^{a_1}
\CO_{k{\bf o}}((d_1+k){\bf o}) \oplus T^{a_2}
\CO_{k{\bf o}}((d_2+k){\bf o}) 
\]
of ${\bf T}$-equivariant $\CO_{\Sigma_2}$-modules and the 
 surjection onto $L_2\otimes_S \CO_{\Sigma_2}$ is 
the natural projection onto the second direct summand. 
Then note that 
\[ 
T^{a_1} \Gamma( \CO_{k{\bf o}}((d_1+k){\bf o})) \simeq 
T^{a_1+d_1+k} \oplus_{s=0}^{k-1} T^{-s}, \quad 
T^{a_2}\Gamma( \CO_{k{\bf o}}((d_2+k){\bf o})) \simeq 
T^{a_2+d_2+k} \oplus_{s=0}^{k-1} T^{-s}
\]
and the torus weight of the monomial $x_2^{l-d_2-k}$ in the second direct summand is 
\[
a_2+d_2+k -l= a_1 +2k -e +d_2-l \geq a_1+k -e \geq a_1 +k +2.  
\]
At the same time the highest torus weight in the character decomposition of the first summand is 
\[
a_1 + d_1+k \leq a_1 +k-1 
\]
since $\chi(L_1) = d_1 +1 \leq 0$ by stability. 
This implies that there is a unique torus invariant lift of 
$\varphi_2: F_2 \to L_2\otimes_S\CO_{\Sigma_2}(\Sigma_2)$ 
to a morphism $\varphi: F \to F_1\otimes_S\CO_{\Sigma_2}(\Sigma_2)$. 

If the opposite holds, $j <k$, 
the extension \eqref{eq:F1extBB} is isomorphic 
to 
\[ 
0\to  T^{a_1}
\CO_{k{\bf o}}(d_1{\bf o}) {\buildrel f_1\over \longto } 
T^{a_1} \CO_{j{\bf o}}(d_1{\bf o}) \oplus T^{a_2}
\CO_{(2k-j){\bf o}}(d_2{\bf o}) {\buildrel f_2\over \longto} 
T^{a_2} \CO_{k{\bf o}}(d_2{\bf o}) 
\to 0
\] 
The two maps are given by 
\[ 
f_1(u) = (p_1(u), x_2^{k-j}u), \qquad f_2(u_1,u_2)= p_2(u_2)-x^{k-j}u_1
\]
where $p_1: \CO_{k{\bf o}}(d_1{\bf o})\twoheadrightarrow 
\CO_{j{\bf o}}(d_1{\bf o})$ and 
$p_2: \CO_{(2k-j){\bf o}}(d_2{\bf o})\twoheadrightarrow 
\CO_{k{\bf o}}(d_2{\bf o})$ 
are the natural projections. Then note the isomorphisms 
\[ 
T^{a_1} \Gamma( \CO_{j{\bf o}}((d_1+k){\bf o})) \simeq 
T^{a_1+d_1+k} \oplus_{s=0}^{j-1} T^{-s}, \quad 
T^{a_2}\Gamma( \CO_{(2k-j){\bf o}}((d_2+k){\bf o})) \simeq 
T^{a_2+d_2+k} \oplus_{s=0}^{2k-j-1} T^{-s}.
\]
The element $x_2^{l-d_2-k}\in \Gamma(\CO_{k{\bf o}}((d_2+k){\bf o}))$ has a natural lift to $\Gamma(\CO_{(2k-j){\bf o}}((d_2+k){\bf o}))$ with torus weight
\[
a_2+d_2+k -l= a_1 +2k -e +d_2-l \geq a_1+k -e \geq a_1 +k +2.  
\]
At the same time, the highest torus weight in the character decomposition of $T^{a_1} \Gamma( \CO_{j{\bf o}}((d_1+k){\bf o}))$ 
is again
\[
a_1 + d_1 +k \leq a_1 +k -1. 
\]
Therefore the natural lift is again the unique torus invariant lift. 

In conclusion, the extension class of a stable torus equivariant 
sheaf $F$ 
is determined by the conditions 
\be\label{eq:stabcondCC}
\chi(L_1)\leq 0,\qquad \chi(L_2)-\chi(L_1)\leq e\leq -2, \qquad 0\leq l \leq k-1, \qquad k-l +\chi(L_2) \geq 1.
\ee
The torus weights $a_1, a_2$ are related to the numerical invariants by 
\be\label{eq:torweightsC}
a_2 = \chi(L_3)-\chi(L_2)-k+l, \qquad a_1 = a_2 -k +e. 
\ee

To conclude this case, we note that the necessary stability conditions \eqref{eq:stabcondCC} are also sufficient. Using the stability lemma, 
suppose 
$F_1'\subset F_1$ is a nonzero proper saturated subsheaf. Then 
$\ch_1(F_1') =\Sigma_1$, hence $F_1'$ is scheme theoretically supported on $\Sigma_1$. Since $F_1$ is a nontrivial extension 
\[
0\to L_1 \to F_1 \to L_2 \to 0
\]
parameterized by a nonzero class in ${\rm Hom}_S(L_2, 
L_1(\Sigma_1))$, it follows that $F_1'\subset L_1$. Since the quotient $F_1/F_1'$ must be pure of dimension one, this implies 
that $F_1'=L_1$ and $F/F_1'= L_2$. Then one obtains the stability 
condition 
\[ 
\chi(L_1) + \chi(F_2) -\chi(I_2) \leq 0
\] 
which has been already listed in \eqref{eq:stabcondCC}. 

\subsection{The equivariant tangent space}
Using the results of Section \ref{sheavestohiggs},
an extension 
\[
0 \to F_1 \to F\to F_2 \to 0
\]
as above determines by pushforward an exact sequence of 
irregular parabolic Higgs bundles 
\[ 
0\to \CE_1 \to \CE \to \CE_2 \to 0.
\]
Moreover $\CE$ is stable if and only if $F$ is stable as a sheaf on $S$. 
The underling equivariant bundles of the Higgs bundles $\CE_1, \CE_2$ are 
\[ 
E_1 = T^{a_1}\CO_C(d_1\infty) \oplus T^{a_2} \CO_C(d_2\infty), 
\qquad 
E_2 = \CO_C(d_3\infty)
\]
respectively.  The flags of $\CE_1, \CE_2$ at $p$ are trivial while the 
extension $\CE$ has a nontrivial flag $E_{1,D}\subset E_D$ of type 
$(2,1)$. The Higgs field of $\CE$ can be explicitely determined 
from the construction of $F$, but its expression will not be needed in the following. 

The isomorphism of moduli spaces constructed in Section \ref{spectralsect} 
implies that all torus fixed points in the moduli space of stable irregular parabolic Higgs bundles must be extensions of this form. 
In particular for the rank three examples considered here, all 
fixed loci are isolated points.

Suppose $[\CE]$ is an isolated fixed point with  underlying equivariant  vector bundle $E$ and flag $E_D^\bullet$ over $D=np$. 
Using the deformation theory results of Section \ref{deftheory}, one has the following 
identity in the representation ring of the torus ${\bf T}$
\[ 
\bal 
T_{[\CE]} = & \big({\bf 1} + T^{k+1}\big) -H^0(Hom(E,E)) + H^1(Hom(E,E))  +
AP{\rm Hom}(E_D^\bullet, E_D^\bullet) \\
& + T^{k}\left(H^0(Hom(E,M\otimes_C E)) - H^1(Hom(E,
M\otimes_C E)) - 
ASP{\rm Hom}(E_D^\bullet, (M\otimes E)_D^\bullet)\right).\\
\eal 
\] 
where $k=n-2$ and $M \simeq \CO_C(kp)$ is equipped with the natural 
${\bf T}$-equivariant structure. Moreover,
\[
 AP{\rm Hom}(E_D^\bullet, E_D^\bullet) -T^k ASP{\rm Hom}(E_D^\bullet, (M\otimes_C E)_D^\bullet) = - \sum_{a=1}^\ell 
{\rm Hom}(E_{D}^a, E_{D}^a). 
\]
For the examples under consideration, $E$ is an extension 
\[ 
0 \to E_1 \to E \to E_2 \to 0
\]
of equivariant bundles on $C$. Hence one can substitute the direct 
sum 
\[
E_1\oplus E_2 = T^{a_1}\CO_C(d_1\infty) \oplus T^{a_2}\CO_C(d_2\infty) \oplus \CO_{C}(d_3\infty)
\]
for $E$ in the above formula. Then, by straightforward $\check{\rm C}$ech cohomology computations, 
the following relations hold in the representation ring of ${\bf T}$
\[
\bal 
& -H^0(Hom(E,E)) + H^1(Hom(E,E))  = -
\sum_{\substack{1\leq i,j,\leq 3 \\ d_i \leq d_j}} T^{a_j-a_i} \sum_{s=0}^{d_j-d_i} T^s + 
\sum_{\substack{1\leq i,j,\leq 3\\ d_i > d_j}} T^{a_j-a_i} 
\sum_{s=1}^{d_i-d_j-1} T^{-s},\\
 & T^{k}\left(H^0(Hom(E,M\otimes_C E)) - H^1(Hom(E,M \otimes_C E)) \right)  = \\
& {}\qquad\qquad \qquad \qquad \qquad \qquad\qquad \qquad \sum_{\substack{1\leq i,j,\leq 3 \\ d_i \leq d_j+k}}
T^{a_j-a_i+k} \sum_{s=-k}^{d_j-d_i} T^{s} - 
\sum_{\substack{1\leq i,j,\leq 3\\ d_i > d_j+k}}
T^{a_j-a_i+k} \sum_{s=-k-1}^{d_i-d_j-1} T^{-s}.\\
\eal 
\]
Furthermore the flag over $D$ is given by $E_{1,D}\subset E_D$,
hence has successive quotients 
\[ 
E_{1,D} =T^{a_1} \CO_D \oplus T^{a_2} \CO_D, \qquad 
E_D/E_{1,D} = \CO_D.
\]
Therefore  one also has 
\[
\bal
 AP{\rm Hom}(E_D^\bullet, E_D^\bullet) & -T^k ASP{\rm Hom}(E_D^\bullet, (M\otimes_C E)_D^\bullet) = - (3+ T^{a_1-a_2}+T^{a_2-a_1})
\sum_{s=0}^{k+1} T^s.\\
\eal
\]

\subsection{Examples}\label{Pexamples} 
Two concrete examples will be computed explicitly below for 
rank three irregular Higgs bundles with a flag of type $(2,1)$. 

\subsubsection{$n=5$} 
Using conditions \eqref{eq:stabcondA}, \eqref{eq:stabcondB} and \eqref{eq:stabcondC}, one obtains 
five type I fixed points listed in Table 1 and 
and one type II fixed point listed in Table 2. 
The Poincar\'e polynomial is $1+2v^2+3v^4$. By comparison with 
Section \ref{rkthreeB}, this is the same as 
$P_{(2,1),5}(1,v)$. 
\begin{table}
\centering
\begin{tabular}{|c|c|c|c|c|c|c|}
\hline
$\chi(L_1)$ & $\chi(L_2)$ & $l_1$ & $l_2$ & $a_1$ & $a_2$ & $T_{[\CE]}$ \\
\hline
-1 & 0 & 1 & 0 & 1& -1 &  $T^{-1}+T+T^3+T^5$ \\
\hline
-1& 0 &0 & 2& 0& 1 & $T^{-2}+T^{-1}+T^5+T^6$  \\
\hline
0 & 0 & 0 & 1 & -2 & -1 & $T^{-2}+T^{-1}+T^5+T^6$  \\
\hline
0 & 0 & 0 & 2 & -2 & 0 &  $T^{-1}+T+T^3+T^5$ \\
\hline
0 & 0 & 1 & 2 & -1 & 0 & $T+2T^2 + T^3$\\
\hline
\end{tabular}
\label{Table1}
\caption{Type $I$ fixed points for $n=5$.}
\end{table}
\begin{table}
\centering
\begin{tabular}{|c|c|c|c|c|c|c|}
\hline
$\chi(L_1)$ & $\chi(L_2)$ & $l_1$ & $l_2$ & $a_1$ & $a_2$ & $T_{[\CE]}$ \\
\hline
0 & -2 & 0 & -2 & -3 & 2 & $T^{-2}+T^{-1}+T^5+T^6$  \\
\hline
\end{tabular}
\caption{Type $II$ fixed points for $n=5$.}
\label{Table2}
\end{table}

\subsubsection{$n=6$}
 In this case there are fifteen type I fixed points listed in Table 3 and five type II fixed points listed in Table 4.
The Poincar\'e polynomial is 
\[ 
1 + 2v^2 + 5v^4+6v^6+6v^8. 
\]
By comparison with 
Section \ref{rkthreeB}, this is the same as 
$P_{(2,1),6}(1,v)$. 

\begin{table}
\centering
\begin{tabular}{|c|c|c|c|c|c|c|}
\hline
$\chi(L_1)$ & $\chi(L_2)$ & $l_1$ & $l_2$ & $a_1$ & $a_2$ & $T_{[\CE]}$ \\
\hline
-1 &  -1 &  2 &  0 &  2 &  0 & $T^{-3} + 2T^{-1}+T+T^4+2T^6+T^8$\\
\hline
-2 &  0 &  1 &  0 &  2 &  -1 & 
$T^{-2}+T^{-1}+T+T^2+T^3+T^4+T^6+T^7$ \\
\hline
-2 &  0 &  0 &  3 &  1 & 2 & 
 $T^{-3}+2T^{-2}+T^{-1}+T^6+2T^7+T^8$\\ 
\hline
-1 &  0 &  0 &  3 &  -1 & 1 & 
$T^{-3}+T^{-2}+T^{-1}+T+T^4+T^6+T^7+T^8$\\ 
\hline
-1 &  0 &  0 &  2 &  -1 &  0 & 
$T^{-3}+T^{-2}+2T^{-1}+2T^6+T^7+T^8$\\ 
\hline
-1 & 0 &  1 &  0 &  0 &  -2 & $T^{-3}+T^{-2}+T^{-1}+T+T^4+T^6+T^7+T^8$\\ 
\hline
-1 &  0 &  1 &  3 &  0 &  1 & $T^{-2}+T^{-1}+T+T^2+T^3+T^4+T^6+T^7$\\
\hline
-1 & 0 &  2 &  0 &  1 &  -2 & 
$2T^{-1}+T+T^2+T^3+T^4+2T^6$\\
\hline
-1 &  0 &  2 &  1 & 1 & -1 & $T^{-1}+2T+T^2+ T^3+2T^4+T^6$\\ 
\hline
0 &  0 &  0 &  1 &  -3 &  -2& $T^{-3}+2T^{-2}+T^{-1}+T^6+2T^7+T^8$\\
\hline
0 &  0 &  0 &  2 &  -3 &  -1 & $T^{-3}+2T^{-1}+T+T^4+2T^6+T^8$\\ 
\hline
0 &  0 &  0 &  3 &  -3 &  0 & $T^{-2}+T^{-1}+T+T^2+T^3+T^4+T^6+T^7$\\
\hline
0 &  0 &  1 & 2 &  -2 &  -1& $T^{-2}+T^{-1}+T+T^2+T^3+T^4+T^6+T^7$\\
\hline
0 &  0 &  1 &  3 &  -2 &  0& $T^{-1}+2T + T^2+T^3+2T^4+T^6$\\
\hline
0 &  0 &  2 &  3 &  -1 &  0& $T+3T^2+3T^3+T^4$\\
\hline
\end{tabular}
\caption{Type $I$ fixed points for $n=6$.}
\label{Table3}
\end{table}
\begin{table}
\centering
\begin{tabular}{|c|c|c|c|c|c|c|}
\hline
$\chi(L_1)$ & $\chi(L_2)$ & $l_1$ & $l_2$ & $a_1$ & $a_2$ & $T_{[\CE]}$ \\
\hline
0 & -2 & 0 & -2 &-5 & 1 & 
$T^{-7}+ T^{-3}+T^{-2}+T^{-1}+T^6+T^7+T^8+T^{12}$\\ 
\hline
0 & -2 & 1 &-2 & -4& 2 &  $T^{-3}+T^{-2}+T^{-1}+T+T^4+T^6+T^7+T^8$\\ 
\hline
-1& -3& 0 & -2& -2& 4 & $T^{-7}+ T^{-3}+T^{-2}+T^{-1}+T^6+T^7+T^8+T^{12}$\\
\hline
0 & -3 & 3& -2& -3 &3 & $T^{-3}+T^{-2}+T^{-1}+T+T^4+T^6+T^7+T^8$ \\
\hline
0 & -3 & 3& -3 & -4 &3 & $T^{-3}+T^{-2}+2T^{-1}+2T^6+T^7+T^8$\\
\hline
\end{tabular}
\caption{Type $II$ fixed points for $n=6$.}
\label{Table4}
\end{table}

\newpage
\appendix

\Appendix{Some technical results}\label{technostuff}

\setcounter{equation}{0}

\

\noindent
Several technical results needed in the paper are proven in this section. Using the notation of Section \ref{setup}, 
suppose the sections $(\xi_1, \ldots, \xi_\ell)$ 
satisfy the genericity condition \eqref{eq:genlocsectA}.
Let $z$ be a local 
coordinate on an open neighborhood $U$ of $p$ in $C$ 
centered at $p$. Then the sections 
 $\xi_i$, $1\leq i \leq \ell$, will have expansions 
\[ 
\xi_{i} = \sum_{k=1}^{n} \lambda_{i,k} z^{-k}dz, \qquad \lambda_{i,k} \in \IC, \quad 1\leq i\leq \ell, \quad 1\leq k \leq n. 
\]
The genericity condition \eqref{eq:genlocsectA} translates into 
\[ 
\lambda_{i,n}\neq 0, 
\quad 1\leq i \leq \ell, \quad 
{\rm and}\quad 
\lambda_{i,n} \neq \lambda_{j,n}, \quad 1\leq i,j \leq \ell, \ i\neq j. 
\]

Now let $(E,E_D^\bullet,\Phi)$ be a 
$\uxi$-parabolic irregular Higgs bundle as in Section 
\ref{setup}. 
Then the following two statements hold.

\subsection{Local trivialization}\label{loctriv}

 {\it There is a  trivialization $E_D \simeq \CO_D^r$ such that 
\[
\Phi_D = \sum_{k=1}^{n} \Lambda_{k} z^{-k} dz
\]
where $\Lambda_{k}$ is the diagonal matrix with entries 
\[
\underbrace{\lambda_{1,k}, \ldots, \lambda_{1,k}}_{m_1},  \ldots, 
\underbrace{\lambda_{2,k}, \ldots, \lambda_{2,k}}_{m_2},\ldots, 
\underbrace{ \lambda_{\ell, k}, \ldots, \lambda_{\ell,k}}_{m_\ell}.
\]}
 
This can be proven by an elementary argument. 
First note that there exists a trivialization $E_D\simeq \CO_D\otimes \IC^r$
such that the submodules $E_{D}^i$ are isomorphic to 
$\CO_D \otimes V_i$, where 
\[
0=V_0 \subset V^1 \subset \cdots \subset V^\ell = \IC^r
\]
is the standard flag of type $(m_1, \ldots, m_\ell)$ in $\IC^r$.
In particular there is a natural splitting 
\[
\IC^r \simeq \bigoplus_{i=1}^\ell V^{i}/V^{i-1}
\] 
which yields a splitting 
\[ 
E_D \simeq \bigoplus_{i=1}^\ell E_D^{i}/E_D^{i-1}.
\]
Using this trivialization, the restriction $\Phi_D$ of the Higgs field has a local expression of the form 
\[
\Phi_D=
{1\over z^n}\left[\begin{array}{ccccc} \lambda_1 I_{m_1} & \phi_{1,2} & \phi_{1,3} & \cdots & \phi_{1,\ell} \\ O & \lambda_2 I_{m_2} & \phi_{2,3} & \cdots & \phi_{2,n} \\ 
\vdots & \vdots & \vdots & \cdots & \vdots \\ 
O & O & O & \cdots & \lambda_\ell I_{m_\ell}\\
\end{array}\right]dz
\]
with 
\[
\lambda_i = \sum_{k=1}^{n} \lambda_{i,k} z^{n-k}. 
\]
Note that the right hand side is an $r\times r$ matrix with coefficients in the ring $R = \IC[z]/(z^n)$. 
In order to prove the statement it suffices to show the existence of an $r\times r$ invertible matrix $U$ with coefficients in $R$ such that 
\be\label{eq:diageqA}
U \Phi_D = z^{-n}\Lambda U dz
\ee
where $\Lambda$ is the diagonal matrix 
\[ 
\Lambda = \left[ \begin{array}{ccccc} \lambda_1 I_{m_1} & O & O & \cdots & O\\ O & \lambda_2 I_{m_2} & O & \cdots & O \\ 
\vdots & \vdots & \vdots & \cdots & \vdots \\ 
O & O & O & \cdots & \lambda_\ell I_{m_\ell}\\
\end{array}\right]. 
\]
In fact it will be shown below that one can find a block upper triangular matrix 
\[ 
U = \left[ \begin{array}{ccccc} I_{m_1} & U_{1,2} & U_{1,3} & \cdots & U_{1,\ell} \\ O & I_{m_2} & U_{2,3} & \cdots & U_{2,n} \\ 
\vdots & \vdots & \vdots & \cdots & \vdots \\ 
O & O & O & \cdots & I_{m_\ell}\\
\end{array}\right]. 
\]
satisfying the above matrix equation. 

Clearly, for $\ell=1$ there is nothing to prove, so one can assume $\ell\geq 2$. 
Then  equation \eqref{eq:diageqA} 
reduces to  an $R$-linear system of $\ell(\ell-1)/2$ equations labelled by pairs 
$(i,j)$ with $1\leq i < j \leq \ell$. For each such pair one 
obtains an equation of the form 
\be\label{eq:diageqB}
\phi_{i,j} + U_{i,i+1} \phi_{i+1,j} + \cdots + U_{i,j-1} \phi_{j-1,j} 
+ \lambda_j U_{i,j} = \lambda_i U_{i,j}. 
\ee
This system can be easily solved be induction using the difference $j-i$ to control the inductive step. Suppose one can solve for all $U_{i,j}$ with $j\leq i+k$ for some $1\leq k \leq \ell-1$. 
Then note that for $(i,j)=(i,k+1)$ equation \eqref{eq:diageqB} 
yields 
\[
(\lambda_{i}-\lambda_{k+1}) U_{i,k+1} = \phi_{i,k+1} + U_{i,i+1} \phi_{i+1,k+1} + \cdots + U_{i,k} \phi_{k,k+1}.
\]
Since $\lambda_i\neq \lambda_{i+1}$ by the genericity assumption, the difference 
$(\lambda_i - \lambda_{i+1})$ is invertible in $R$,
which proves the inductive step. 
To finish the proof note that for $(i,j)=(i,i+1)$, equation 
\eqref{eq:diageqB} specializes to 
\[
(\lambda_i - \lambda_{i+1}) U_{i,i+1} = \phi_{i,i+1}
\]
for any $1\leq i \leq \ell-1$. Again, since 
$(\lambda_i - \lambda_{i+1})$ is invertible in $R$, one can easily solve for $U_{i,i+1}$.

\subsection{Parabolic Higgs subsheaves}\label{parsub}

{\it  Let $F\subset E$  be nonzero proper saturated subsheaf preserved by $\Phi$. Then the submodules $F_{D}^i \subset F_D$ and the quotient sheaves $F_D/F_D^i$, $1\leq i \leq \ell$,  are 
locally free $\CO_D$-modules.}
\bigskip

The question is local, so one can choose a trivialization of 
$E_D$ as in Section \ref{loctriv}. Hence $\Phi_D$ is identified with a diagonal matrix $z^{-n}\Lambda$. By assumption, the quotient $E/F$ is nonzero and locally free. This implies that there is an exact sequence of $\CO_D$-modules 
\[
0\to F_D \to E_D \to E_D/F_D \to 0.
\]
Moreover, $F_D$ is locally free, hence isomorphic to a module 
of the form $\CO_D^{\oplus s}$, where $0<s = {\rm rk}(F)<r$. 
Therefore one obtains an injective morphism of 
$\CO_D$-modules 
\[
\psi: \CO_D^{\oplus s} \to \CO_D^{\oplus r} 
\]
such that ${\rm Coker}(\psi)$ is locally free. Recall also that by choice of trivialization there is a direct sum decomposition 
\be\label{eq:splittingB}
E_D\simeq \bigoplus_{i=1}^\ell E_D^i/E_D^{i-1} \simeq 
\bigoplus_{i=1}^\ell \CO_D^{\oplus m_i}.
\ee
Let $\psi_i:  \CO_D^{\oplus s} \to \CO_D^{\oplus m_i}$, $1\leq i \leq \ell$ be the components of $f$ with respect to this decomposition. Note that $F_D\cap E_D^i \simeq {\rm Im}(\psi_i)$ 
for all $1\leq i \leq \ell$. Therefore one needs to prove that  
 ${\rm Im}(\psi_i)$ is a locally free $\CO_D$-module, or equivalently, that each ${\rm Ker}(\psi_i)$ is locally free for 
each $1\leq i\leq \ell$.

Now recall that $F_D\subset E_D$ must be preserved by $\Phi_D$ which is identified with the diagonal matrix 
$z^{-n}\Lambda$.  Moreover, given the genericity assumption, 
$\Lambda$ is invertible in $\CM_r(R)$, the ring of $r\times r$ matrices with coefficients in $R$. This implies that there must exist an isomorphism $g: F_D \to F_D$ such that 
\[ 
\Phi_D \circ \psi= (\psi\otimes 1_{M_D}) \circ (g \otimes 1_{M_D}). 
\]
In order to write this equation in terms of local expressions, let 
\[
\Psi = (\psi_{j,a}) \in \CM_{r,s}(R)
\]
be the matrix of $\psi$ with respect to the chosen trivializations, where $1\leq j \leq r$ and $1\leq a\leq s$. 
Note that $\Psi$ can be also written in block form 
\[ 
\Psi=\left[ \begin{array}{c} \Psi_1\\ \vdots \\ \Psi_\ell
\end{array} \right]
\]
with respect to the direct sum decomposition \eqref{eq:splittingB}. Let also $G=(g_{a,b})\in \CM_{s}(R)$, $1\leq a,b\leq s$, denote the matrix of $g$, which must be invertible. Then one has
\[ 
\Lambda \Psi = \Psi G.
\]
Finally, since $\Psi$ is injective, $\Psi$ must have maximal rank 
which means that there exists a strictly increasing injective function 
$j:\{ 1, \ldots, s\} \to \{1,\ldots, r\}$ such that the square matrix 
\[ 
A = (\psi_{j(a),b}) \in \CM_{s}(R), \qquad 1\leq a,b\leq s,
\]
is invertible. 

Next note that there exists an ordered partition $s= k_1 + \cdots k_l$ with $k_i \in \IZ_{\geq 0}$ such that the following inequalities hold 
\[
\bal 
1 & \leq j(1) < \cdots < j(k_1) \leq m_1 \\
m_1+1 & \leq j(k_1+1) < \cdots < j(k_1+k_2) \leq m_2 \\
&{}\ \  \vdots \\
m_1+\cdots+m_{\ell-1}+1 & \leq j(k_1+\cdots + k_{\ell-1}+1) < 
\cdots < j(s) \leq r.
\eal
\]
Furthermore let $\Lambda_{(k_1, \ldots, k_\ell)}$ be 
the diagonal $s\times s$ matrix with entries 
\[ 
\underbrace{ \lambda_1, \ldots, \lambda_1}_{k_1}, \ldots, 
\underbrace{ \lambda_\ell, \ldots, \lambda_\ell}_{k_\ell},
 \]
This matrix can be written in block diagonal form 
\[
\Lambda_{(k_1, \ldots, k_\ell)} = \left[ \begin{array}{cccc} 
\lambda_1 I_{k_1} & O & \cdots & O \\
O & \lambda_2 I_{k_2} & \cdots & O\\
\vdots & \vdots & \vdots \\ 
O & O & \cdots & \lambda_\ell I_{k_\ell}\\
\end{array}\right]. 
\]
By convention such a block is empty if $k_i=0$. 
Then note that the matrix relation $\Lambda \Psi = \Psi G$ implies 
\[
\Lambda_{(k_1, \ldots, k_\ell)} A = AG.
\]
Since $A$ is invertible, this further yields 
\[
G = A^{-1} \Lambda_{(k_1, \ldots, k_\ell)} A.
\]
Next note that relation $\Lambda \Psi = \Psi G$ also implies 
\[
\lambda_i \Psi_i = \Psi_i G 
\]
for each $1\leq i \leq \ell$. The last two relations yield in turn
\[
\Psi_i A^{-1}\left(\lambda_i I_s - \Lambda_{(k_1, \ldots, k_\ell)} 
\right) = 0.
\]
Finally, note that the matrix $\lambda_i I_s - \Lambda_{(k_1, \ldots, k_\ell)}$ has block form 
\[ 
\left[ \begin{array}{cccc} 
(\lambda_i -\lambda_1) I_{k_1} & O & \cdots & O \\
O & \lambda_2 I_{k_2} & \cdots & O\\
\vdots & \vdots & \vdots \\ 
O & O & \cdots & (\lambda_i - \lambda_\ell) I_{k_\ell}\\
\end{array}\right]. 
\]
with respect to the partition $s= k_1 +\cdots + k_\ell$. 
In particular the $i$-th diagonal block is zero. Using the ordered 
direct sum decomposition 
\[ 
 \CO_D^{\oplus s} \simeq \bigoplus_{i=1}^\ell 
\CO_D^{k_i}
\]
this implies that for each $1\leq i \leq \ell$ the submodule 
\[
\CO_D^{\oplus k_1}\oplus \cdots \oplus \CO_D^{\oplus k_{i-1}} \oplus 
\CO_D^{\oplus k_{i+1}} \oplus \cdots \oplus \CO_{D}^{\oplus k_\ell} 
\]
is contained in ${\rm Ker}(\psi_i)$. 
In order to conclude the proof it will be shown below that 
${\rm Ker}(\psi_i) \cap \CO_D^{\oplus k_i} = 0$ hence 
\[
{\rm Ker}(\psi_i) = \CO_D^{\oplus k_1}\oplus \cdots \oplus \CO_D^{\oplus k_{i-1}} \oplus 
\CO_D^{\oplus k_{i+1}} \oplus \cdots \oplus \CO_{D}^{k_\ell} 
\]
is a free $\CO_D$-module. 
Note that by restriction to the $i$-th direct summand $\CO_D^{\oplus k_i}\subset \CO_D^{\oplus s}$, the matrix 
$\Psi_i A^{-1}$ determines a morphism 
\[
\gamma_i:\CO_D^{\oplus k_i}\to \CO_D^{\oplus m_i}\subset \CO_D^{\oplus r} 
\]
Moreover, by construction, 
\[
\gamma_i(u_a) = v_{j(k_1+\cdots+k_{i-1}+a)}
\]
for all $1\leq a\leq k_i$, where $u_a$, $1\leq a\leq k_i$, and  $v_j$, $1\leq j \leq r$ are the standard generators 
of $\CO_D^{\oplus k_i}$, $\CO_D^{\oplus r}$ respectively. 
Therefore ${\rm Ker}(\psi_i) \cap \CO_D^{\oplus k_i}$ is indeed 
zero. 

In conclusion, the above local computation shows that the kernel of the composition of morphisms 
\[ 
F_D^i \hookrightarrow E_D^i \twoheadrightarrow E_D^i/E_D^{i-1} 
\]
is a locally free $\CO_D$-module for each $1\leq i \leq \ell$. 
This implies that its image, $F_D^i/F_D^{i-1}$ is also locally free 
since it is a $\CO_D$-module with a finite locally free resolution. Using the exact sequences 
\[ 
0 \to F_D^i/F_D^{i-1} \to F_D/F_D^{i-1} \to F_D/F_D^{i} \to 0
\]
it follows inductively that the quotients $F_D/F_D^i$ are locally free for all $1\leq i \leq \ell$. 

\subsection{A pushforward result}\label{pwd}
The third technical result proven next is needed for spectral construction in Section \ref{sheavestohiggs}.
 Recall that $T_\uxi$ denotes the surface constructed in Section \ref{surface} by blowing up the total space of the coefficient line bundle $M$ along the images of the sections $\xi_1, \ldots, \xi_\ell$ over $D=np$. The holomorphic 
symplectic surface $S_\uxi$ is the complement of an anticanonical divisor on $T_\uxi$ which includes all irreducible exceptional divisors of the blow-up except the last ones, $\Xi_{n,i}$, $1\leq i \leq n$. The natural projection from $S_\uxi$ to $C$ is denoted by 
$\pi_{S}:S_\uxi\to C$. The statement needed in Section \ref{sheavestohiggs} is the following.
\bigskip

{\it Let $F$ be a pure dimension one sheaf 
on $T_\uxi$ with compact support contained in $S_\uxi$. 
Then $\pi_{S*}(F\otimes_{S_\uxi} 
\CO_{k\Xi_{n,i}})$ is a locally free $\CO_{kp}$-module for all $1\leq i \leq \ell$ and for any positive integer $k\geq 1$.}
\bigskip 

Since $F$ is pure of dimension one, its scheme theoretic support, $\Xi_F$ is a divisor on $T_\uxi$, which is, by assumption compact. 
In particular $\Xi_F$ cannot have a component along $\Xi_{n,i}$, since the intersection $\Xi_{n,i}\cap S_\uxi$ is a non-compact divisor on $S_\uxi$. 
Hence $F\otimes_{S_\uxi} 
\CO_{k\Xi_{n,i}}$ has zero dimensional support. 
Let $(x,y)$ be affine coordinates 
in an open neighborhood of $\Xi_{n,i}$ in $S_\uxi$ such that 
$\Xi_{n,i}$ is given by $x=0$.  Let  $f(x,y)=0$ the 
defining equation of $\Xi_F$ 
in the chosen open neighborhood of $\Xi_{n,i}$. Since $\Xi_F$ does not have a component along $\Xi_{n,i}$, it follows that
$f(x,y)$ is not a multiple of $x$. 
The scheme theoretic intersection between the divisor 
$k \Xi_{n,i}$ and $\Xi_F$ is isomorphic to the spectrum of the ring 
\[
B = \IC[x,y]/(f) \otimes_{\IC[x,y]} \IC[x,y]/(x^k). 
\]
Let also $A= \IC[x]/(x^k)$ and note that there are isomorphisms
\[
B \simeq \IC[x,y]/(x^k, f) \simeq A[y]/({\bar f}) 
\]
where ${\bar f} \in A[y]$ is the image of $f$ under the natural 
projection. 

Next, let $\Gamma_F$ be the space of sections of F over the chosen open subset, which is a 
$\IC[x,y]/(f)$-module as well as a $\IC[x,y]$-module. 
Then the space of sections 
of $F\otimes_{S_\uxi} \CO_{k \Xi_{n,i}}$ is 
\[
\Upsilon_F=\Gamma_F \otimes_{\IC[x,y]} \IC[x,y]/(x^k) \simeq \Gamma_F/x^k\Gamma_F.
\]
Since $F\otimes_{S_\uxi} \CO_{k \Xi_{n,i}}$ is scheme theoretically supported on ${\rm Spec}(B)$, this is a $B$-module, hence also an $A$-module. In order to prove the above statement it suffices to prove that $\Upsilon_F$ is isomorphic to a free $A$-module.

Recall that any finitely generated $A$-module is isomorphic to a 
direct sum of free and cyclic $A$-modules. Moreover any cyclic $A$-module $A/(x^l)$, $1\leq l \leq k-1$ has an infinite free resolution of the form 
\be\label{eq:freeresA}
\cdots  {\buildrel x^l\over \longto } A {\buildrel x^{k-l}\over \longto} A 
{\buildrel x^l\over \longto } 
A   {\buildrel x^{k-l}\over \longto} \cdots  {\buildrel x^{k-l}\over \longto}A {\buildrel x^l\over \longto} A
\ee
and any $A$-module $M_A$ is isomorphic to a free one if and only if 
\[ 
{\rm Tor}_j^A(A/(x^l), M_A)=0
\]
for all $j\geq 1$, $1\leq l\leq k-1$.

Taking a tensor product of the complex 
\eqref{eq:freeresA} with $\Upsilon_F$ over $A$ yields a 
complex of $A$-modules 
\[
\cdots  {\buildrel x^{k-l}\over \longto}  \Gamma_F/x^k\Gamma_F
{\buildrel x^l\over \longto } 
 \Gamma_F/x^k\Gamma_F {\buildrel x^{k-l}\over \longto} \cdots  {\buildrel x^{k-l}\over \longto}
\Gamma_F/x^k\Gamma_F {\buildrel x^l\over \longto} \Gamma_F/x^k\Gamma_F.
\]
Since $F$ is pure of dimension one, and its support does not a have a component along $\Xi_{n,i}$,  the annihilator 
${\rm Ann}(x^l)\subset \Gamma_F$ must be zero for all $1\leq l \leq k-1$. Otherwise 
$F$ would have a nontrivial zero dimensional subsheaf. This implies easily that the above complex of $A$-modules is exact, hence 
\[ 
{\rm Tor}^A_j(A/(x^l), \Upsilon_F) = 0 
\]
 for all $1\leq l\leq k-1$ and all $j\geq 1$.  
Therefore $\Upsilon_F$ is isomorphic to a free $A$-module, proving the claim. 
\bigskip

\Appendix{Examples}\label{examples}
\setcounter{equation}{0}
\smallskip

\noindent
Several polynomials $P_{\mu,n}(u,v)$ obtained from equation 
\eqref{eq:refBPSa} are listed 
below for some examples. In all regular examples the results are in agreement with the formula of Hausel., Mereb and Wong 
\eqref{eq:HMWb}.

\subsection{Regular rank two examples}\label{rktwo}

\begin{itemize}
\item $P_{(1^2),3}(u,v)=1$. 
\item $P_{(1^2),4}(u,v)=u^2v^2+uv^2+1$. 
\item $P_{(1^2),5}(u,v)=u^4v^4+u^3v^4+u^2v^4+u^2v^2+uv^2+1$.
\item  $P_{(1^2),6}(u,v)=u^6v^6+u^5v^6+u^4v^6+u^3v^6+u^4v^4+u^3v^4+u^2v^4+
u^2v^2+uv^2+1$.
\item $P_{(1^2),7}(u,v)= {u}^{8}{v}^{8}+{u}^{7}{v}^{8}+{u}^{6}{v}^{8}+{u}^{5}{v}^{8}+{u}^{6}{v}^{6}+{u}^{4}{v}^{8}+{u}^{5}{v}^{6}+{u}^{4}{v}^{6}+{u}^{3}{v}^{6}+{u}^{4}{v}^{4}+{u}^{3}{v}^{4}+{u}^{2}{v}^{4}
\mbox{}+{u}^{2}{v}^{2}+  u {v}^{2}+1$. 
\end{itemize}

\subsection{Regular rank three examples}\label{rkthreeA}

\begin{itemize}
\item $P_{(1^3),3}(u,v)=u^2v^2+2\, uv^2+1$.
\item $P_{(1^3),4}(u,v)={u}^{8}{v}^{8}+2\,{u}^{7}{v}^{8}+3\,{u}^{6}{v}^{8}+4\,{u}^{5}{v}^{8}+{u}^{6}{v}^{6}+2\,{u}^{4}{v}^{8}+3\,{u}^{5}{v}^{6}+5\,{u}^{4}{v}^{6}+4\,{u}^{3}{v}^{6}+{u}^{4}{v}^{4}+3\,{u}^{3}{v}^{4}
\mbox{}+3\,{u}^{2}{v}^{4}+{u}^{2}{v}^{2}+2\, u {v}^{2}+1$.
\item $P_{(1^3),5}(u,v)=  {u}^{14}{v}^{14}+2\,{u}^{13}{v}^{14}+3\,{u}^{12}{v}^{14}+4\,{u}^{11}{v}^{14}+{u}^{12}{v}^{12}+5\,{u}^{10}{v}^{14}+3\,{u}^{11}{v}^{12}\\
\mbox{}+6\,{u}^{9}{v}^{14}+5\,{u}^{10}{v}^{12}+4\,{u}^{8}{v}^{14}+7\,{u}^{9}{v}^{12}+2\,{u}^{7}{v}^{14}+{u}^{10}{v}^{10}+9\,{u}^{8}{v}^{12}+3\,{u}^{9}{v}^{10}\\
\mbox{}+8\,{u}^{7}{v}^{12}+6\,{u}^{8}{v}^{10}+4\,{u}^{6}{v}^{12}+9\,{u}^{7}{v}^{10}+{u}^{8}{v}^{8}+9\,{u}^{6}{v}^{10}+3\,{u}^{7}{v}^{8}+6\,{u}^{5}{v}^{10}+6\,{u}^{6}{v}^{8}\\
\mbox{}+7\,{u}^{5}{v}^{8}+{u}^{6}{v}^{6}+5\,{u}^{4}{v}^{8}+3\,{u}^{5}{v}^{6}+5\,{u}^{4}{v}^{6}+4\,{u}^{3}{v}^{6}+{u}^{4}{v}^{4}+3\,{u}^{3}{v}^{4}+3\,{u}^{2}{v}^{4}+{u}^{2}{v}^{2}+2\,  u  {v}^{2}\\
\mbox{}+1$.
\end{itemize}

\subsection{Rank three examples with $(m_1,m_2)=(2,1)$}\label{rkthreeB}
\begin{itemize}
\item $P_{(2,1),5}(u,v) = {u}^{4}{v}^{4}+{u}^{3}{v}^{4}+{u}^{2}{v}^{4}+{u}^{2}{v}^{2}+ u {v}^{2}+1$. 
\item $P_{(2,1),6}(u,v) = {u}^{8}{v}^{8}+{u}^{7}{v}^{8}+2\,{u}^{6}{v}^{8}+{u}^{5}{v}^{8}+{u}^{6}{v}^{6}+{u}^{4}{v}^{8}+2\,{u}^{5}{v}^{6}+2\,{u}^{4}{v}^{6}+{u}^{3}{v}^{6}+{u}^{4}{v}^{4}+2\,{u}^{3}{v}^{4}
+2\,{u}^{2}{v}^{4}+{u}^{2}{v}^{2}+ u {v}^{2}+1$. 
\item $P_{(2,1),7}(u,v) = {u}^{12}{v}^{12}+{u}^{11}{v}^{12}+2\,{u}^{10}{v}^{12}+2\,{u}^{9}{v}^{12}+{u}^{10}{v}^{10}+2\,{u}^{8}{v}^{12}+2\,{u}^{9}{v}^{10}+{u}^{7}{v}^{12}\\
\mbox{}+3\,{u}^{8}{v}^{10}+{u}^{6}{v}^{12}+3\,{u}^{7}{v}^{10}+{u}^{8}{v}^{8}+2\,{u}^{6}{v}^{10}+2\,{u}^{7}{v}^{8}+{u}^{5}{v}^{10}+4\,{u}^{6}{v}^{8}+3\,{u}^{5}{v}^{8}+{u}^{6}{v}^{6}\\
\mbox{}+2\,{u}^{4}{v}^{8}+2\,{u}^{5}{v}^{6}+3\,{u}^{4}{v}^{6}+2\,{u}^{3}{v}^{6}+{u}^{4}{v}^{4}+2\,{u}^{3}{v}^{4}+2\,{u}^{2}{v}^{4}+{u}^{2}{v}^{2}+ u {v}^{2}+1$.
\end{itemize}

\bibliography{FM_ref.bib}

\begin{thebibliography}{10}

\bibitem{refCS}
M.~Aganagic and S.~Shakirov.
\newblock {Knot Homology and Refined Chern-Simons Index}.
\newblock {\em Commun. Math. Phys.}, 333(1):187--228, 2015.

\bibitem{Md_vert_I}
H.~Awata and H.~Kanno.
\newblock {Instanton counting, Macdonald functions and the moduli space of
  D-branes}.
\newblock {\em JHEP}, 05:039, 2005.

\bibitem{Md_vert_II}
H.~Awata and H.~Kanno.
\newblock {Refined BPS state counting from Nekrasov's formula and Macdonald
  functions}.
\newblock {\em Int. J. Mod. Phys.}, A24:2253--2306, 2009.

\bibitem{Wild_curves}
O.~Biquard and P.~Boalch.
\newblock Wild non-abelian {H}odge theory on curves.
\newblock {\em Compos. Math.}, 140(1):179--204, 2004.

\bibitem{Wild_merom}
P.~{Boalch}.
\newblock {Wild character varieties, meromorphic Hitchin systems and Dynkin
  diagrams}.
\newblock {arXiv:1703.10376}.

\bibitem{Quasi_hamiltonian}
P.~Boalch.
\newblock Quasi-{H}amiltonian geometry of meromorphic connections.
\newblock {\em Duke Math. J.}, 139(2):369--405, 2007.

\bibitem{HK_wild}
P.~{Boalch}.
\newblock {Hyperkahler manifolds and nonabelian Hodge theory of (irregular)
  curves}.
\newblock {\em ArXiv e-prints}, Mar. 2012.
\newblock Arxiv:1203.6607.

\bibitem{Poisson_Riemann}
P.~Boalch.
\newblock Poisson varieties from {R}iemann surfaces.
\newblock {\em Indag. Math. (N.S.)}, 25(5):872--900, 2014.

\bibitem{Braiding_stokes}
P.~P. Boalch.
\newblock Geometry and braiding of {S}tokes data; fission and wild character
  varieties.
\newblock {\em Ann. of Math. (2)}, 179(1):301--365, 2014.

\bibitem{Counting_Hitchin_II}
P.-H. Chaudouard.
\newblock Sur le comptage des fibr\'es de {H}itchin.
\newblock {\em Ast\'erisque}, (369):223--284, 2015.

\bibitem{Counting_Hitchin_I}
P.-H. Chaudouard and G.~Laumon.
\newblock Sur le comptage des fibr\'es de {H}itchin nilpotents.
\newblock {\em J. Inst. Math. Jussieu}, 15(1):91--164, 2016.

\bibitem{DAHA_links}
I.~{Cherednik} and I.~{Danilenko}.
\newblock {DAHA approach to iterated torus links}.
\newblock arXiv:1509.08351.

\bibitem{CKK}
J.~Choi, S.~Katz, and A.~Klemm.
\newblock {The refined BPS index from stable pair invariants}.
\newblock 2012.
\newblock arXiv:1210.4403.

\bibitem{Par_ref}
W.-y. Chuang, D.-E. Diaconescu, R.~Donagi, and T.~Pantev.
\newblock {Parabolic refined invariants and Macdonald polynomials}.
\newblock {\em Commun. Math. Phys.}, 335(3):1323--1379, 2015.
\newblock arXiv:1311.3624.

\bibitem{wallpairs}
W.-y. Chuang, D.-E. Diaconescu, and G.~Pan.
\newblock {Wallcrossing and Cohomology of The Moduli Space of Hitchin Pairs}.
\newblock {\em Commun.Num.Theor.Phys.}, 5:1--56, 2011.

\bibitem{BPSPW}
W.-Y. Chuang, D.-E. Diaconescu, and G.~Pan.
\newblock B{PS} states and the {$P=W$} conjecture.
\newblock In {\em Moduli spaces}, volume 411 of {\em London Math. Soc. Lecture
  Note Ser.}, pages 132--150. Cambridge Univ. Press, Cambridge, 2014.

\bibitem{hodgechar}
M.~A.~A. de~Cataldo, T.~Hausel, and L.~Migliorini.
\newblock Topology of {H}itchin systems and {H}odge theory of character
  varieties: the case {$A_1$}.
\newblock {\em Ann. of Math. (2)}, 175(3):1329--1407, 2012.

\bibitem{Hodge_maps}
M.~A.~A. de~Cataldo and L.~Migliorini.
\newblock The {H}odge theory of algebraic maps.
\newblock {\em Ann. Sci. \'Ecole Norm. Sup. (4)}, 38(5):693--750, 2005.

\bibitem{GV_details}
M.~Dedushenko and E.~Witten.
\newblock Some details on the {G}opakumar-{V}afa and {O}oguri-{V}afa formulas.
\newblock {\em Adv. Theor. Math. Phys.}, 20(1):1--133, 2016.

\bibitem{TH_II}
P.~Deligne.
\newblock Th\'eorie de {H}odge. {II}.
\newblock {\em Inst. Hautes \'Etudes Sci. Publ. Math.}, (40):5--57, 1971.

\bibitem{TH_III}
P.~Deligne.
\newblock Th\'eorie de {H}odge. {III}.
\newblock {\em Inst. Hautes \'Etudes Sci. Publ. Math.}, (44):5--77, 1974.

\bibitem{modADHM}
D.~E. Diaconescu.
\newblock Moduli of {A}{D}{H}{M} sheaves and local {D}onaldson-{T}homas
  {T}heory.
\newblock {\em J. Geom. Phys.}, (62):763--799, 2012.

\bibitem{DHS}
D.-E. Diaconescu, Z.~Hua, and Y.~Soibelman.
\newblock {HOMFLY polynomials, stable pairs and motivic Donaldson-Thomas
  invariants}.
\newblock {\em Commun. Num. Theor. Phys.}, 6:517--600, 2012.

\bibitem{DSV}
D.~E. Diaconescu, V.~Shende, and C.~Vafa.
\newblock {Large N duality, lagrangian cycles, and algebraic knots}.
\newblock {\em Commun. Math. Phys.}, 319:813--863, 2013.

\bibitem{Mtop}
R.~Dijkgraaf, C.~Vafa, and E.~Verlinde.
\newblock {M-theory and a topological string duality}.
\newblock 2006.
\newblock hep-th/0602087.

\bibitem{Ind_obj}
G.~{Dobrovolska}, V.~{Ginzburg}, and R.~{Travkin}.
\newblock {Moduli spaces, indecomposable objects and potentials over a finite
  field}.
\newblock ArXiv:1612.01733.

\bibitem{Dir_images}
R.~{Donagi}, T.~{Pantev}, and C.~{Simpson}.
\newblock {Direct Images in Non Abelian Hodge Theory}.
\newblock arXiv:1612.06388.

\bibitem{Super_evol}
P.~Dunin-Barkowski, A.~Mironov, A.~Morozov, A.~Sleptsov, and A.~Smirnov.
\newblock Superpolynomials for torus knots from evolution induced by
  cut-and-join operators.
\newblock {\em J. High Energy Phys.}, (3):021, front matter+85, 2013.

\bibitem{Mot_conn_higgs}
R.~Fedorov, A.~Soibelman, and Y.~Soibelman.
\newblock {Motivic classes of moduli of Higgs bundles and moduli of bundles
  with connections}.
\newblock to appear.

\bibitem{Volume_refined}
H.~Fuji, S.~Gukov, P.~Su\l~kowski, and H.~Awata.
\newblock Volume conjecture: refined and categorified.
\newblock {\em Adv. Theor. Math. Phys.}, 16(6):1669--1777, 2012.

\bibitem{y_genus_higgs}
O.~Garc\'\i~a Prada and J.~Heinloth.
\newblock The {$y$}-genus of the moduli space of {${\rm PGL}_n$}-{H}iggs
  bundles on a curve (for degree coprime to {$n$}).
\newblock {\em Duke Math. J.}, 162(14):2731--2749, 2013.

\bibitem{Mot_chains}
O.~Garc\'\i~a Prada, J.~Heinloth, and A.~Schmitt.
\newblock On the motives of moduli of chains and {H}iggs bundles.
\newblock {\em J. Eur. Math. Soc. (JEMS)}, 16(12):2617--2668, 2014.

\bibitem{rankthreepar}
O.~Garc{\'{\i}}a-Prada, P.~B. Gothen, and V.~Mu{\~n}oz.
\newblock Betti numbers of the moduli space of rank 3 parabolic {H}iggs
  bundles.
\newblock {\em Mem. Amer. Math. Soc.}, 187(879):viii+80, 2007.

\bibitem{GV_II}
R.~Gopakumar and C.~Vafa.
\newblock {{M} theory and topological strings {I}{I}}.
\newblock arXiv:9812127.

\bibitem{Ref_knots_Hilb}
E.~Gorsky and A.~Negut.
\newblock Refined knot invariants and {H}ilbert schemes.
\newblock {\em J. Math. Pures Appl. (9)}, 104(3):403--435, 2015.

\bibitem{Link_ref_vert}
S.~Gukov, A.~Iqbal, C.~Kozcaz, and C.~Vafa.
\newblock {Link homologies and the refined topological vertex}.
\newblock {\em Commun. Math. Phys.}, 298:757--785, 2010.

\bibitem{Seq_BPS}
S.~Gukov, S.~Nawata, I.~Saberi, M.~Sto\v~si\'c, and P.~Su\l~kowski.
\newblock Sequencing {BPS} spectra.
\newblock {\em J. High Energy Phys.}, (3):004, front matter+160, 2016.

\bibitem{HLRV}
T.~Hausel, E.~Letellier, and F.~Rodriguez-Villegas.
\newblock Arithmetic harmonic analysis on character and quiver varieties.
\newblock {\em Duke Math. J.}, 160(2):323--400, 2011.

\bibitem{Arithmetic_wild}
T.~Hausel, M.~Mereb, and M.~L. Wong.
\newblock {Arithmetic and representation theory of wild character varieties}.
\newblock {\em ArXiv e-prints}.
\newblock Arxiv:1604.03382.

\bibitem{HRV}
T.~Hausel and F.~Rodriguez-Villegas.
\newblock Mixed {H}odge polynomials of character varieties.
\newblock {\em Invent. Math.}, 174(3):555--624, 2008.
\newblock With an appendix by Nicholas M. Katz.

\bibitem{HST}
S.~Hosono, M.-H. Saito, and A.~Takahashi.
\newblock Relative {L}efschetz action and {BPS} state counting.
\newblock {\em Internat. Math. Res. Notices}, (15):783--816, 2001.

\bibitem{Moduli_connections_III}
M.-A. Inaba.
\newblock Moduli of parabolic connections on curves and the {R}iemann-{H}ilbert
  correspondence.
\newblock {\em J. Algebraic Geom.}, 22(3):407--480, 2013.

\bibitem{Moduli_connections_IV}
M.-a. Inaba and M.-H. Saito.
\newblock Moduli of unramified irregular singular parabolic connections on a
  smooth projective curve.
\newblock {\em Kyoto J. Math.}, 53(2):433--482, 2013.

\bibitem{Hopf_revisited}
A.~Iqbal and C.~Kozcaz.
\newblock {Refined Hopf Link Revisited}.
\newblock {\em JHEP}, 04:046, 2012.

\bibitem{Iqbal:2008ra}
A.~Iqbal, C.~Kozcaz, and K.~Shabbir.
\newblock {Refined Topological Vertex, Cylindric Partitions and the U(1)
  Adjoint Theory}.
\newblock {\em Nucl. Phys.}, B838:422--457, 2010.

\bibitem{ref_vert}
A.~Iqbal, C.~Kozcaz, and C.~Vafa.
\newblock {The refined topological vertex}.
\newblock {\em JHEP}, 10:069, 2009.

\bibitem{Non_arch_sheaves}
Y.~{Jiang}.
\newblock {The moduli space of stable coherent sheaves via non-archimedean
  geometry}.
\newblock arXiv:1703.00497.

\bibitem{Ref_N_torus}
M.~{Kameyama} and S.~{Nawata}.
\newblock {Refined large N duality for torus knots}.
\newblock arXiv:1703.05408.

\bibitem{Open_GW}
S.~Katz and C.-C.~M. Liu.
\newblock Enumerative geometry of stable maps with {L}agrangian boundary
  conditions and multiple covers of the disc.
\newblock {\em Adv. Theor. Math. Phys.}, 5(1):1--49, 2001.

\bibitem{spinBH}
S.~H. Katz, A.~Klemm, and C.~Vafa.
\newblock {M-theory, topological strings and spinning black holes}.
\newblock {\em Adv. Theor. Math. Phys.}, 3:1445--1537, 1999.

\bibitem{KKV}
S.~H. Katz, A.~Klemm, and C.~Vafa.
\newblock {M theory, topological strings and spinning black holes}.
\newblock {\em Adv.Theor.Math.Phys.}, 3:1445--1537, 1999.

\bibitem{wallcrossing}
M.~Kontsevich and Y.~Soibelman.
\newblock Stability structures, {D}onaldson-{T}homas invariants and cluster
  transformations.
\newblock arXiv.org:0811.2435.

\bibitem{structures}
M.~Kontsevich and Y.~Soibelman.
\newblock Wall-crossing structures in {D}onaldson-{T}homas invariants,
  integrable systems and mirror symmetry.
\newblock In {\em Homological mirror symmetry and tropical geometry}, volume~15
  of {\em Lect. Notes Unione Mat. Ital.}, pages 197--308. Springer, Cham, 2014.

\bibitem{Higgs_ind_proj_line}
E.~{Letellier}.
\newblock {Higgs bundles and indecomposable parabolic bundles over the
  projective line}.
\newblock arXiv:1609.04875.

\bibitem{Zariski_closures}
E.~Letellier.
\newblock Character varieties with {Z}ariski closures of {${GL}_n$}-conjugacy
  classes at punctures.
\newblock {\em Selecta Math. (N.S.)}, 21(1):293--344, 2015.

\bibitem{Hecke_colored}
X.-S. Lin and H.~Zheng.
\newblock On the {H}ecke algebras and the colored {HOMFLY} polynomial.
\newblock {\em Trans. Amer. Math. Soc.}, 362(1):1--18, 2010.

\bibitem{Moduli_parabolic}
M.~Maruyama and K.~Yokogawa.
\newblock Moduli of parabolic stable sheaves.
\newblock {\em Math. Ann.}, 293(1):77--99, 1992.

\bibitem{HRV_proof}
D.~{Maulik}.
\newblock {Refined stable pair invariants for local curves}.
\newblock to appear.

\bibitem{Homfly_pairs}
D.~Maulik.
\newblock Stable pairs and the {HOMFLY} polynomial.
\newblock {\em Invent. Math.}, 204(3):787--831, 2016.

\bibitem{GV_vanishing}
D.~Maulik and Y.~Toda.
\newblock {Gopakumar-Vafa invariants via vanishing cycles}.
\newblock 2016.
\newblock arXiv:1610.07303.

\bibitem{Wild_harmonic}
T.~Mochizuki.
\newblock Wild harmonic bundles and wild pure twistor {$D$}-modules.
\newblock {\em Ast\'erisque}, (340):x+607, 2011.

\bibitem{Counting_Higgs}
S.~{Mozgovoy} and O.~{Schiffmann}.
\newblock {Counting Higgs bundles}.
\newblock ArXiv:1411.2101.

\bibitem{Membranes_Sheaves}
N.~Nekrasov and A.~Okounkov.
\newblock {Membranes and Sheaves}.
\newblock 2014.
\newblock ArXiv:1404.2323.

\bibitem{ORS}
A.~{Oblomkov}, J.~{Rasmussen}, and V.~{Shende}.
\newblock {The Hilbert scheme of a plane curve singularity and the HOMFLY
  homology of its link}.
\newblock arXiv:1201.2115.

\bibitem{OS}
A.~Oblomkov and V.~Shende.
\newblock The {H}ilbert scheme of a plane curve singularity and the {HOMFLY}
  polynomial of its link.
\newblock {\em Duke Math. J.}, 161(7):1277--1303, 2012.

\bibitem{Coh_ring}
A.~{Oblomkov} and Z.~{Yun}.
\newblock {The cohomological ring of a certain compactified Jacobian}.
\newblock {to appear}.

\bibitem{Geom_reps}
A.~{Oblomkov} and Z.~{Yun}.
\newblock {Geometric representations of graded and rational Cherednik
  algebras}.
\newblock July 2014.
\newblock arXiv: 1407.5685.

\bibitem{stabpairsI}
R.~Pandharipande and R.~P. Thomas.
\newblock Curve counting via stable pairs in the derived category.
\newblock {\em Invent. Math.}, 178(2):407--447, 2009.

\bibitem{Harmonic_metrics}
C.~Sabbah.
\newblock Harmonic metrics and connections with irregular singularities.
\newblock {\em Ann. Inst. Fourier}, 49(4):1265--1291, 1999.

\bibitem{Ind_vb_higgs}
O.~Schiffmann.
\newblock Indecomposable vector bundles and stable {H}iggs bundles over smooth
  projective curves.
\newblock {\em Ann. of Math. (2)}, 183(1):297--362, 2016.

\bibitem{Colored_HL}
S.~Shakirov.
\newblock {Colored knot amplitudes and Hall-Littlewood polynomials}.
\newblock 2013.

\bibitem{Cluster_legendrian}
V.~{Shende}, D.~{Treumann}, H.~{Williams}, and E.~{Zaslow}.
\newblock {Cluster varieties from Legendrian knots}.
\newblock arXiv:1512.08942.

\bibitem{Fukaya_knots}
V.~Shende, D.~Treumann, and E.~Zaslow.
\newblock {Legendrian knots and constructible sheaves}.
\newblock {\em ArXiv e-prints}.
\newblock Arxiv:1402.0490.

\bibitem{Par_DS}
A.~{Soibelman}.
\newblock {The moduli stack of parabolic bundles over the projective line,
  quiver representations, and the Deligne-Simpson problem}.
\newblock arXiv:1310.1144.

\bibitem{Stokes}
G.~G. Stokes.
\newblock On the discontinuity of arbitrary constants that appear as
  multipliers of semi-convergent series.
\newblock {\em Acta Math.}, 26(1):393--397, 1902.
\newblock A letter to the editor.

\bibitem{Birational_irreg}
S.~{Szab{\'o}}.
\newblock {The birational geometry of irregular Higgs bundles}.
\newblock arXiv:1502.02003.

\bibitem{GL_wild}
E.~Witten.
\newblock Gauge theory and wild ramification.
\newblock {\em Anal. Appl. (Singap.)}, 6(4):429--501, 2008.

\bibitem{infparhiggs}
K.~Yokogawa.
\newblock Infinitesimal deformation of parabolic {H}iggs sheaves.
\newblock {\em Internat. J. Math.}, 6(1):125--148, 1995.

\end{thebibliography}
 \bibliographystyle{abbrv}

\

\

\bigskip

\noindent 
Duiliu-Emanuel 
Diaconescu, {\sf
 NHETC, Rutgers University, 
 126 Frelinghuysen Road, Piscataway NJ 08854, USA},
 duiliu@physics.rutgers.edu

 \

 \

\noindent
Ron Donagi, {\sf Department of Mathematics, University of
  Pennsylvania, David Rittenhouse 
Laboratory, 209 South 33rd Street,
  Philadelphia, PA 19104, USA}, donagi@math.upenn.edu

\

\

\smallskip

\noindent
Tony Pantev, {\sf Department of Mathematics, University of
  Pennsylvania, David Rittenhouse 
Laboratory, 209 South 33rd Street,
  Philadelphia, PA 19104, USA}, tpantev@math.upenn.edu

\end{document}